\documentclass{jfm}
\usepackage{latexsym}
\usepackage{natbib} 
\usepackage{graphicx} 
\usepackage{epsfig} 
\usepackage{amssymb} 
\usepackage{amsmath} 
\usepackage{epstopdf} 
\usepackage{xcolor} 
\usepackage{url} 

\newcommand{\pd}[1]{\partial_{#1}}
\newcommand{\dd}[1]{\,\text{d}{#1}}

\newcommand{\Hilbert}[1]{\mathcal{H}\!\left[ #1 \right]}

\newcommand{\btropic}[1]{\!\left<#1\right>}

\newcommand{\Order}[1]{\mathcal{O}( #1 )}

\newcommand{\eps}{\varepsilon}
\newcommand{\Ek}{\textit{E}}
\newcommand{\Ta}{\textit{Ta}}
\newcommand{\Ra}{\textit{Ra}}
\newcommand{\Ro}{\textit{Ro}}
\newcommand{\Nu}{\textit{Nu}}

\newcommand{\eq}[1]{(\ref{#1})}
\newcommand{\eqs}[2]{(\ref{#1})~\&~(\ref{#2})}
\newcommand{\eqss}[2]{(\ref{#1})--(\ref{#2})}

\title[Journal of Fluid Mechanics]{Rapidly Rotating Wall-Mode Convection}

\author[G.M. Vasil, \textit{et al.}]{Geoffrey M.\ Vasil$^{1}$\footnote{{\bf Corresponding author}: Geoffrey M.\ Vasil, \url{gvasil@ed.ac.uk}}, Keaton J.\ Burns$^{2,3}$, Daniel Lecoanet$^{4,5}$, Jeffrey S.\ Oishi$^{6}$, Benjamin P.\ Brown$^{7}$ \& Keith Julien$^{8}$} 

\affiliation{
$^{1}$School of Mathematics \&  the Maxwell Institute for Mathematical Sciences, 
University of Edinburgh, EH9 3FD, UK \\ 
$^{2}$Department of Mathematics,  Massachusetts Institute of Technology, Cambridge MA 02139, USA\\ 
$^{3}$Center for Computational Astrophysics,  Flatiron Institute,  New York NY 10010, USA\\
$^{4}$Department of Engineering Sciences  and Applied Mathematics,  Northwestern University, Evanston IL 60208, USA \\
$^{5}$CIERA, Northwestern University, Evanston IL 60201, USA \\ 
$^{6}$Department of Mechanical Engineering, University of New Hampshire,  Durham NH 03824, USA\\ 
$^{7}$Department of Astrophysical \& Planetary Sciences, 
University of Colorado Boulder, Boulder CO 80309, USA\\ 
$^{8}$Department of Applied Mathematics, 
University of Colorado Boulder, Boulder CO 80309, USA}

\date{\today}
\pagerange{\pageref{firstpage}--\pageref{lastpage}}

\begin{document}

\label{firstpage}
\maketitle

\begin{abstract}

In the rapidly rotating limit, we derive a balanced set of reduced equations governing the strongly nonlinear development of the convective wall-mode instability in the interior of a general container. The model illustrates that wall-mode convection is a multiscale phenomenon where the dynamics of the bulk interior diagnostically determine the small-scale dynamics within Stewartson boundary layers at the sidewalls. The sidewall boundary layers feedback on the interior via a nonlinear lateral heat-flux boundary condition, providing a closed system. Outside the asymptotically thin boundary layer, the convective modes connect to a dynamical interior that maintains scales set by the domain geometry. In many ways, the final system of equations resembles boundary-forced planetary geostrophic baroclinic dynamics coupled with barotropic quasi-geostrophic vorticity. The reduced system contains the results from previous linear instability theory but captured in an elementary fashion, providing a new avenue for investigating wall-mode convection in the strongly nonlinear regime. We also derive the dominant Ekman-flux correction to the onset Rayleigh number for large Taylor number, $\Ra \approx 31.8 \,\Ta^{1/2} - 4.43 \,\Ta^{5/12} + \mathcal{O}(\Ta^{1/3})$ for no-slip boundaries. However, we find that the linear onset in a finite cylinder differs noticeably compared to a Cartesian channel. We demonstrate some of the reduced model's nonlinear dynamics with numerical simulations in a cylindrical container. 

\end{abstract}

\section{\label{S: Intro} Introduction}

After more than a century of vigorous investigation, Rayleigh-B\'{e}nard convection remains a quintessential distilled model for convective instability, pattern formation, and turbulence in buoyancy-driven flows \citep{chassignet_cenedese_verron_2012}.  
Likewise, rotationally influenced Rayleigh-B\'{e}nard convection serves a similar classic role in modelling many natural environments such as oceanic, stellar, and planetary systems \citep{ecke_shishkina_2023}.  

Four non-dimensional control parameters characterise Rayleigh-B\'{e}nard convection in a closed rotating vessel.  
Historically, the Rayleigh number, Taylor number, Prandtl number, and aspect ratio provide the most common choice,
\begin{eqnarray}
\label{eq: intro control parameters}
\Ra \equiv \frac{g \alpha  \, \Delta T H^{3}}{\nu \kappa}, 
\quad 
\Ta \equiv \frac{4\Omega^{2} H^{4}}{\nu^{2}},
\quad
\sigma \equiv \frac{\nu}{\kappa}, \quad \Gamma \equiv \frac{L}{H}.
\end{eqnarray}
These measure the strength of the characteristic buoyancy force, the strength of rotation, the fluid's dissipative properties, and the geometry. 
Here, $g$ gives the gravitational acceleration, $\alpha = - \rho^{-1}(\partial \rho/\partial T)|_{T_{0}, P_{0}}$ is the thermal expansion coefficient, $\Delta T$ is the vertical temperature contrast, $\nu$ represents the kinematic viscosity, $\kappa$ represents the thermal diffusivity, $\Omega$ gives the background rotation rate, $H$ represents the container depth, and $L$ gives a characteristic lateral extent (or diameter) of the container.  
The Ekman and convective Rossby numbers also provide useful equivalents to $\Ta$ and $\Ra$.  
That is,  
\begin{eqnarray}
\Ek \equiv \Ta^{-1/2}  = \frac{\nu}{2 \Omega  H^{2}},\quad \Ro \equiv \sqrt{\frac{\Ra }{\sigma \Ta}}  =  \frac{\sqrt{g \alpha \Delta T/ H}}{2 \Omega},
\end{eqnarray}
where $\Ek$ measures the size of viscous boundary layers, and $\Ro$ measures the relative influence of buoyancy and rotation.  
In rapidly rotating systems,  both $\Ek\ll1$ and $\Ro \ll 1$.  

A significant partition in Rayleigh-B\'{e}nard convection modelling exists between \textsl{large-aspect-ratio} and \textsl{finite-aspect-ratio} systems.  
In large-aspect-ratio systems, sidewall boundaries play no significant role in the dynamics. Theoretical efforts in this area focus on linear stability analyses and numerical simulations with periodic boundary conditions, a useful simplification in many instances.  
\citet{chandrasekhar_1953} first solved the linear stability theory for large-aspect-ratio rotating Rayleigh-B\'{e}nard convection; \citet{chandrasekhar_1961} discusses the subject in thorough detail. 

For moderate Prandtl number, \citet{niiler_bisshopp_1965} further clarified the influence of flow boundary conditions on the top and bottom plates in the rapidly rotating limit. 
For the present discussion, the most relevant result of these efforts is that in the rapidly rotating limit (large $\Ta$) with no-slip boundary conditions, the critical onset Rayleigh number scales as 
\begin{eqnarray}
\label{eq: Ra_c bulk}
\Ra_{\,\text{c},\text{bulk}} \approx 8.70\times\Ta^{\,2/3} - 9.63\times\Ta^{\,7/12} + \Order{\Ta^{\,1/2\,}} \quad \text{as} \quad \Ta \to \infty,  
\end{eqnarray}
assuming $\sigma = \mathcal{O}(1)$ \citep[see][]{clune_knobloch_1993,dawes_2001}.  The $\Order{\Ta^{7/12}}$ term in \eq{eq: Ra_c bulk} arises because of Ekman pumping effects correlating with leading-order convective modes \citep{zhang_roberts_1998}. \citet{buell_catton_1983} first attempted to solve the linear stability problem in a closed cylinder, finding notable decreases in the critical Rayleigh number compared to the extended-layer modes considered in \citet{chandrasekhar_1953}. They also pointed out the presence of a wall-localised mode, implying a tendency for instability at a reduced Rayleigh number for any aspect ratio. However, their analysis only considered steady onset, precluding the preferred types of propagating modes discovered later.
The laboratory work of \citet{nakagawa_frenzen_1955} gave the first qualitative stability agreement and flow visualisation for rotating convection.  

Pioneering experimental studies, with an approximately 2:1 aspect ratio, found interesting discrepancies from the large-aspect ratio theory \citep{rossby_1969}. 
Without a means to visualise the interior flow, heat transport data was the primary quantitative outcome of these experiments.  
In discussing the convective onset in water, Rossby noted 

\smallskip
\begin{quote}
\textit{We find excellent agreement between theory and experiment for the critical Rayleigh number at all Taylor numbers less than} $5\times10^{4}$; \textit{beyond this the fluid becomes unstable at lower Rayleigh numbers. At a Taylor number $=10^{8}$, for example, the measured critical Rayleigh number is about one-third the expected value. We do not understand why this should be. It is quite reproducible; \textit{i.e.}, if one changes the depth of the fluid, the instability will occur at the same Rayleigh number for a given Taylor number}. \citep[\textit{p.} 322--323]{rossby_1969}
\end{quote}
\smallskip

Motivated by the theory of \citet{veronis_1968} for subcritical phenomena, Rossby conjectured at the possibility of finite-amplitude effects and speculated about the possible modes for the ``subcritical'' instability. 
\citet{pfotenhauer_1987} found similar onset phenomena in cryogenic helium and speculated that the lower onset state was related to linear stability calculations of the static wall mode from \citet{buell_catton_1983}. Inspired by precision laboratory work in the 1990s \citep{zhong_etal_1991,ecke_etal_1992,ning_ecke_1993,zhong_etal_1993}, several almost simultaneous efforts helped clarify our theoretical understanding of Rossby's paradox: sidewalls lead to distinct modes of linear instability at lower Rayleigh numbers than in a large-aspect-ratio system.

\citet{goldstein_etal_1993,goldstein_etal_1994} conducted the first comprehensive linear stability analysis for a cylindrical vessel with a range of $\Ra,\,\Ta,\,\sigma,\,\Gamma$. 
This work demonstrated that in a rapidly rotating finite-aspect ratio system, the onset of convection occurs via two distinct mode sets.  
The first type comprises non-propagating modes supported within the bulk and corresponds to the periodic modes in Chandrasekhar's large-aspect-ratio analyses and Buell's \& Catton's finite-domain modes. 
These solutions remain stationary in the rotating coordinate frame.  

\citet{goldstein_etal_1993} also found a second type of instability that proceeds via a set of faster precessing modes (retrograde with respect to system rotation). 
In contrast to the bulk dynamics, these waves exist in a thin boundary layer attached to the container's sidewall.
For large enough background rotation rates, the wall-localised modes emerge as the preferred pattern of instability in a regime subcritical to the bulk-mode instability. In the rapidly rotating limit, 
\begin{eqnarray}
\label{eq: Ra_c wall}
\Ra_{\,\text{c},\text{wall}} \approx 31.8\times \Ta^{\,1/2} - 4.43 \times \Ta^{\,5/12} + \Order{\Ta^{\,1/3\,}} \quad \text{as}\quad \Ta \to \infty.
\end{eqnarray}
\citet{herrmann_busse_1993} derived the leading-order term in this expression. Assuming stress-free top and bottom boundary conditions, \citet{liao_etal_2006} computed an $\Order{\Ta^{\,1/3}}$ correction distinguishing between different sidewall velocity boundary conditions. The $\Order{\Ta^{\,5/12}}$ correction term also arises from no-slip top and bottom boundary conditions. As far as we know, this term is absent from the literature. In \S\ref{S: Linear correction}, we account for Ekman pumping \citep[similar to][]{zhang_roberts_1998} to calculate this term, along with the other associated corrections. 

\begin{figure}
\begin{center}
\vspace{+0.10in}
\includegraphics[width=12cm]{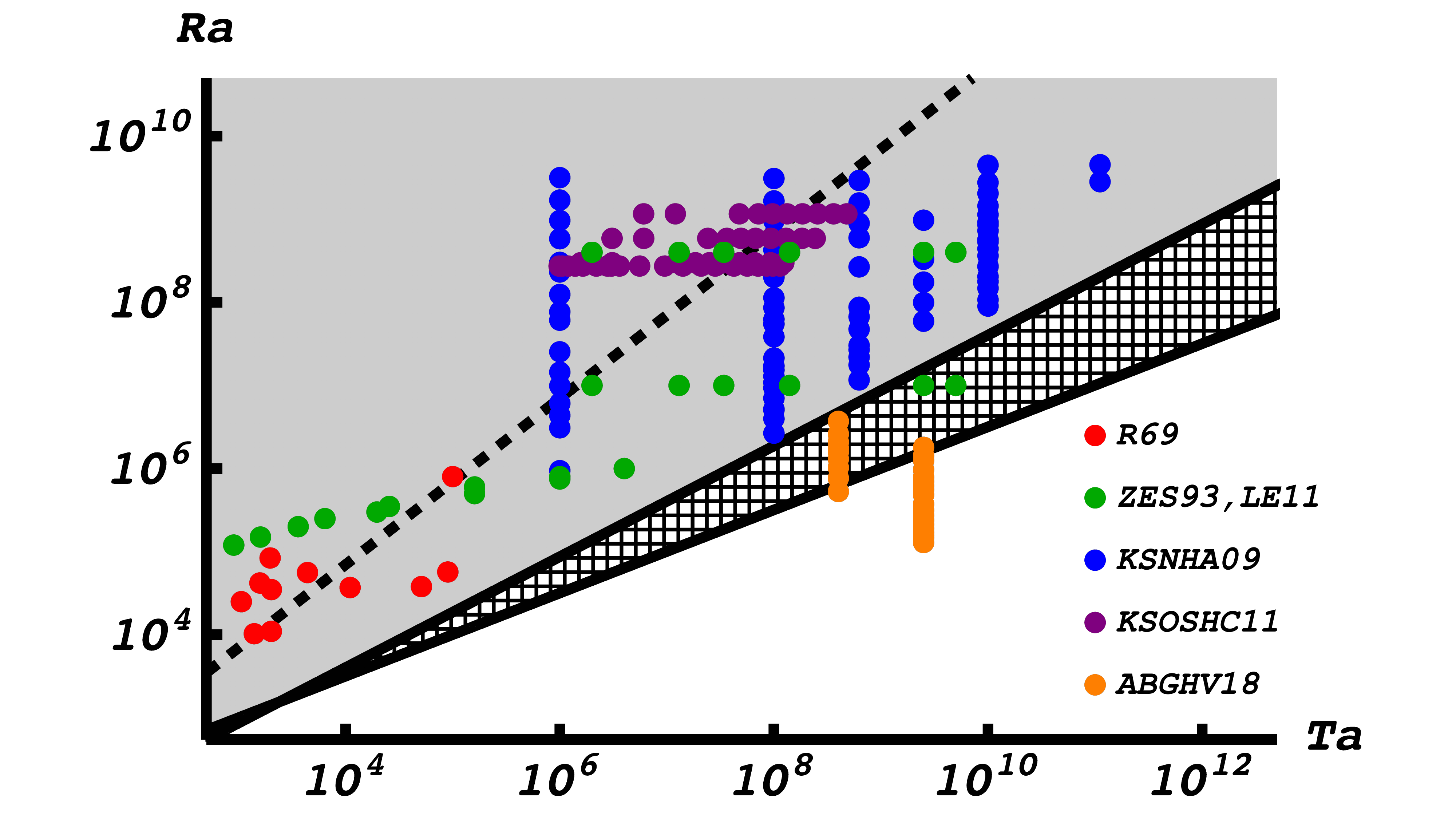}
\caption{Approximate stability regions and experimental parameters. The respective sources are all laboratory experiments: 
\textit{R69}: \citep{rossby_1969};  \textit{ZES93}: \citep{zhong_etal_1993}; 
\textit{LE11}: \citep{liu_ecke_2011};
\textit{KSNHA09}: \citep{king_etal_2009};
\textit{KSOCHC11}: \citep{kunnen_etal_2011}; \textit{ABGHV18}: \citep{aurnou_etal_2018}. In each case, the current author gathered the information from the respective sources by hand. All values are approximate and intended to get an idea of the relative ranges of data. The cross-hatched region shows the wall-mode regime between the asymptotic stability scalings $\Ra_{\,\text{bulk}} \approx 8.7 \, \Ta^{2/3}$  and $\Ra_{\,\text{wall}} \approx 31.8 \, \Ta^{1/2}$. The 31.8 coefficient is roughly an upper bound; the actual value somewhat lower in a finite-aspect-ratio cylinder (see fig.~\ref{fig: critical cylinder parameters}). Except for $\textit{ABGHV18}$, all experiments use moderate Prandtl numbers, approximately between 4-7. The steep dashed line shows the convective Rossby number threshold $\Ro \approx 1$ for Prandtl $\approx 7$. The $\textit{ABGHV18}$ experiment used gallium with Prandtl $\approx 0.025$ with markedly different onset behaviour, but still well below $\Ro = 1$.   
\label{parameter space}}
\end{center}
\end{figure}

Along with the linear stability theory, the advent of practical optical shadowgraph techniques allowed much more detailed studies of the flow structures contained in experiments \citep{boubnov_golitsyn_1986,kuo_cross_1993,zhong_etal_1993}.  
A large number of studies --- for example, \citet{rossby_1969,
boubnov_golitsyn_1990,
ning_ecke_1993,
julien_etal_1996,
sakai_1997,
liu_ecke_1997,
hart_etal_2002,
ahlers_etal_2009,
king_etal_2009,
kunnen_etal_2009,
niemela_etal_2010,
zhong_ahlers_2010,
zhan_etal_2009,
weiss_ahlers_2011,
liu_ecke_2011,
stevens_2011} --- examine the heat-transport and flow properties in rotating Rayleigh-B\'{e}nard systems significantly above the onset of bulk convection (see fig.~\ref{parameter space} for a sample of experiment parameter ranges). 
Alternatively, we focus on the parameter range below onset for bulk convection, where wall-mode convection exists in isolation. Most recently, \cite{aurnou_etal_2018} investigated the regime before bulk onset for $\Ta \approx 10^{10}$. We point out that in the limit $\Ta\rightarrow\infty$ wall-mode convection can reach values of $\Ra/\Ra_{\,\text{c},\text{wall}}={\cal O}(\Ta^{1/6})$
before exciting the bulk mode. 
This indicates the potential development of strongly nonlinear dynamics in the cross-hashed region of fig.~\ref{parameter space}. 

More recently, several studies have examined the cross-hatched region and its vicinity with direct numerical simulations  \citep{kunnen_etal_2011,horn_schmid_2017,zhang_etal_2020,favier_knobloch_2020,ecke_etal_2022,dewit_etal_2023,ravichandran_wettlaufer_2024,zhang_etal_2024}. 
One significant result from these efforts is that wall mode dynamics can undergo various degrees of wall separation and even generate interior turbulence below the threshold for bulk convection onset \citep{albaiz_etal_1990,zhong_etal_1991,marques_lopez_2008,li_etal_2008,rubio_etal_2009,favier_knobloch_2020,zhang_etal_2024}. The generation of boundary zonal flows has also been an area of interest \citep{zhang_etal_2020,dewit_etal_2020,zhang_etal_2021, wedi_etal_2022,ecke_etal_2022}. 
With the rich interplay of wall-localised and interior dynamics, many questions remain. What is genuinely wall-generated behaviour versus what is wall-catalysed bulk dynamics? Without getting into a regime of $\Ta$ far exceeding current simulations and experiments, there is not enough dynamic range to disentangle various phenomena.

Within the wall-dominated cross-hatch region, the convective Rossby number satisfies
$\Ro = \Order{\Ek^{\,1/2}}$ indicating strong rotational influence. 
\citet{sprague_etal_2006, julien_knobloch_2007} show that regimes of low $\Ek, \Ro$ exhibit balanced dynamics for bulk convection and are prime candidates for dynamical reductions capturing nonlinear evolution through the application of multiple-scale perturbation theory. 
In this paper, we derive a systematic dynamical reduction for wall-mode convection.

\subsection{\label{S: Outline} Outline} 

We organise the paper as follows: 
\begin{itemize}
\item \S\ref{S: Model Setup} outlines the parameter choices and model equation setup.
\item \S\ref{S: Asymptotics} derives the fundamental equations, including
\begin{itemize}
\item 
\S\ref{S: Interior} interior balances,
\item \S\ref{S: Stewartson} sidewall thermal Stewartson layers,
\item \S\ref{S: Ekman} Ekman layer effects. 
\end{itemize}
\item \S\ref{S: Summary} gives a concise summary of all derived results in \eqss{eq: thermal wind}{eq: vertical stress-free}.
\item
\S\ref{S: linear theory} describes linear theory, including 
\begin{itemize}
\item
\S\ref{S: Semi-infinite} for Cartesian coordinates near a semi-infinite domain,
\item \S\ref{S: Linear correction} the leading-order correction to the critical Rayleigh number, 
\item \S\ref{S: Direct numerical} direct-numerical validation of critical parameters,
\item \S\ref{S: Finite cylinder} within a closed finite cylinder, including the tall-aspect-ratio limit,
\item \S\ref{S: baroclinic instability} a local baroclinic instability. 
\end{itemize}

\item
\S\ref{S: Simulations} reports on nonlinear results in cylindrical geometry, including 
\begin{itemize}
\item
 \S\ref{sec: numerical setup} setup of numerical implementation of reduced equations,
\item
\S\ref{sec: Weakly nonlinear theory} weakly nonlinear theory determining key diagnostic parameters,
\item
\S\ref{sec: Fully nonlinear} results of fully nonlinear simulations.
\end{itemize}

\item
\S\ref{S: Conclusions} gives conclusions and discusses generalisations, including  
\begin{itemize}
\item
\S\ref{S: Double diffusion} double diffusion, 
\item 
\S\ref{S: Magnetism} magnetism.
\end{itemize}
\end{itemize}

\section{\label{S: Model Setup} Model Setup}

We start our multiscale asymptotic analysis from the Boussinesq equations \citep{spiegel_veronis_1960}. We pose and analyse the system in a local Cartesian coordinate system $(x,y,z)$ along a flat vertical wall, with $x$ increasing into the wall at $x=\Gamma$, the coordinate $y$ tangential along the wall, and $z$ vertical. We choose this coordinate system to easily translate to more general situations, e.g. an upright cylinder, $(x,y,z) \to (r,\theta,z)$. 

We pick our non-dimensionalisation to agree mostly with the standard thermal diffusion scaling. For box depth, $H$, and thermal diffusivity, $\kappa$, we scale $\text{length} \sim H$, $\text{velocity} \sim \kappa / H$, and $\text{time} \sim H^{2} / \kappa$. 

For the temperature and pressure, the known properties of wall-mode dynamical balances \citep{herrmann_busse_1993} imply a slightly non-standard normalisation. We define a linear reference temperature profile,
\begin{eqnarray}
T_{\text{ref}}(z) = T_{\text{top}} + \Delta T\, (1-z/H),
\end{eqnarray}
where the domain comprises $0\leq z/H \leq 1$ and $T_{\text{top}}$ and $T_{\text{bot}}=T_{\text{top}}+\Delta T$ are the fixed temperatures at the top and bottom boundary. We use the temperature perturbation, $\Theta = T - T_{\text{ref}}(z)$ as a dynamical variable. Similarly, we evolve the \textsl{kinematic} pressure perturbation, $p = P/\rho$, where $P$ is the dynamic pressure and $\rho$ is the mass density. We also subtract the hydrostatic reference profile balancing buoyancy from the background $T_{\text{ref}}(z)$.  From now on, we refer to $p$ simply as ``pressure''.

It is common to scale $p \sim \nu \kappa / H^{2}$ and $\Theta \sim \Delta T$. However, we pick 
\begin{eqnarray}
p \sim 2\Omega \, \kappa , \qquad  \Theta \sim \frac{2\Omega\,\kappa}{g\,\alpha \, H},
\end{eqnarray}
which allow the pressure to balance buoyancy and rotation; the temperature maintains thermal wind balance.

These scalings naturally introduce a ``reduced'' Rayleigh number as the buoyancy control parameter,
\begin{eqnarray}
\label{eq: R=Ra Ek} R \ \equiv \ \Ra\,\Ek  =  \sigma \frac{\Ro^{2}}{\Ek}  =   \frac{g\alpha \Delta T H}{2\Omega \kappa} =\Order{1}\quad \text{as} \quad \Ek \to 0.
\end{eqnarray}

\begin{figure}
\begin{center}
\vspace{+0.10in}
\includegraphics[width=10cm]{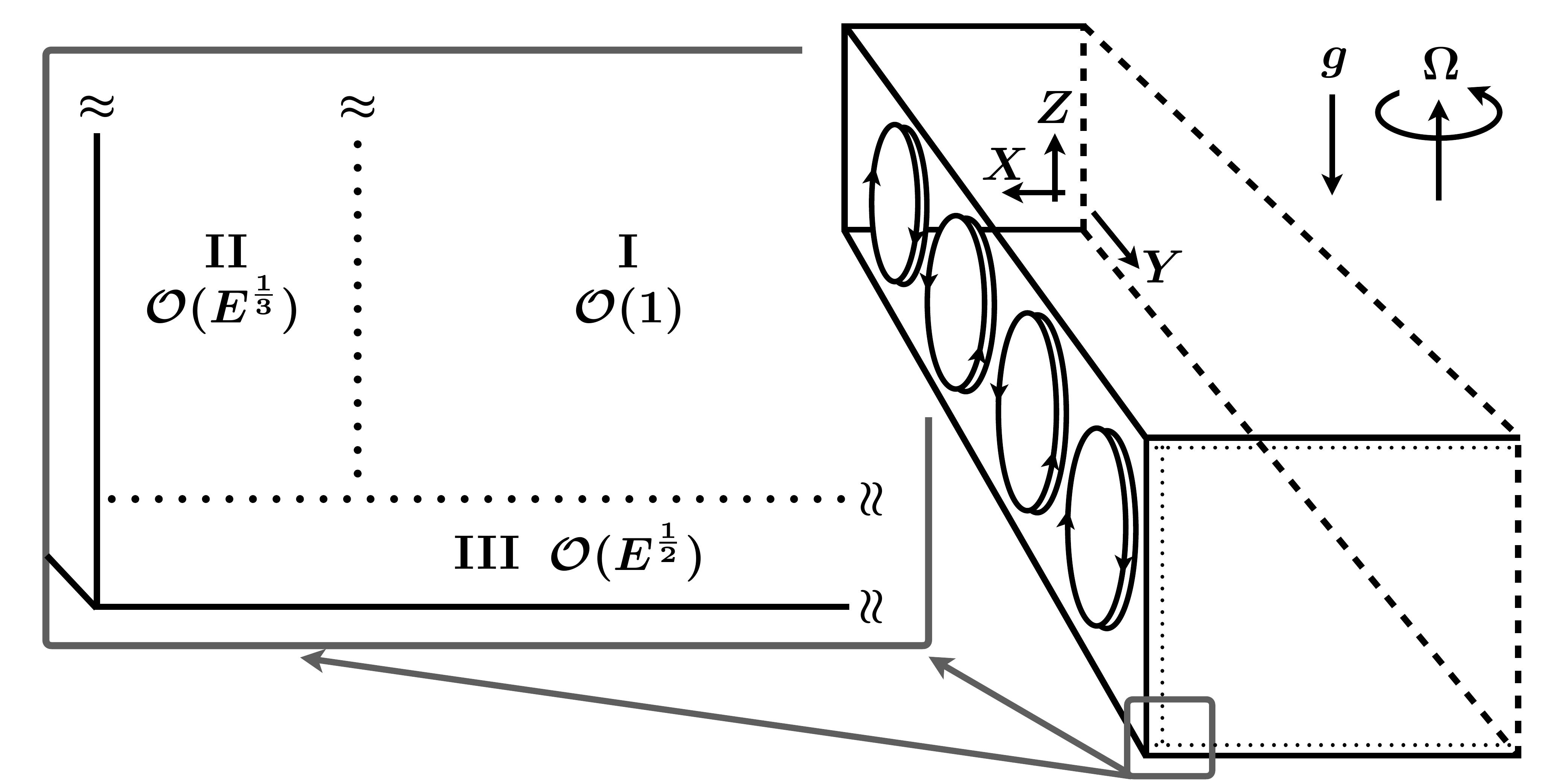}
\caption{Schematic diagram of the basic domain. Each region, I--III, comprises different length, time, and amplitude scalings in terms of powers of $\eps = \Ek^{\,1/3}$. The asymptotic analysis considers each region separately and connects each through a series of matching conditions, eventually being left entirely with $\Order{1}$ dynamics in the bulk (Region I). We refer to Region II as the Stewartson sidewall layer and Region III as the Ekman layer (at both the top and bottom of the domain). We conduct the asymptotic analysis in Cartesian coordinates $(x,\,y,\,z)$ without loss of generality. After obtaining the final results, it is straightforward to use general coordinates and re-pose the system in cylindrical polar coordinates with $x \to r$, $y \to \phi$ and $z \to z$.
\label{basic-diagram}}
\end{center}
\end{figure}

Our non-dimensional system takes the form, 
\begin{eqnarray}
\frac{1}{\sigma}  \frac{D u}{Dt} &+& \frac{\pd{x} p - v}{\Ek}  = \nabla^{2} u, \label{eq: boussinesq-u} 
\\
\frac{1}{\sigma}  \frac{D v}{Dt} &+& \frac{\pd{y} p + u}{\Ek}  = \nabla^{2} v, \label{eq: boussinesq-v}
\\
\frac{1}{\sigma}  \frac{D w}{Dt} &+& \frac{\pd{z} p - \Theta}{\Ek}  = \nabla^{2} w, \label{eq: boussinesq-w}
\end{eqnarray}
\begin{eqnarray}
\label{eq: divergence}
\pd{x} u + \pd{y} v + \pd{z} w  = 0,
\label{eq: boussinesq-p}
\end{eqnarray}
\begin{eqnarray}
\label{eq: boussinesq-Theta}
\frac{D \Theta}{Dt} - R \, w   = \nabla^{2} \Theta,
\end{eqnarray}
where
\begin{eqnarray}
\frac{D}{Dt}  = \pd{t} + u \, \pd{x} + v \, \pd{y} + w \, \pd{z} \,  , \qquad \nabla^{2}  = \pd{x}^{2} + \pd{y}^{2} + \pd{z}^{2}.
\end{eqnarray}
The system models the dynamics of a weakly compressible fluid in an upright container with uniform height $0 \le z \le 1$ and arbitrary horizontal geometry. The system rotates at a fixed rate about an axis anti-aligned with gravity; see, e.g., fig.~\ref{basic-diagram}. The triplet $(u,v,w)$ represents the fluid velocities in the $(x,y,z)$ directions respectively. The quantity $\Theta$ represents the temperature departure from a constant reference value, and $p$ gives the dynamic pressure. 

In this simplified geometry, we address both no-slip (\textit{NS}) and stress-free (\textit{SF}) boundary conditions on all walls:
\begin{eqnarray}
\mbox{(\textit{z-NS}): }  u = v = w = 0, \mbox{ at } z=0,1,
\end{eqnarray}
\begin{eqnarray}
\label{eq: SF-z}
\mbox{(\textit{z-SF}): }  \pd{z}u = \pd{z}v = w = 0, \mbox{ at } z=0,1,
\end{eqnarray}
\begin{eqnarray}
\mbox{(\textit{x-NS}): }  u = v = w = 0, \mbox{ at } x = \Gamma,
\end{eqnarray}
\begin{eqnarray}
\mbox{(\textit{x-SF}): } u = \pd{x} v = \pd{x} w = 0, \mbox{ at } x = \Gamma. &&
\end{eqnarray}
Differences between no-slip and stress-free conditions on the sidewalls produce no leading-order effect on the final result, while the top/bottom boundaries show important differences.  

Given that $\Theta$ is the perturbation from the reference profile, $T_{\text{ref}}(z)$, the thermal boundary conditions on both plates are
\begin{eqnarray}
\label{eq: top-bottom Theta} \Theta|_{z=0}  =  \Theta|_{z=1}  =  0.  
\end{eqnarray}
The thermal sidewall condition influences the dynamics more than any other. We use perfectly insulating sidewalls, 
\begin{eqnarray}
\label{eq: insulating side-wall} \pd{x}\Theta|_{x=\Gamma} = 0.
\end{eqnarray}
Relaxing the insulating condition to include finite conduction stabilises the wall mode because of heat losses. In the infinite conductivity limit, the container walls short-circuit the instability. In that case, the critical $\Ra \sim \mathcal{O}(\Ek^{\,-4/3})$ along with the bulk convection. This regime nonetheless contains rich wall-mode dynamics with investigation ongoing \citep[e.g.,][]{ravichandran_wettlaufer_2024,ecke_etal_2024}. It is also possible to match the final asymptotic equations to external thermal field with arbitrary heat capacity and conductivity. A large conductivity will increase the critical Rayleigh number.

\section{\label{S: Asymptotics} Multiscale Asymptotic Analysis}

We aim to find reduced equations capturing the leading-order nonlinear behaviour of the wall-mode instability in the limit of an infinitesimal Ekman number, $\Ek \ll 1$. We have already discussed the relevant pressure amplitude and Rayleigh number scaling. We now describe the \textit{a priori} known properties of the boundary layer thicknesses and flow amplitudes. 

\medskip
\noindent(\textit{i}): A thermally forced side-wall Stewartson boundary layer \citep{stewartson_1957,barcilon_pedlosky_1967a,barcilon_pedlosky_1967b, hashimoto_1976,fein_1978,albaiz_etal_1990,pedlosky_2009,kunnen_etal_2013} drives the dynamics from within a characteristic layer thickness 
\begin{eqnarray}
\dd{x} \sim \Ek^{\,1/3}.
\end{eqnarray} 
\noindent(\textit{ii}) For no-slip boundary conditions, top and bottom Ekman layers \citep{greenspan_1969,pedlosky_1987} influence the dynamics from within a  characteristic layer thickness 
\begin{eqnarray}
\dd{z} \sim \Ek^{\,1/2}.
\end{eqnarray}
\noindent(\textit{iii}) Relative to the thermal diffusion velocity normalisation $\kappa/H$, the fluid velocities near the wall carry a large amplitude
\begin{eqnarray}
\left(u,v,w\right) \sim \Ek^{\,-1/3},
\end{eqnarray}
while the interior velocities remain $\mathcal{O}(1)$ in these units. 

Throughout this paper, the Prandtl number remains a fixed order-unity number and does not scale with $\Ek$ in any way. Allowing $\sigma$ to scale with $\Ek$ can lead to several interesting possibilities \citep{dawes_2001,zhang_etal_2007,zhang_liao_2009}, but this remains beyond our current scope.

Assumptions (\textit{i})~\&~(\textit{ii}) introduce multiple scales, where all variables depend on both small and large length scales. For example, with the pressure 
\begin{eqnarray} 
p = p\!\left(\frac{x}{\Ek^{\,1/3}},\,x,\,y,\,\frac{z}{\Ek^{\,1/2}},\,z,\,t\right). 
\end{eqnarray}

For the duration of the asymptotic derivation, we use capital-letter variables to denote coordinates varying on the bulk-interior scale, \textit{i.e.}, $(X, Y, Z) = (x^{*}, y^{*}, z^{*})/H$, where (e.g.) $x^{*}$ is a \textsl{physical} coordinate. Also, for the duration of the asymptotic analysis, the lower-case variables represent boundary layers. In the horizontal directions, $(x,y) = \Ek^{\,-1/3}\,(x^{*},y^{*})/H$ and $z = \Ek^{\,-1/2} z^{*}/H$. After establishing the final asymptotic equations, all remaining coordinates will revert to \textsl{bulk-only}. We will, therefore, restore all coordinates to simply $(x,y,z)$ only, representing bulk physical scales. 

Multiple-scale theory dictates \citep{kevorkian_cole_1981} that each argument varies independently of the others. This assumption allows redefining the derivative operators
\begin{eqnarray}
\pd{x} \to \frac{1}{\Ek^{\,1/3}}\pd{x} + \pd{X},
\quad
\pd{y} \to  \pd{Y},\quad
\pd{z} \to \frac{1}{\Ek^{\,1/2}}\pd{z} + \pd{Z}.
\end{eqnarray}
The capital-letter coordinates $(X, Y, Z)$ now label position within the ``large-scale'' interior (outside boundary layers). The small-letter coordinates $(x, z)$ label position within the Stewartson and Ekman boundary layers, respectively. We exclude asymptotically small-scale dynamics in the $Y$-direction along the sidewall. In the bulk-convection regime (\textit{i.e.}, $R \sim \Ek^{\,-1/3}$) both horizontal directions adopt $\Order{\Ek^{\,1/3 }}$ length scales \citep{julien_knobloch_2007}.

The $\Order{\Ek^{\,1/3}}$ length scale controls the overall system, suggesting the expansion parameter, 
\begin{eqnarray}
\label{eq: eps} \eps \equiv \Ek^{\,1/3}.
\end{eqnarray}
The Ekman layers on the top and bottom thus require half-integer powers $\Ek^{\,1/2}=\eps^{3/2}$.

Eq.~\eq{eq: eps} and the above scalings of pressure and velocity with $\Ek$ imply the following asymptotic series for the dynamical variables
\begin{eqnarray}
\label{eq: u expansion} 
u &=& \frac{u_{-1}}{\eps} + \frac{u_{-1/2}}{\eps^{1/2}} + u_{0} + \Order{\eps^{1/2}} \\ 
v &=& \frac{v_{-1}}{\eps} + \frac{v_{-1/2}}{\eps^{1/2}} + v_{0} + \Order{\eps^{1/2}} \\ 
w &=& \frac{w_{-1}}{\eps} + \frac{w_{-1/2}}{\eps^{1/2}} + w_{0} + \Order{\eps^{1/2}} \\ 
p &=& p_{0} + \eps^{1/2}\,p_{1/2} + \eps\, p_{1} + \Order{\eps^{3/2}} \\ 
\label{eq: Theta expansion} 
\Theta &=& \Theta_{0} + \eps^{1/2}\, \Theta_{1/2} + \eps\, \Theta_{1} + \Order{\eps^{3/2}}.
\end{eqnarray}

Defining pure interior variables aids in the following analysis. Capitol-letter variables (e.g., $U, V, W, P$) denote quantities depending only on large-scale interior coordinates $(X, Y, Z)$. These connect to the boundary layer variables via a matching principle. For example, with the leading-order pressure,
\begin{eqnarray}
P_{0}(X,Y,Z) \equiv \lim_{|x|,\vert z\vert  \to \infty} p_{0}(x,X,Y,z,Z),  
\end{eqnarray}
and so on. The analysis in \S\ref{S: Interior}--\ref{S: Ekman} shows that the temperature is an interior variable, removing the need to start with a lowercase variable. 

The three following subsections determine the leading-order dynamics. We examine each region shown in fig.~\ref{basic-diagram} and connect each through appropriate matching conditions. 

We comment on a key aspect to keep in mind throughout the derivation. In our final equation, the only explicit time dependence comes from interior temperature and barotropic vorticity evolution; the full system contains many more time derivatives in the momentum equation.  

In many systems, it is common for time-dependence to be subdominant in some part of the domain. In particular, it is common for boundary layers to have direct diagnostic constraints (compared to prognostic evolution). In boundary-layer theory, the idea is often that the time-evolving interior provides the boundary layer with a state that
does not satisfy boundary conditions at the physical wall. The thinness of the boundary
layer implies that diffusion timescales are extremely fast and can relax an arbitrary far-
field to the required balance virtually instantaneously, which obviates
the need for explicit $\partial_{t}$.

\subsection{\label{S: Interior} Region-I (Interior)}

Given the scaling of \eqss{eq: boussinesq-u}{eq: boussinesq-p}, we can read off the leading-order interior dynamical equations by inspection. That is, for the bulk interior, all variables retain their original amplitude scalings and the Coriolis and hydrostatic terms are the only remaining in \eqss{eq: boussinesq-u}{eq: boussinesq-p} as $\Ek \to 0$.
The bulk dynamics then only contains $\Order{\eps^{0}}$ amplitudes and the only time derivative occurs in the temperature equation \eq{eq: boussinesq-Theta}. It is possible to derive this explicitly by considering the leading-order balances of the full asymptotic expansion.  
From the momentum equations and continuity,
\begin{eqnarray}
U_{0} = -\pd{Y}P_{0}, \quad V_{0} = \pd{X} P_{0}, \quad W_{0} = 0, \quad \Theta_{0} =  \pd{Z} P_{0}. \label{eq: geo-hydro balance}
\end{eqnarray}
The pressure gives the horizontal velocities and temperature in a \textit{thermal-wind} balance (\textit{i.e.}, a joint hydrostatic and geostrophic balance). 

The temperature equation retains its original nonlinear form (except for vertical advection),
\begin{eqnarray}
(\pd{t} + U_{0} \pd{X} + V_{0}\pd{Y}) \, \Theta_{0}  = ( \pd{X}^{2} + \pd{Y}^{2} + \pd{Z}^{2} )\, \Theta_{0}. \label{eq: reduced heat}
\end{eqnarray}
The combined equations \eqs{eq: geo-hydro balance}{eq: reduced heat} almost form a closed system, with two important caveats.

The first major caveat is that the temperature evolution equation contains no obvious sources. Defining the horizontal velocity vector, $U_{\bot} = (U_{0}, V_{0})$, the available thermal energy evolution,
\begin{eqnarray}
 \frac{d}{dt} \int_{0}^{1}\! \! \int_{\!\mathcal{A}} \frac{\Theta_{0}^{2}}{2} \dd{X}\! \dd{Y} \! \dd{Z}  & + & \int_{0}^{1} \! \! \int_{\partial \mathcal{A}} \left[ \frac{\Theta_{0}^{2}}{2} \, U_{\bot} - \Theta_{0} \nabla_{\bot} \Theta_{0} \right] \cdot \hat{n} \, \dd{\ell} \dd{Z}  =   \nonumber \\ 
&-& \int_{0}^{1}\! \! \int_{\!\mathcal{A}} \left[ |\nabla_{\bot}\Theta_{0}|^{2} + |\pd{Z}\Theta_{0}|^{2}\right] \dd{X}\! \dd{Y} \!\dd{Z} \ \le \ 0, \label{eq: total thermal energy}
\end{eqnarray}
where $\dd{\ell}$ is the integration measure along the boundary, $\partial \mathcal{A}$ and $\hat{n}$ is the outward unit normal vector. Eq.~\eq{eq: total thermal energy} shows that the boundary terms must force the system strongly enough to overcome persistent dissipation. Without the boundary terms (like in conventional bulk convection), the available thermal energy would monotonically decrease. Driving \eq{eq: reduced heat} requires forcing from the sidewall boundary layers, which we explain in \S\ref{S: Stewartson}.

The second caveat is that \eq{eq: reduced heat} 
determines the evolution only for the temperature. Obtaining the velocities requires the pressure, and obtaining the pressure requires solving hydrostatic balance \eq{eq: geo-hydro balance}.  The difficulty comes from not knowing the depth-independent, or \textsl{barotropic}, component of the pressure,
\begin{eqnarray}
\label{eq: average}
\btropic{ P_{0} } \equiv \int_{0}^{1}P_{0} \, \text{d}Z.  
\end{eqnarray}
We derive an equation for $\btropic{ P_{0} }$ by considering the top and bottom boundaries together with their velocity boundary conditions in \S\ref{S: Ekman}.

\subsection{\label{S: Stewartson} Region-II (thermal Stewartson Layer)}

In the Stewartson sidewall layers, focusing on $X=\Gamma$ and $- \infty < x \le 0$, the physical container boundary corresponds to $x=0$ and $X=\Gamma$, while the boundary of the interior corresponds to $X=\Gamma$ and $x \to - \infty$.
The leading-order momentum equation \eq{eq: boussinesq-u} in the $X$-direction gives  
\begin{eqnarray}
&&
u_{-1} 
 = u_{-1/2} 
 = 0.
\end{eqnarray}
The leading-order diffusion of temperature from \eq{eq: boussinesq-Theta} gives
\begin{eqnarray}
\label{eq: leading-order Theta diffusion}
\pd{x}^{2}\Theta_{0}  = 0,
\end{eqnarray}
which implies a constant leading-order temperature across the boundary layer. Strictly speaking, \eq{eq: leading-order Theta diffusion} contains a linearly growing solution in $x$. The possibility of unbounded amplitudes as $|x| \to \infty$ 
preclude this solution. Therefore, the leading-order temperature exists as a pure interior dynamical variable,
\begin{eqnarray}
\label{eq: leading-order Theta dependence}
\Theta_{0} = \Theta_{0}(X,Y,Z).  
\end{eqnarray} 
While \eq{eq: leading-order Theta dependence} implies that $\Theta_{0}$ does not vary \textsl{across} the boundary layer, there exist temperature fluctuations in the directions along the wall both in $Y$ and $Z$. The interior thermally forces the boundary layer via tangential gradients. 

Given a temperature field in the interior, the leading-order momentum and continuity equations follow
\begin{eqnarray}
\label{eq: leading-order u0}
\pd{x}p_{0} - v_{-1}  &=& 0, \\ 
\label{eq: leading-order v-1}
\pd{Y} p_{0} + u_{0}  - \pd{x}^{2}v_{-1} &=& 0, \\ 
\label{eq: leading-order w0}
\pd{Z} p_{0} - \pd{x}^{2} w_{-1} 
&=& \Theta_{0}, \\ 
\label{eq: leading-order p0}
\pd{x}u_{0} 
+ \pd{Y}v_{-1} 
+ \pd{Z}w_{-1} 
&=& 0.
\end{eqnarray}
The system is similar to the Stewartson layer balances derived in other work \citep{stewartson_1957,heijst_1983,kunnen_etal_2013}. The main difference is that \eqss{eq: leading-order u0}{eq: leading-order p0} have enough derivatives to satisfy all velocity boundary conditions at the physical wall, and thermal anomalies force the whole boundary layer from the interior. There is no need for the classical $\Ek^{\, 1/4}$ exterior layer needed typically needed to satisfy all velocity boundary conditions.

The interior forces the vertical velocity via eq.~\eq{eq: leading-order w0}. Leading-order balances also imply $\pd{x}^{2} \Theta_{1/2} = 0$. To see how the boundary layer feeds back to the interior, the next-order temperature equation for $\Theta_{1}$,
\begin{eqnarray}
\label{eq: next-order T0}
v_{-1} \, \pd{Y} \Theta_{0} + w_{-1} \, ( \pd{Z}\Theta_{0} - R)  = \pd{x}^{2} \Theta_{1}.
\end{eqnarray}

The crucial connection between the interior and boundary layer comes from the full form of the insulating thermal boundary condition at the physical sidewall, 
\begin{eqnarray}
\label{eq: full boundary condition} \pd{X}\Theta_{0} + \pd{x}\Theta_{1}  = 0 \quad \text{at} \quad x=0,\ X=\Gamma.
\end{eqnarray} 
This couples $\Theta_{0}$, from the interior, with $\Theta_{1}$, from \eq{eq: next-order T0}.
Integrating \eq{eq: next-order T0} over boundary layer coordinate,
\begin{eqnarray}
x\in (-\infty, 0\, ] = \mathbb{R}^{-},
\end{eqnarray}
gives
\begin{eqnarray}
q_{Y} \, \pd{Y} \Theta_{0} + q_{Z} \, (\pd{Z}\Theta_{0} - R) = \pd{x}\Theta_{1}\big|_{x=0} = -\pd{X}\Theta_{0}\big|_{X=\Gamma}
\label{eq: nonlinear Theta BC}
\end{eqnarray}
where the $\pd{x}\Theta_{1}\big|_{x\to -\infty}$ term vanishes because $v_{-1}$ and $w_{-1}$ decay exponentially away from $x=0$ (see below), and there is no linearly growing homogeneous solution in $x$ with bounded amplitude as $x\to -\infty$. Eq~\eq{eq: nonlinear Theta BC} expresses energy conservation between the boundary layer and the interior.   

The total momentum fluxes within the sidewall layers are
\begin{eqnarray}
q_{Y}  =  \int_{\mathbb{R}^{-}} \! \!  v_{-1}  \dd{x}, \qquad 
q_{Z}  =  \int_{\mathbb{R}^{-}} \! \! w_{-1} \dd{x}.
\end{eqnarray}
These purely surface quantities depend on the tangent directions, $(Y, Z)$. They couple to the $\Order{\eps^{0}}$ dynamics because while $v_{-1}, \, w_{-1} \sim \Order{\eps^{-1}}$, the boundary layer width, $\dd{x} \sim \Order{\eps}$. The tangential velocities act exactly like Dirac-$\delta$ distributions at the boundary of the bulk, $X=\Gamma$.

To further constrain $q_{Y}$ and $q_{Z}$, we integrate the continuity equation
\begin{eqnarray}
\pd{Y}q_{Y} + \pd{Z}q_{Z} = u_{0}\big|_{x=-\infty} = U_{0}\big|_{X=\Gamma}. 
\label{eq: non-zero normal flux BC}
\end{eqnarray}
The integration limit $u_{0}\big|_{x=0} = 0$ because of impenetrability at the physical container wall. Eq~\eq{eq: non-zero normal flux BC} expresses mass conservation between the boundary layer and the interior.

To close the system, we need another relation between $q_{Y}$ and $q_{Z}$, which comes from directly solving \eqss{eq: leading-order u0}{eq: leading-order p0}.
We represent the horizontal velocities strictly in terms of the pressure
\begin{eqnarray}
u_{0} &=& -\pd{Y}p_{0} + \pd{x}^{3}p_{0}, \\ 
v_{-1} &=& \pd{x}p_{0}.
\end{eqnarray}
The following coupled system relates the pressure and vertical velocity:
\begin{eqnarray}
\label{eq: (w0,p0) 1}
\pd{Z}w_{-1} + \pd{x}^{4}p_{0} &=& 0, \\
\label{eq: (w0,p0) 2}
- \pd{x}^{2}w_{-1} + \pd{Z}p_{0} &=& \Theta_{0},
\end{eqnarray}
which collapses into a single system for  the vertical velocity,
\begin{eqnarray}
\label{eq: funny-laplacian} \left(\pd{Z}^{2} + \pd{x}^{6}\right) w_{-1}  = 0.
\end{eqnarray}

\subsubsection{Bulk onset comparison}

Eq.~(\ref{eq: funny-laplacian}) should look familiar; it is half of the traditional asymptotic stability condition in the rapidly rotating regime when $\Ra \sim \Ek^{\,4/3}$. In terms of the $R$ and $\eps$ parameters, the full bulk leading-order balance,
\begin{eqnarray}
\label{eq: bulk funny-laplacian} \left(\pd{Z}^{2} + \pd{x}^{6}\right) w_{-1}  = \eps \, R  \, \pd{x}^{2} w_{-1} .
\end{eqnarray}
If $\eps \, R \sim \Order{1}$, the left-hand side balances with $\pd{x}^{2} \to -k^{2}$,  $\pd{Z}^{2} \to -\pi^{2}$ and we have the traditional stability condition for rotating convection \citep{chandrasekhar_1961}, 
\begin{eqnarray}
\text{Bulk onset:} \quad \eps\, R_{\rm c} \sim  \frac{\pi^{2} + k^{6}}{k^{2}} \ge \frac{3 \pi^{4/3}}{2^{2/3}} \quad \text{most unstable at} \quad k_{\rm c} = \frac{\pi^{1/3}}{2^{1/6}}.
\label{eq:standard rotating convection Rc kc}
\end{eqnarray}
Here, $R_{\rm c}$ is the critical reduced Rayleigh number and $k_{\rm c}$ is the critical wavenumber for instability.

Equation~\ref{eq:standard rotating convection Rc kc} represents a triple balance between vortex stretching (\textit{i.e.},  $\pd{Z}^{2}$) and diffusion (\textit{i.e.}, $\pd{x}^{6}$) on the left, with buoyancy on the right. However, if $R \sim \Order{1}$, the right-hand side of \eq{eq: bulk funny-laplacian} drops out, and the bulk no longer supports harmonic $x$ dependence. In that case, the only possible solutions happen with exponential $x$ dependence, which can only remain bounded near a wall. 

\subsubsection{Boundary layer solutions}

We solve the system for no-slip physical sidewall boundary conditions,
\begin{eqnarray}
&& u_{0}  = v_{-1}  = w_{-1}  = 0 \qquad \text{at} \qquad x=0.
\end{eqnarray}
We could alternatively use stress-free conditions; however, doing so gives exactly the same final relationship between $q_{Y}$ and $q_{Z}$. We also impose $w_{-1} = 0$ at $Z=0,1$.  

\begin{figure}
\begin{center}
\vspace{+0.10in}
\includegraphics[width=10cm]{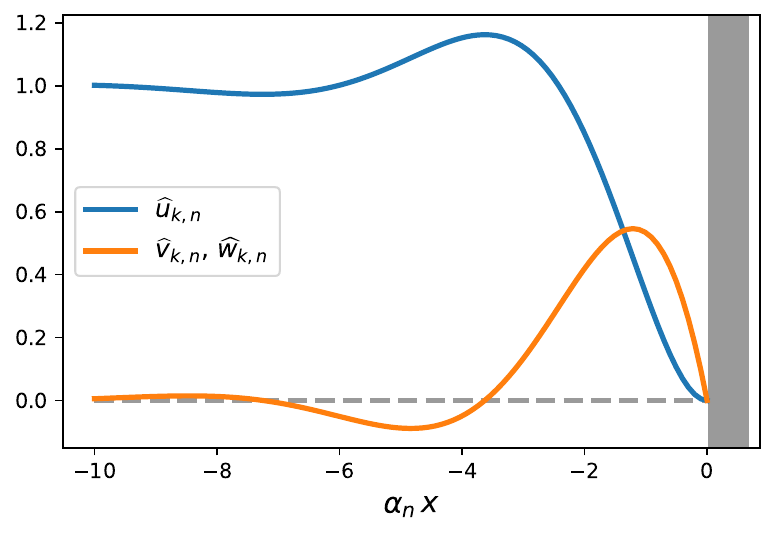}
\caption{Thermal Stewartson boundary layer solutions for no-slip tangential velocities: The blue line shows $1-\mathcal{X}(x)$. The orange line shows $\mathcal{X}'(x)$. The orange line compares well with Fig~2b from \cite{buell_catton_1983} for $\Ta \approx 1400^{2}$.
\label{boundary layers}}
\end{center}
\end{figure}

Solving \eqs{eq: (w0,p0) 1}{eq: (w0,p0) 2} requires Fourier decomposing the velocities, pressure, and temperature in the $Z$ direction.
\begin{eqnarray}
\Theta_{0} &=& \sum_{n\ge 1}\widehat{\Theta}_{n}(X,Y)\,\sin(n \pi Z),  \\ 
p_{0} &=&  \sum_{n\ge 0}\hat{p}_{n}(x,X,Y)\, \cos(n \pi Z), \\ 
u_{0}  &=&  \sum_{n\ge 0}\hat{u}_{n}(x,X,Y)\cos(n \pi Z), \\ 
v_{-1} &=& \sum_{n\ge 0}\hat{v}_{n}(x,X,Y)\, \cos(n \pi Z),  \\ 
w_{-1} &=& \sum_{n\ge 1}\hat{w}_{n}(x,X,Y)\, \sin(n \pi Z).
\end{eqnarray}
Solving for the Fourier coefficients, which decay as $x\to -\infty$ and satisfy either no-slip or stress-free boundary conditions at $x=0$, gives
\begin{eqnarray}
\label{eq: Fourier pn}
\hat{p}_{n} &=& \widehat{P}_{n} + \widehat{Q}_{n} \, \mathcal{X}(\alpha_{n} x)  , \\ 
\label{eq: Fourier un}
\hat{u}_{n} &=& \widehat{U}_{n}\left( 1 -   \mathcal{X}(\alpha_{n} x) \right), \\ 
\label{eq: Fourier vn}
\hat{v}_{n} &=& \alpha_{n} \, \widehat{Q}_{n} \,\mathcal{X}'(\alpha_{n} x), \\ 
\label{eq: Fourier wn}
\hat{w}_{n} &=& \alpha_{n} \, \widehat{Q}_{n} \,\mathcal{X}'(\alpha_{n} x),
\end{eqnarray}
where $\alpha_{n} = (n\pi)^{1/3}$. The boundary layer profile function,
\begin{eqnarray}
\mathcal{X}(x)  =  \frac{e^{ \frac{x}{2} } \cos( \tfrac{\sqrt{3}\,x}{2}  + \delta ) } {\cos(\delta)},
\end{eqnarray}
where the phase $\delta = \pm \pi/6$ depends on the tangential boundary conditions. 
In both cases, 
\begin{eqnarray}
\mathcal{X}''(x) - \mathcal{X}'(x) + \mathcal{X}(x)  = 0, \qquad  \mathcal{X}(0) = 1,
\end{eqnarray}
which satisfies the normal velocity, $\hat{u}_{n} = 0$ at $x=0$. For the tangential velocities, 
\begin{eqnarray}
\text{No-slip:} \qquad  &\hat{v}_{n} = \hat{w}_{n} = 0& \quad \iff \quad  \delta = +\frac{\pi}{6},  \qquad \mathcal{X}'(0)  = 0, \quad \\ 
\text{Stress-free:} \qquad  &\pd{x}\hat{v}_{n} = \pd{x}\hat{w}_{n} = 0 & \quad \iff \quad  \delta = -\frac{\pi}{6},  \qquad \mathcal{X}''(0)  = 0. 
\end{eqnarray}
Fig.~\ref{boundary layers} shows the velocity amplitudes for the no-slip case. 

The capitol-letter variables, $\widehat{U}_{n}$, $\widehat{P}_{n}$ and $\widehat{\Theta}_{n}$ are the Fourier transforms of the interior bulk quantities that satisfy thermal wind balance at $X=\Gamma$, 
\begin{eqnarray}
\widehat{U}_{n}  = - \pd{Y} \widehat{P}_{n}, \qquad - n \pi \widehat{P}_{n}  = \widehat{\Theta}_{n}.
\end{eqnarray}
The overall amplitude satisfies, 
\begin{eqnarray}
(\pd{Y} + n \pi )\, \widehat{Q}_{n}  = \widehat{U}_{n},
\end{eqnarray}
which is the Fourier-space version of mass conservation at the bulk boundary \eq{eq: non-zero normal flux BC}.

There are a couple of notable comments regarding \eqss{eq: Fourier pn}{eq: Fourier wn}.  The first is that both $\hat{v}_{n}$ and $\hat{w}_{n}$ exist strictly within the boundary layer and do not continue into the bulk interior. The second comment is that $\hat{p}_{n}$ and $\hat{u}_{n}$ both contain components that are independent of $x$. These terms take precisely the same form as the interior geostrophic and hydrostatic balance found in \S\ref{S: Interior}.

The most important comment regarding \eqs{eq: Fourier vn}{eq: Fourier wn} is that the coefficients of $\hat{v}_{k,n}$ and $\hat{w}_{k,n}$ take identical form,  
\begin{eqnarray}
\hat{v}_{n}  = \hat{w}_{n},
\end{eqnarray}
and 
\begin{eqnarray}
\widehat{Q}_{n}  = \int_{\mathbb{R}^{-}} \! \hat{v}_{n}\dd{x}  = \int_{\mathbb{R}^{-}} \! \hat{w}_{n}\dd{x}.
\end{eqnarray}
That is, the $\widehat{Q}_{n}$ amplitude gives the Fourier coefficients for \textit{both} $q_{Y}$ and $q_{Z}$, closing the system of equations.
However, we still need to collect all the relevant information and pose the system in a local abstract form, not in Fourier space. 

Finally, regarding the bulk, all relevant quantities do not depend on the value of $\delta = \pm \pi/6$. This observation accords with the findings from \cite{liao_etal_2006} that the leading-order stability is independent of the sidewall velocity boundary conditions. 

\subsubsection{\label{S: Hilbert} Hilbert transform}

While the Fourier amplitudes of $v_{-1}$, $q_{Y}$, $w_{-1}$ and $q_{Z}$ are all identical, the local functions are not equal because $v_{-1}$ and $q_{Y}$ are Fourier cosine series in $Z$, but $w_{-1}$ and $q_{Z}$ are Fourier sine series. We need an operator to transform between the different parities, which naturally leads to the Hilbert transform, $\mathcal{H}$, in the $Z$ direction. For the present analysis, the utility of the Hilbert transform lies in its action on a Fourier sine and cosine basis. For $n \ne 0$,
\begin{eqnarray}
\Hilbert{\,\sin(n \pi Z)\,}  =   -\cos (n \pi Z), \qquad
\Hilbert{\cos(n \pi Z)}  =  + \sin(n \pi Z),
\end{eqnarray}
which alters the parity of the basis elements in the same manner as a derivative but without the multiplication by the wavenumber, $n \pi$. Additionally, the Hilbert transform possesses the following useful properties 
\begin{eqnarray}
\label{eq: properties} \mathcal{H}^{2}[f]  = -f + \left< f\right>, \quad \pd{Z}\Hilbert{f}  = \Hilbert{\pd{Z}f}.
\end{eqnarray}
The Benjamin-Ono equation \citep{benjamin_1967,ono_1975} for small-amplitude surface gravity waves over an infinitely deep interior employs the Hilbert transform for reasons similar to the current context.   In particular, \eq{eq: properties} implies 
\begin{eqnarray}
\mathcal{H}\pd{Z} \equiv |\pd{Z}|,
\end{eqnarray}
for trigonometric functions.
Therefore, we may factor the Laplacian-like operator in \eq{eq: funny-laplacian}
\begin{eqnarray}
\left(\pd{Z}^{2} + \pd{x}^{6}\right) =  \left(|\pd{Z}| + \pd{x}^{3}\right)\!\left(-|\pd{Z}| + \pd{x}^{3}\right),
\end{eqnarray}
which separates modes that decay as $x \to \infty$ versus $x \to -\infty$.

Our final closure for the boundary momentum fluxes is 
\begin{eqnarray}
q_{Y}  = - \Hilbert{ q_{Z} },
\end{eqnarray}
which means they share the same Fourier coefficients but with different vertical parity.

\subsubsection{\label{S: Barotropic} Barotropic mode}

Finally, we derive properties of the barotropic mode in the Stewartson layer. We first take the vertical average of the horizontal velocities
\begin{eqnarray}
&&\btropic{v_{-1}} = \pd{x}\btropic{p_{0}},  \\ 
&& \btropic{u_{0}} = -\pd{Y} \btropic{p_{0}} + \pd{x}^{3}\btropic{p_{0}}, \\ 
\label{eq: barotropic divergence}
&&
\pd{x}\btropic{u_{0}} 
+ \pd{Y}\btropic{v_{-1}}  = \pd{x}^{4}\btropic{p_{0}} 
 = 0. 
\end{eqnarray}
Therefore, $\btropic{p_{0}} = \btropic{P_{0}}$ within the sidewall layer, with $\btropic{u_{0}} = -\pd{Y} \btropic{P_{0}}$. Impenetrability at the physical container boundary, along with fixing the overall pressure gauge condition, implies   
\begin{eqnarray}
\btropic{P_{0}}  = 0 \qquad \text{at} \qquad x=0, \ X=\Gamma,
\end{eqnarray}
which is the only sidewall boundary condition needed for no-slip top and bottom boundary conditions.  

Free-slip top and bottom boundaries also require the next-order barotropic momentum balances and continuity in the sidewall layer,
\begin{eqnarray}
\btropic{f} + \pd{Y} \btropic{p_{1}} + \btropic{u_{1}} &=& \pd{x}^{2} \btropic{v_{0}},   \\
\pd{x} \btropic{p_{1}} + \pd{X} \btropic{P_{0}}  - \btropic{v_{0}} &=& 0, \\
\pd{x}\btropic{u_{1}} + \pd{X}\btropic{U_{0}} + \pd{Y} \btropic{v_{0}} &=& 0,
\end{eqnarray} 
where
\begin{eqnarray}
\btropic{f}  = \frac{\pd{x} \btropic{u_{0}\, v_{-1}} + \pd{Y}\btropic{v_{-1}\, v_{-1}}}{\sigma}.
\end{eqnarray}
The tangential solution in the sidewall layer,
\begin{eqnarray}
\btropic{v_{0}} &=&  \pd{X}\btropic{P_{0}} - \int_{-\infty}^{x} (x'-x) \, \btropic{\,f(x')\,} \dd{x'}.
\end{eqnarray}
A no-slip boundary condition, $\btropic{v_{0}} = 0$, at $x=0$, $X=\Gamma$ implies 
\begin{eqnarray}
\sigma \, \pd{X}\btropic{P_{0}} &=& \int_{\mathbb{R}^{-}} \!\! x \, \left[\, \pd{x} \btropic{u_{0}\, v_{-1}} + \pd{Y}\btropic{v_{-1}\, v_{-1}}\, \right] \dd{x} \\ \nonumber  
&=&  \ \btropic{\, U_{0} \, q_{Y} } \int_{\mathbb{R}^{-}} \!\! x \, \pd{x} \left[ (1-\mathcal{X}(x))\, \mathcal{X}'(x) \right] \dd{x} + \\ 
& &  2\,\btropic{\, q_{Y} \, \pd{Y} q_{Y} } \int_{\mathbb{R}^{-}} \!\! x \, \mathcal{X}'(x)^{2} \dd{x}. 
\end{eqnarray}
Computing the coefficients, 
\begin{eqnarray}
\int_{\mathbb{R}^{-}}\!\! x \, \pd{x} \left[ (1-\mathcal{X}(x))\, \mathcal{X}'(x) \right] \dd{x} = - \frac{1}{2}, \qquad 
\int_{\mathbb{R}^{-}} \!\! x \, \mathcal{X}'(x)^{2} \dd{x} = - \frac{3}{4}.
\end{eqnarray}
Both coefficients are independent of $\delta = \pm \pi/6$.
Putting everything together,
\begin{eqnarray}
\label{eq: SF BC}
\pd{X}\btropic{P_{0}}   = - \frac{ 3\btropic{\, q_{Y}\, \pd{Y} q_{Y} } - \btropic{\, q_{Y} \,\pd{Y} P_{0}  }}{2\sigma} \qquad \text{at} \qquad  X=\Gamma.
\end{eqnarray}

\subsection{\label{S: Ekman} Region-III (Ekman Layers)}

We still need to fix the barotropic pressure, $\btropic{P_{0}}$, in the interior, which requires considering the vertical vorticity along with the top and bottom boundary conditions. 
Geostrophic balance relates the vertical vorticity and the pressure,
\begin{eqnarray}
\zeta_{0} \equiv \pd{X}V_{0} - \pd{Y} U_{0}   = \left(\pd{X}^{2} + \pd{Y}^{2}\right) P_{0}.
\end{eqnarray}
Taking the horizontal curl of the momentum equations and using the continuity equation,
\begin{eqnarray}
\label{eq: zeta-2} \frac{1}{\sigma}\left(\pd{t} + U_{0} \pd{X} + V_{0} \pd{Y} \right)\zeta_{0} - \left(\pd{X}^{2} + \pd{Y}^{2} \right) \zeta_{0}  = \pd{Z}W_{3} + \pd{Z}^{2} \zeta_{0}.  
\end{eqnarray}
Notably, the leading-order rotational terms enforce 
\begin{eqnarray}
\pd{Z}W_{i}  = 0 \quad \text{for} \quad i = 0,\, \tfrac{1}{2}, \,1, \, \tfrac{3}{2}, \, 2,\, \tfrac{5}{2}.
\label{eq: Dw=0}
\end{eqnarray}
The first non-constant vertical velocity is $W_{3}$. 

\subsubsection{\label{S: Stress-free} Stress-free boundaries}

For stress-free boundaries \eq{eq: SF-z}, applying the averaging operator to \eq{eq: zeta-2} gives
\begin{eqnarray}
\label{eq: barotropic evolution}
\mbox{(\textit{z-SF}): }\quad   \pd{t}\btropic{\zeta_{0}} + \pd{X}\btropic{ U_{0} \,\zeta_{0}} + \pd{Y}\btropic{ V_{0} \,  \zeta_{0} }  =  \sigma\left(\pd{X}^{2} + \pd{Y}^{2}\right)\btropic{ \zeta_{0} }.
\end{eqnarray}
The $\pd{Z}$ terms on the right-hand side vanish because of impenetrability and vanishing stress, 
\begin{eqnarray}
& W_{3}(Z=0) = W_{3}(Z=1) = 0, \\
& \pd{Z}\zeta_{0}(Z=0) = \pd{Z}\zeta_{0}(Z=1) = 0.
\end{eqnarray}
Eq.~(\ref{eq: barotropic evolution}) determines the missing depth-independent pressure component needed to obtain the velocities in \eq{eq: reduced heat}. The depth-dependent pressure (hence temperature) forces \eq{eq: barotropic evolution} through the nonlinear baroclinic interactions, e.g., 
$\btropic{ U_{0} \zeta_{0}} - \btropic{ U_{0}} \btropic{ \zeta_{0}}$, and the two systems are coupled.  
We solve \eq{eq: barotropic evolution} together with boundary conditions $\btropic{P_{0}}=0$ and \eq{eq: SF BC}.

\subsubsection{\label{S: No-slip} No-slip boundaries}

For no-slip top and bottom boundaries, one expects an Ekman-pumping effect of the order \citep{zhang_roberts_1998}
\begin{eqnarray}
\label{eq: W Ek}
W_{\text{Ekman}} \sim \Ek^{\,1/2}\, \zeta_{0} = \eps^{3/2} \, \zeta_{0}.
\end{eqnarray}
Together, \eqs{eq: Dw=0}{eq: W Ek} imply a depth-independent Ekman velocity,   
\begin{eqnarray}
\label{eq: W5} 
\pd{Z}W_{3/2}  = 0.
\end{eqnarray}
The Ekman layers determine the barotropic pressure.

Given an interior flow, the Ekman boundary layer response follows as a well-known exercise \citep{greenspan_1969,pedlosky_1987}.  Region-III lies beyond the sidewall Stewartson layer, hence $\pd{x} \to 0$. Within the top and bottom boundary layers,
\begin{eqnarray}
\label{eq: Ekman pressure}
\pd{z}P_{0}  &=& 0, \\
- v_{0} 
+ \pd{X}P_{0} 
&=& 
\pd{z}^{2}u_{0}, \\
u_{0} 
+ \pd{Y}P_{0}
&=& 
\pd{z}^{2}v_{0}, \\ 
\label{eq: Ekman divergence}
\pd{X} u_{0} 
+ \pd{Y}v_{0} 
+ \pd{z} w_{3/2} &=& 0.  
\end{eqnarray}
The solutions to \eqss{eq: Ekman pressure}{eq: Ekman divergence} are well known. For $Z=0$ and the boundary layer coordinate $0 \le z < \infty$. For $Z=1$ and the boundary layer coordinate $-\infty < z \le 0$. In both cases, 
\begin{eqnarray}
\left[
\begin{array}{c}
 u_{0} \\
 v_{0} \\
\end{array}
\right] = \left[
\begin{array}{rr}
 \text{se}(z) & \text{ce}(z) \\
 -\text{ce}(z) & \text{se}(z) \\
\end{array}
\right]\cdot \left[
\begin{array}{c}
 \pd{X}P_{0} \\
 \pd{Y}P_{0} \\
\end{array}
\right],
\end{eqnarray}
where $P_{0}$ is evaluated at $Z=0$ or $Z=1$ depending on the boundary layer and the functions,
\begin{eqnarray}
\text{se}(z)  = e^{- \frac{|z|}{\sqrt{2}}} \sin\left(\tfrac{|z|}{\sqrt{2}}\right), \qquad \text{ce}(z)  = 1- e^{- \frac{|z|}{\sqrt{2}}} \cos\left(\tfrac{|z|}{\sqrt{2}}\right). 
\end{eqnarray}
Both $\text{se}(0)=\text{ce}(0)=0$.

For the vertical velocity at either boundary, 
\begin{eqnarray}
w_{3/2} =  \text{sign}(z)\,  \frac{\zeta_{0}}{\sqrt{2}} \left(\text{ce}(z) - \text{se}(z) \right).
\end{eqnarray}
The Ekman pumping velocity must connect to the interior as $|z| \to \infty$,
\begin{eqnarray}
\lim_{|z| \to \infty} \! w_{3/2}  = \text{sign}(z) \frac{\zeta_{0}}{\sqrt{2}}. \label{Ekman flux condition}
\end{eqnarray}
However, in the interior $\pd{Z}W_{3/2} = 0$. Therefore,
\begin{eqnarray}
\zeta_{0}(X,Y,Z=0) + \zeta_{0}(X,Y,Z=1)  = 0,
\end{eqnarray}
which indirectly fixes the barotropic pressure. In terms of the local pressure, 
\begin{eqnarray}
\label{Harmonic-def} P_{0}|_{Z=0} + P_{0}|_{Z=1} = 2\,\Phi(X,Y),
\end{eqnarray}
where $\Phi(X, Y)$ represents a two-dimensional harmonic function (the factor of two provides a later simplification). The Stewartson layer supports no barotropic pressure, and $\Phi$ ensures $\left<P_{0}\right>$ vanishes at the sidewall. 

Given hydrostatic balance, $\pd{Z}P_{0}(Z) = \Theta_{0}(Z)$, the pressure decomposes into barotropic and baroclinic components, 
\begin{eqnarray}
\label{barotropic-baroclinic decomp}
P_{0}(Z)  = \btropic{P_{0}} +  \int_{0}^{Z} \! Z_{0}\, \Theta_{0}(Z_{0})\dd{Z_{0}} -   \int_{Z}^{1} \! (1-Z_{0})\, \Theta_{0}(Z_{0})\dd{Z_{0}}.
\end{eqnarray}
Applying \eq{Harmonic-def}, the barotropic pressure is  
\begin{eqnarray}
\label{eq: mean P - Phi}
\btropic{P_{0}}  = 
\Phi(X,Y) 
- \int_{0}^{1} \! \left(Z-\frac{1}{2}\right)\Theta_{0}(X,Y,Z) \dd{Z}.
\end{eqnarray}
The Stewartson layer equations imply $\btropic{P_{0}}= 0$ within the side wall layer, thus fixing the harmonic function, $\Phi(X, Y)$ when applying \eq{eq: mean P - Phi} at $X=\Gamma$. 

While the Ekman layers in the bulk dictate leading-order dynamics of the barotropic mode, the Ekman layers within the sidewall Stewartson layers generate a lower-order passive response. That is, it is possible to compute the effects of Ekman pumping after the fact (e.g \S\ref{S: Linear correction}), but we can press ahead with calculating the leading-order dynamics if we are not interested in the correction terms.

There is, however, one notable aspect. The balances in the sidewall naturally produce velocities $u_{0}, v_{-1}, w_{-1}$. That is, the tangential velocities are $\Order{\Ek^{\,-1/3}}$ compared to $\Order{1}$ for the normal component. However, the normal component achieves its largest amplitudes in the form of an Ekman response. 

Like before, $\pd{z}p_{0}  = 0$. Now, the horizontal velocities are isotropic,
\begin{eqnarray} 
 - v_{-1} + \pd{x}p_{0} &=& \pd{z}^{2} \, u_{-1}, \\  
u_{-1} &=& \pd{z}^{2}\, v_{-1} .
\end{eqnarray}
The Ekman flux is now larger than within the bulk, but is still subdominant compared to the $w_{-1}$ convection; \textit{i.e.}, $\pd{x} u_{-1} + \pd{z} w_{-1/2} =0$,
\begin{eqnarray}
w_{-1/2} =  \text{sign}(z)\,  \frac{\pd{x}^{2}p_{0}}{\sqrt{2}} \left(\text{ce}(z) - \text{se}(z) \right).
\end{eqnarray}

The $w_{-1/2}$ does not automatically satisfy the no-slip condition at $x=0$, which requires an $\Order{\Ek^{\,1/2}}$ corner layer \citep{kerswell_barenghi_1995} that remains passive with respect to the leading-order dynamics. Solving the corner layer requires a somewhat technical direct numerical calculation \citep{burns_etal_2022}. While the sidewall Ekman response remains passive in the fully asymptotic regime, we should expect noticeable corrections to heat transport as Rayleigh number transitions out of a strongly rotationally dominated (asymptotic) regime \citep{stellmach_etal_2014,julien_etal_2016}.

\section{\label{S: Summary} System summary}

To summarise our multiscale asymptotic analysis, we here report the final equations in dimensional form, with all explicit parameter values restored. All equations and variables are either purely bulk quantities or supported on the bulk boundary. All boundary layers have been eliminated and act only backstage. We, therefore, drop the notation of small-case and capital letters distinguishing between different scales. All coordinates are lowercase, and all variables are in their dimensional physical form. The relevant physical variables are velocity, $u$, temperature, $T$, vertical vorticity, $\zeta  = \hat{z} \cdot \nabla \times u$, and wall fluxes, $q_{\ell},\, q_{z}$. 

\subsection{Geometry and operators}

The container is now a general upright direct product of a $z$ direction with a smooth but otherwise arbitrary horizontal area,  
\begin{eqnarray}
\mathcal{V}  = \mathcal{A} \times [\, 0,H \, ],
\end{eqnarray}
with (e.g.) Cartesian horizontal coordinates $(x,y) \in \mathcal{A}$, and vertical coordinate $z \in [0,H]$, with local unit vector, $\hat{z}$. The boundary of the horizontal area is $\partial \mathcal{A}$ with the local outward normal vector, $\hat{n}$. In the case of a general smooth $\partial \mathcal{A}$, the analysis in \S\ref{S: Stewartson} would follow the same reasoning with $x$ representing a local signed-distance coordinate \citep{hester_vasil_2023}.  
We represent the coordinate along the tangential direction to $\partial \mathcal{A}$ as $\ell$, with local unit vector, $\hat{\ell} = \hat{z} \times \hat{n}$ assuming a right-handed coordinate system. For short-hand, we denote derivatives, $\pd{z} = \hat{z} \cdot \nabla$, $\pd{n} = \hat{n} \cdot \nabla$, and  $\pd{\ell} = \hat{\ell} \cdot \nabla$. In the two most relevant cases,
\begin{eqnarray}
\text{Cartesian:} \quad && \pd{n}  = \pd{x}, \quad \pd{\ell}  = \pd{y}, \\
\text{Cylindrical:} \quad && \pd{n}  = \pd{r}, \quad \pd{\ell}  = \frac{1}{r}\,\pd{\phi}.
\end{eqnarray}
We define the respective barotropic average and Hilbert transform, 
\begin{eqnarray}
\btropic{p}  = \frac{1}{H}\int_{0}^{H} \! \! p(z') \dd{z'} , \qquad 
\Hilbert{q}(z) =  \text{p.v.} \btropic{ q(z')\, \cot(  \tfrac{\pi(z'-z)}{H}) },
\end{eqnarray}
where p.v.~denotes the principle value in the Hilbert transform integral. 

\subsection{Dynamics \label{S: summary dynamics}}

For $(x,y) \in \mathcal{A}$,
\begin{eqnarray}
2 \Omega \, \pd{z}u  = g \alpha\, \hat{z} \times \nabla T, \qquad \pd{t} T + u \cdot \nabla T  = \kappa  \nabla^{2}  T, \label{eq: thermal wind}
\end{eqnarray}
\begin{eqnarray}
T\big|_{z=0} = T_{\text{bot}}, \qquad T\big|_{z=H} = T_{\text{top}}.
\end{eqnarray}
For $(x,y) \in \partial \mathcal{A} $,
\begin{eqnarray}
q_{\ell}  = - \Hilbert{ q_{z}}, \qquad \pd{\ell}  q_{\ell} + \pd{z}  q_{z}  = u_{n}, \qquad (q_{\ell}  \,\pd{\ell} + q_{z}\,  \pd{z}) \, T  = - \kappa\,  \pd{n} T \label{wall-conservation},
\end{eqnarray}
\begin{eqnarray}
q_{z} = 0 \quad \text{for} \quad z = 0,H.
\end{eqnarray}
For vertical no-slip,
\begin{eqnarray}
\zeta\big|_{z=0} + \zeta\big|_{z=H}  = 0.
\end{eqnarray}
For vertical stress-free,
\begin{eqnarray}
\pd{t}\btropic{\zeta} + \nabla \cdot\, \btropic{ u \,\zeta}  = \nu\, \nabla^{2} \btropic{\zeta}, \qquad -2\nu\,   \btropic{u_{\ell}} \! \big|_{\partial \mathcal{A}}  = \btropic{ \, q_{\ell}  \left( 4\,\pd{\ell}q_{\ell} + \pd{z}q_{z} \right)\,}. \label{eq: vertical stress-free}
\end{eqnarray}

\subsection{\label{S: Heat transport} Heat transport}

Heat transport is an essential diagnostic in convection-dominated systems, with much work for laboratory and natural systems still ongoing   \citep{doering_etal_2019,schumacher_sreenivasan_2020, vasil_etal_2021,lohse_shishkina_2024}. 
Rapid rotation adds many additional nuances \citep{julien_etal_2012,stellmach_etal_2014,julien_etal_2016}.
In studies of bulk convection, heat transport from wall modes tends to confound measurements, and efforts have attempted to quantify and/or limit their influence \citep{ecke_etal_2022,terrien_etal_2023,zhang_etal_2024}.

At first glance, the reduced wall-mode equations do not appear to contain any means for vertical heat transport; there is no bulk vertical velocity. The resolution occurs when taking into account the sidewall heat fluxes. All the vertical heat transport happens in the sidewall layer, which manifests within the bulk via $q_{z}$.

Defining the horizontal average, $\overline{T}$, the mean thermal equation,
\begin{eqnarray}
\pd{t}\overline{T} +  \pd{z} \left[  \frac{1}{|\mathcal{A}|}\oint_{\partial \mathcal{A}}   \! q_{z} T   \dd{\ell} - \kappa \, \pd{z} \overline{T}\right]  = 0.
\end{eqnarray}
The integrated boundary term follows straightforwardly from integrating the bulk equation and applying \eq{wall-conservation}. When the system is in a statistically steady state, the Nusselt number is a global constant with  
\begin{eqnarray}
\Nu = \frac{1}{|\mathcal{A}|}\oint_{\partial \mathcal{A}}   \! q_{z} \Theta   \dd{\ell} - \kappa \, \pd{z} \overline{T},
\end{eqnarray}
where in this case $\Theta = T - \overline{T}$.

When considering the boundary terms in \eq{eq: total thermal energy}, we also now find the nontrivial wall-modes ``power integral'' \citep{howard_1963} that regulates overall input and dissipation,
\begin{eqnarray}
\frac{1}{2} \frac{d}{dt} \btropic{ \overline{\Theta^{2}} }  = \frac{ 1 }{|\mathcal{A}|} \oint_{\partial \mathcal{A}} \btropic{ q_{z} \,\pd{z}\overline{T}\,  \Theta }  \dd{\ell} - \kappa \btropic{\overline{|\nabla \Theta|^{2}}}.
\end{eqnarray}
In both cases, it is clear the vertical velocity behaves precisely as a Dirac-$\delta$ distribution supported on the boundary; e.g., in local Cartesian coordinates with the wall at $x=0$,
\begin{eqnarray}
w(x,y,z) = q_{z}(y,z) \, \delta(x),
\end{eqnarray}
where the $\delta$-function replaces the detailed behaviour within the sidewall layer.

\section{\label{S: linear theory} Linear Instability}

From now on, we revert to the same non-dimensionalisation used throughout the asymptotic analysis. However, we do not use mixtures of lowercase and capital letters to distinguish between boundary layer and bulk quantities. All variables and coordinates are bulk-only; hence, we use their standard nomenclature. 

For the temperature perturbations linearised around a vertical gradient, $T = R\, (1-z) + \theta(x,y,z)$,
\begin{eqnarray}
& \pd{t}\theta =  \nabla^{2}\theta , \qquad \theta =  \pd{z}p, \\ 
& R\, q_{z}  =  \pd{n} \theta , \qquad 
 \pd{\ell}q_{\ell} + \pd{z}q_{z} = - \pd{\ell}p, \qquad  
 q_{\ell} = -  \Hilbert{q_{z}}.
\end{eqnarray}
The linearised equations collapse into a single equation and set of boundary conditions for the temperature perturbations. 
\begin{eqnarray}
& \pd{t}\theta = \nabla^{2}\theta \qquad \text{in} \qquad \mathcal{A} \times [\,0,1\,], \\ 
& \left(\pd{\ell} + |\pd{z}| \right)  |\pd{z}|\,  \pd{n} \theta =  R\,\pd{\ell} \theta \qquad \text{on} \qquad \partial \mathcal{A} \times [\,0,1\,] , 
\end{eqnarray}
where $|\pd{z}| =  \mathcal{H}\, \pd{z}$. For the top and bottom boundary conditions, 
\begin{eqnarray}
\theta\big|_{z=0}  = \theta\big|_{z=1}  = 0.
\end{eqnarray}

\subsection{\label{S: Semi-infinite} Semi-infinite channel}

This section reproduces the final results in \citet{herrmann_busse_1993} with the advantage that now all aspects are physically apparent. We assume a semi-infinite domain with the wall-normal coordinate, $- \infty < x \le 0$ and harmonic dependence along the wall, $\sim e^{i k y}$, with wavenumber $k$.

The following exponential profile satisfies the boundary condition at $x=0$,
\begin{eqnarray}
\theta &=& e^{\,\beta \, x   + i\, (k y +\omega t) } \sin(\pi z) + \text{c.c.}  \qquad \text{where} \qquad \beta = \frac{k (k+i \pi ) R }{\pi  \left(k^{2}+\pi^{2}\right)}.
\end{eqnarray}
Assuming $R>0$, the temperature perturbations decays as $x\to -\infty$.  The bulk evolution equation gives the dispersion relation for the complex-valued frequency 
\begin{eqnarray}
\omega &=& \frac{2k^{3}R^{2} }{\pi (\pi^{2}+k^{2})^{2} }  
- i\, \frac{k^{2} (k^{2}-\pi^{2}) R^{2} - \pi^{2} (\pi^{2}+k^{2})^{3}}{\pi^{2} (\pi^{2}+k^{2})^{2}}.
\end{eqnarray}
The real-valued growth rate, 
\begin{eqnarray}
\gamma = -\mathfrak{Im}(\omega) = \frac{k^{2} (k^{2}-\pi^{2}) R^{2} - \pi^{2} (\pi^{2}+k^{2})^{3}}{\pi^{2} (\pi^{2}+k^{2})^{2}}.
\end{eqnarray}
The growth rate is positive for 
\begin{eqnarray}
R \ \ge \ R_{\text{c}}(k)  = \frac{\pi (k^{2}+\pi^{2})^{3/2}}{k (k^{2}-\pi^{2})^{1/2} }  \ge R_{\text{c}} (k_{\text{c}}) = \pi^{2}\sqrt{6\sqrt{3}} \ \approx \ 31.8167. \label{R vs k}
\end{eqnarray} 
\cite{zhang_etal_2024} fit a quadratic curve through their stability data near the minimum of the form, 
\begin{align}
    \frac{R_{\text{c}}(k)}{R_{\text{c}}(k_{\text{c}})} = 1 + \xi^{2} ( k - k_{\text{c}})^{2} + \mathcal{O}((k - k_{\text{c}})^{3} ).
\end{align}
They find $\xi \approx 0.18$ for $\Ek = 10^{-6}$. Equation \eq{R vs k} implies $\xi = \sqrt{2 - \sqrt{3}}/\pi \approx 0.164769$. 

The respective critical wavenumber and frequency are 
\begin{eqnarray}
k_{\text{c}}=\pi \sqrt{2+\sqrt{3}} \approx 6.0690, \qquad \omega_{\text{c}} = 2 \pi^{2}\sqrt{3 (2 + \sqrt{3})} \approx 66.0487.
\end{eqnarray}
The sign convention for frequency means that $\omega > 0$ moves retrograde to the background rotation, $\Omega$. The respective phase and group speeds, 
\begin{eqnarray}
\frac{\omega_{\text{c}}}{k_{\text{c}}} = 
2 \sqrt{3} \, \pi \approx 10.8828, \qquad \frac{\partial \omega}{\partial k}\Big|_{k_{\text{c}},R_{\text{c}}} = 
2 (\sqrt{3} - 2)\,\pi \approx -1.68357,
\end{eqnarray}
within $\approx\,$20\% of the group velocity of $-2.1$ reported in the laboratory experiments of \cite{ning_ecke_1993} for $\Ek=10^{-3}$.
Notably, the group speed is always prograde above onset. 

Even though we remove the explicit Stewartson boundary layer, at onset, the temperature perturbations remain quite localised near the wall (on the bulk scale, $x$), where 
\begin{eqnarray}
\beta_{\text{c}}  = \pi \sqrt{3+2 \sqrt{3}} + i \, \pi\,3^{1/4}  \ \approx \ 7.9873 + 4.1345 \, i. 
\end{eqnarray}
In physical units, the critical wavelength along the wall is approximately $\lambda_{y} \approx 1.04\, H$ and the $e$-folding length away from the wall is $\lambda_{x} \approx 0.13\, H$.

For large $R$ the growth rate, 
\begin{eqnarray}
\gamma \sim  \frac{R^{2}}{\pi^{2}} -\frac{k^{4}+3 R^{2}}{k^{2}} + \Order{1} \qquad \text{as} \qquad R \to \infty. 
\end{eqnarray}
The fastest-growing wavenumber, $k$ maximises the 2nd term with $k \sim 3^{1/4} \sqrt{R}$ leaving 
\begin{eqnarray}
\gamma \sim \frac{R^{2}}{\pi^{2}} - 2 \sqrt{3} R + \Order{1} \qquad \text{as} \qquad R \to \infty. 
\end{eqnarray}
The severe dependence of $\gamma$ on $R$ makes nonlinear simulations very stiff at high values of criticality. For example, when $R = 5 \, R_{\text{c}}(k_{\text{c}})$, $\gamma \approx 2500$. Fast timescales exist for the propagation as well. As $R\to \infty$ the phase and group speeds are exactly opposite, 
\begin{eqnarray}
    c_{p} \sim \frac{2 R}{\sqrt{3} \pi} - \frac{4\pi}{3}, \qquad c_{g} \sim -\frac{2 R}{\sqrt{3} \pi} - 4\pi.
\end{eqnarray}

Reintroducing physical parameters implies the dimensional growth rate in the large-$R$ limit,
\begin{eqnarray}
\gamma^{*} \sim \frac{(g\alpha \Delta T)^{2}}{(2\Omega)^{2} \kappa },
\end{eqnarray}
which is notably independent of the container size.

\subsection{\label{S: Linear correction} Ekman-pumping correction}

In \S\ref{S: Intro}, we report the critical Rayleigh numbers, including the leading-order corrections due to Ekman pumping effects in \eqs{eq: Ra_c bulk}{eq: Ra_c wall}. The $\mathcal{O}(\Ta^{7/12})$ correction to the bulk $\Ra$ has an interesting history. The $\mathcal{O}(\Ta^{5/12})$ correction to the wall $\Ra$ has not, to our knowledge, been calculated before.

In his famous book, \cite{chandrasekhar_1961} calculated the leading-order contribution for bulk modes (unaware of wall modes at the time) using sound asymptotic methods, 
\begin{eqnarray}
\Ra \sim (27 \pi^{4}/4)^{1/3} \, \Ta^{\,2/3} , \qquad \text{as} \qquad \Ta \to \infty.
\end{eqnarray}
However, when comparing to numerical calculations up to $\Ta \approx 10^{12}$ he noticed a discrepancy for no-slip boundary conditions that appeared like an offset in the numerical proportionality factor, commenting  \textit{``... the discussion is not `fine' enough to account for the differences in the constants of proportionality ...''} \citep[][\S 27\textit{d} ``The origin of the $T^{\frac{2}{3}}$-law'']{chandrasekhar_1961}. The resolution of the issue is that the relative leading-order correction is $\propto \Ta^{-1/12}$, which produces a noticeable change for realistic numerical and laboratory parameter values. Ekman pumping manifests especially clearly in heat-transport enhancement \citep{king_etal_2009,stellmach_etal_2014,julien_etal_2016}. The upshot is that Chandrasekhar's discussion was `fine' enough; it simply did not go far enough. 

The issue became fully resolved when \cite{zhang_roberts_1998} showed the leading-order correction results from considering the Ekman pumping boundary conditions of the form we use in \S\ref{S: Ekman} \eq{Ekman flux condition}, together with a leading-order interior equation of the form as \eq{eq: bulk funny-laplacian}. The critical bulk Rayleigh number and wavenumber,
\begin{eqnarray}
\Ra &\sim & (27 \pi^{4}/4)^{1/3} \, \Ta^{\,2/3}  -   (8192\,\pi^{4})^{1/6} \Ta^{\,7/12}, \\
k &\sim & (\pi^{2}/2)^{1/6} \Ta^{1/6} - (27 \pi^{4}/4)^{-1/3}  \Ta^{1/12},
\end{eqnarray}
as $\Ta \to \infty$.
We use analogous techniques to find the equivalent correction for wall modes. 

The Ekman flux in the bulk is much too small to produce a significant correction. However, in the sidewall,
\begin{eqnarray}
\label{eq: next-order u}
\pd{x}p_{1/2} - v_{-1/2}  &=& 0, \\ 
\label{eq: next-order v}
\pd{Y} p_{1/2} + u_{1/2}  - \pd{x}^{2}v_{-1/2} &=& 0, \\ 
\label{eq: next-order w}
\pd{Z} p_{1/2} - \pd{x}^{2} w_{-1/2} 
&=& \Theta_{1/2}, \\ 
\label{eq: next-order p}
\pd{x}u_{1/2} 
+ \pd{Y}v_{-1/2} 
+ \pd{Z}w_{-1/2} 
&=& 0.
\end{eqnarray}
The boundary condition is the key difference from the leading order,  
\begin{eqnarray}
    w_{-1/2} = \mp \frac{\pd{x}^{2}p_{0}}{\sqrt{2}} \qquad \text{at} \qquad Z = 0, 1.
\end{eqnarray}
Projecting the system on to the relevant $2\cos(n\pi Z)$ and $2\sin(n\pi Z)$ functions,
\begin{eqnarray}
\label{eq: next-order u hat}
\pd{x}\hat{p}_{n,1/2} - \hat{v}_{n,-1/2}  &=& 0, \\ 
\label{eq: next-order v hat}
\pd{Y} \hat{p}_{n,1/2} + \hat{u}_{n,1/2}  - \pd{x}^{2}\hat{v}_{n,-1/2} &=& 0, \\ 
\label{eq: next-order w hat}
- \alpha_{n}^{3} \, \hat{p}_{n,1/2} - \pd{x}^{2} \hat{w}_{n,-1/2} 
&=& \widehat{\Theta}_{n,1/2}, \\  
\label{eq: next-order p hat} 
\pd{x}\hat{u}_{n,1/2}
+ \pd{Y}\hat{v}_{n,-1/2}
+ \alpha_{n}^{3} \hat{w}_{n,-1/2}
&=& \! \! \! \! \sum_{m+n = \text{even}} \! \! \! \!  \sqrt{2}\,\alpha_{m}^{2} \widehat{Q}_{m} \mathcal{X}''(\alpha_{m} x),
\end{eqnarray}
where recalling $\alpha_{n} = (n\pi)^{1/3}$.
The right-hand side results from integrating by parts in $Z$ and using the Ekman boundary conditions,
\begin{eqnarray}
    2 \int_{0}^{1}\cos(n \pi Z) \, \pd{Z} w(Z) \dd{Z} = n \pi\, \hat{w}_{n} + 2\left[(-1)^{n} w(1) -  w(0) \right],
\end{eqnarray}
where by definition 
\begin{eqnarray}
    \hat{w}_{n} = 2 \int_{0}^{1}\sin(n\pi Z)\,  w(Z) \dd{Z}.
\end{eqnarray}
The solution to the system is straightforward but messy. However, we only need the relationship between $q_{Y}$ and $q_{Z}$ to close the system. 

The final result in general form (dropping the $X, Y, Z$ bulk notation),
\begin{eqnarray}
q_{\ell} + \Hilbert{q_{z}} = -\sqrt{\eps} \, \mathcal{E} [q_{\ell}] \qquad \text{where} \qquad \sqrt{\eps} = \Ek^{\,1/6}. \label{eq: ekman flux correction}
\end{eqnarray}
The right-hand side contains the highly nonlocal operator 
\begin{eqnarray}
\widehat{\mathcal{E} [q]}_{n} = \frac{2\sqrt{2}}{\pi^{2/3}}\sum_{m+n=\text{even}} \frac{m}{(n m)^{1/3} (m+n)} \, \widehat{q}_{m}.
\end{eqnarray}

We can now continue to find the $\Order{\eps^{1/2}}$ perturbation to $R, \omega,k$ from the semi-infinite stability problem: 
\begin{eqnarray}
& \pd{t}\theta_{1/2} + \pd{\tau} \theta_{0} =  \nabla^{2}\theta_{1/2} , \\ & \theta_{1/2} =  \pd{z}p_{1/2}, \\ 
& R_{0}\, q_{z,1/2} + R_{1/2}\, q_{z,0}  =  \pd{x} \theta_{1/2} , \\ 
& 
 \pd{y}q_{y,1/2} + \pd{z}q_{z,1/2} = - \pd{y}p_{1/2} , \\
&  q_{y,1/2} + \Hilbert{q_{z,1/2}} = -\frac{\sqrt{2}}{\pi^{2/3}} q_{y,0}.
\end{eqnarray}
The system has an extra $R_{1/2}$ term along with an $\Order{\eps^{1/2}}$ timescale, $\pd{\tau}$. For a given wavenumber along the wall, $k$, 
\begin{eqnarray}
R_{1/2} = -\frac{\sqrt{2} \pi^{1/3 } k \left(k^{2}-3 \pi^{2}\right) \sqrt{k^{2}+\pi^{2}}}{ \left(k^{2}-\pi^{2}\right)^{3/2}}, \qquad 
\omega_{1/2} = \frac{2\sqrt{2} \,\pi^{1/3} k \left(k^{2}+\pi^{2}\right)^{2}}{\left(k^{2} - \pi^{2}\right)^{2}} .
\end{eqnarray}
Optimising the combined reduced Rayleigh number, $R \sim R_{0}(k) + \sqrt{\eps}\, R_{1/2}(k)$, over $k$ yields
\begin{eqnarray}
\label{asymptotic Rc}
& R_{\rm c}  \sim  
\sqrt{6\sqrt{3}} \, \pi^{2} - \sqrt{\eps}\, \sqrt{4\sqrt{3}-6}\, \pi^{4/3} 
\ \approx \  31.8167 - 4.4329 \sqrt{\eps}, \\ 
& k_{\rm c} \sim   
\sqrt{2 + \sqrt{3}}\, \pi + \sqrt{\eps}\,\frac{5\,\left(1+\sqrt{3}\right)}{6} \,\pi^{1/3} 
\ \approx \  6.06909 + 3.33445 \sqrt{\eps}, \\ 
& \omega_{\rm c} \sim   
2 \sqrt{3 (2+\sqrt{3})} \, \pi^{2} + \sqrt{\eps}\, \frac{\left(23+13 \sqrt{3}\right)}{3} \pi^{4/3} 
\ \approx \  66.0487 + 69.8097 \sqrt{\eps},
\label{asymptotic omegac}
\end{eqnarray}
as $\eps \to 0$. The next section computes the critical onset parameters directly.

\subsection{\label{S: Direct numerical} Direct  numerical calculations}

We validate the asymptotic correction from \S\ref{S: Linear correction} with high-precision direct numerical calculations of the critical Rayleigh number, wave number and precession frequency over a range of very small but finite Ekman numbers. We solve a linearised version of \eqss{eq: boussinesq-u}{eq: boussinesq-Theta} in a finite Cartesian channel assuming $y$-direction Fourier dependence with a complex-valued growth rate, \textit{i.e.},
\begin{eqnarray}
\label{full linear u}
\frac{1}{\sigma}  (\gamma + i\, \omega)\, u &+& \frac{\pd{x} p - v}{\Ek}  = (\pd{x}^{2} + \pd{z}^{2} - k^{2})\, u, 
\\
\frac{1}{\sigma}  (\gamma + i\, \omega) \, v &+& \frac{ik\, p + u}{\Ek}  = (\pd{x}^{2} + \pd{z}^{2} - k^{2})\, v, 
\\
\frac{1}{\sigma}  (\gamma + i\, \omega)\, w &+& \frac{\pd{z} p - \theta}{\Ek}  = (\pd{x}^{2} + \pd{z}^{2} - k^{2})\, w, 
\end{eqnarray}
\begin{eqnarray}
\pd{x} u + ik\,  v + \pd{z} w  = 0,
\end{eqnarray}
\begin{eqnarray}
(\gamma + i\, \omega)\, \theta - R \, w   = (\pd{x}^{2} + \pd{z}^{2} - k^{2})\, \theta,
\end{eqnarray}
Nominally, the system has no-slip boundary conditions on all walls. For the top and sides of the domain,
\begin{eqnarray}
u \ = \  v \ = \  w \ = \  0  \qquad \text{at} \qquad z = 1, \quad \text{and} \quad x = 0,\, \Gamma.
\end{eqnarray}
However, vertical reflection symmetry, $z \to 1-z$, implies we can save computational cost by restricting to only the upper-half domain with mid-plane conditions,
\begin{eqnarray}
u \ = \  v \ = \  p \ = \ \pd{z} \theta \ = \ 0  \qquad \text{at} \qquad z=\frac{1}{2}.
\label{full linear mid-plane bc}
\end{eqnarray}
The perfectly insulating thermal boundary condition at $x=\Gamma$ implies $\pd{x} \theta = 0$. We impose perfectly conducting conditions at all other walls, $\theta = 0$ at $x=0$ and $z=1$. The conducting condition at $x=0$ isolates the dynamics only to the $x=\Gamma$ boundary with quasi-exponential decay into the interior. 

We use Dedalus to implement a continuous Galerkin spectral-element solver for \eqss{full linear u}{full linear mid-plane bc} with a $3 \times 3$ array of elements scaled to capture the extreme boundary-layer behaviour for $\Ek \ll 1$. Each subdomain uses a tensor-product discretisation with between 20-120 polynomial modes in each direction, chosen so that the solutions in each element are resolved to a truncation error of $\approx 10^{-6}$.

We find the minimum onset value for $R$ as a function of $k$ at fixed $E$ by solving the simultaneous equations for the growth rate, 
\begin{eqnarray}
\gamma(k,R) = \pd{k} \gamma(k,R) = 0.
\label{gamma dgamma eq}
\end{eqnarray}
We solve these equations numerically by computing $\gamma(k,R)$ on a $3\times 3$ grid of $(k,R)$ input values in the vicinity of the leading-order solution and interpolating with a local polynomial approximation of the form 
\begin{eqnarray}
\gamma(k,R) \approx \gamma_{0,0} + \gamma_{1,0}\, k + \gamma_{2,0}\, k^{2} + \gamma_{0,1}\, R.
\end{eqnarray}
We update the approximate solution to \eq{gamma dgamma eq} from the polynomial fit via,
\begin{eqnarray}
k \approx - \frac{\gamma_{1,0}}{2 \gamma_{2,0}}, \qquad 
R \approx   \frac{\gamma_{1,0}^{2}}{4 \gamma_{0,1} \gamma_{2,0}} -\frac{\gamma_{0,0}}{\gamma_{0,1}}.
\end{eqnarray}
The true solution need not lie exactly within the original $3\times 3$ grid; the polynomial approximation can extrapolate slightly outside the fitting range.  
We perform the procedure with both Chebyshev and Legendre polynomial discretisations to estimate the numerical uncertainty in the critical parameters, which are in the 4th digit or smaller.
The asymptotic results are independent of $\sigma$. We set the Prandtl number $\sigma = 1$ throughout. We use both $\Gamma = 2, 4$ over the whole range of $\Ek$. The results agree to within the other uncertainties of the calculation. 

\begin{figure}
\begin{center}
\vspace{+0.10in}
\includegraphics[width=\textwidth]{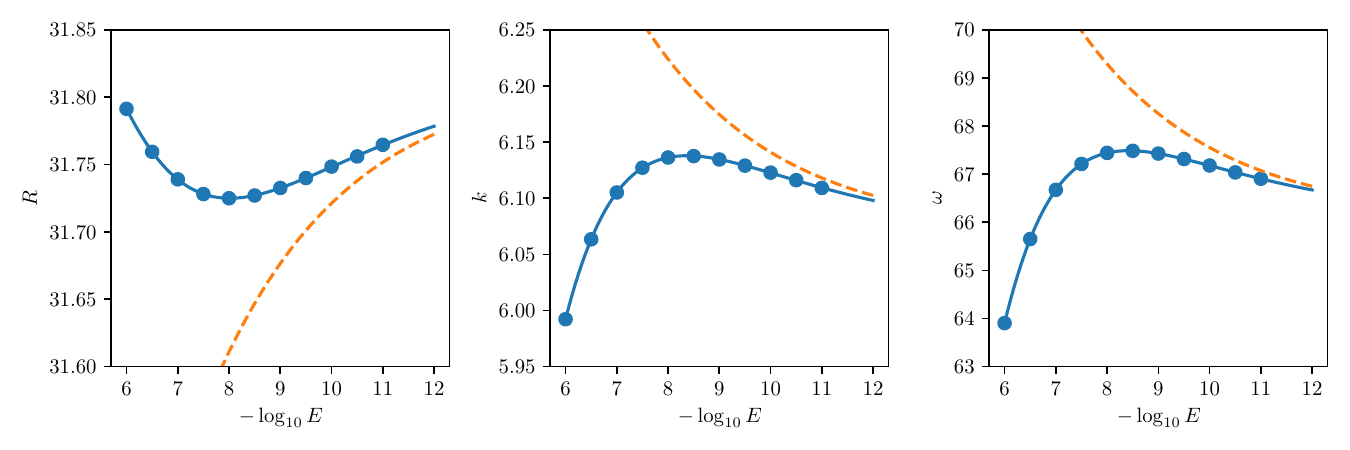}
\caption{Plots of critical parameters versus Ekman number for $6 \le -\log_{10} \Ek \le 11$ from the spectral-element calculations of \eqss{full linear u}{full linear mid-plane bc}. From left to right, the panels show scaled Rayleigh number, $R$, tangential wave number, $k$, and precession frequency, $\omega$.  
Table \ref{fit parameters} gives the coefficients of the best-fit expansions in powers of $E^{1/6}$. The orange dashed lines show the asymptotic predictions in \eqss{asymptotic Rc}{asymptotic omegac}.
\label{fig: DNS R vs E}}
\end{center}
\end{figure}

We scan over a range of powers of Ekman number with $- \log_{10} \Ek \ge 6$. These go to 11. 
Table~\ref{tab:to 11} reports the onset critical parameters, $(R, k, \omega)$, from the scans over $\Ek$.
We perform five-term fits for the critical parameters in powers of $\eps = \Ek^{\,1/3}$ as
\begin{eqnarray}
    R(\eps) & \approx & R_{0} + R_{1/2} \, \eps^{1/2} + R_{1}\, \eps + R_{3/2} \, \eps^{3/2} + R_{2}\, \eps^{2}, \\ 
    k(\eps) & \approx & k_{0} + k_{1/2} \, \eps^{1/2} + k_{1}\, \eps + k_{3/2} \, \eps^{3/2} + k_{2}\, \eps^{2}, \\ 
    \omega(\eps) & \approx & \omega_{0} + \omega_{1/2} \, \eps^{1/2} + \omega_{1}\, \eps + \omega_{3/2} \, \eps^{3/2} + \omega_{2}\, \eps^{2}.
\end{eqnarray}
Fig.~\ref{fig: DNS R vs E} shows plots over the values covered along with the respective bit-fit curves. Table~\ref{fit parameters} shows the best-fit parameters. The $R_{0}, \,R_{1/2}$, $k_{0}, \,k_{1/2}$ and $\omega_{0}, \,\omega_{1/2}$ all agree with the predictions in \S\ref{S: Linear correction} to the number of reported digits. 

\begin{table}
\begin{center}
\begin{tabular}{ |c|c|c|c| } 
   \ $-\log_{10}\Ek$ \ & \ $R$ \ &\ $k$ \ & \ $\omega$ \\ 
 \hline
  \hline
6.0 & \, 31.7913 \, & \, 5.9922 \, & \, 63.905 \, \\
6.5 & \, 31.7594 \, & \, 6.0634 \, & \, 65.652 \, \\
7.0 & \, 31.7390 \, & \, 6.1051 \, & \, 66.673 \, \\
7.5 & \, 31.7280 \, & \, 6.1271 \, & \, 67.211 \, \\
8.0 & \, 31.7249 \, & \, 6.1362 \, & \, 67.440 \, \\
8.5 & \, 31.7270 \, & \, 6.1375 \, & \, 67.483 \, \\
9.0 & \, 31.7326 \, & \, 6.1345 \, & \, 67.427 \, \\
9.5 & \, 31.7399 \, & \, 6.1290 \, & \, 67.314 \, \\
10.0 & \, 31.7484 \, & \, 6.1227 \, & \, 67.179 \, \\
10.5 & \, 31.7560 \, & \, 6.1160 \, & \, 67.036 \, \\
11.0 & \, 31.7646 \, & \, 6.1091 \, & \, 66.904 \, \\
 \hline  
 $\infty$ & \, 31.8167... \, & \, 6.0691... \, & \, 66.049... \,  \\ 
 \hline
 \hline
\end{tabular}
\end{center}
\caption{Direct spectral element calculations of the stability threshold at various Ekman number.  Values of $R$, $k$, and $\omega$ have estimated respective uncertainties of 4e-4, 3e-4, and 3e-3, with asymptotic analytical results reported to the same number of digits. \label{tab:to 11}} 
\end{table} 

\begin{table}
\begin{center}
\begin{tabular}{ |c|c|c|c|c|c| } 
  \, $\Order{\eps^{p}}$ \, &  0 & 1/2 & 1 & 3/2 & 2 \\ 
 \hline
  \hline
 $R$ \, & \, 31.817(2) \, & \, -4.5(2) \, & \, \ 65(8) \, & \, -253(102) \, & \, 258(450) \, \\ 
 \hline
 $k$ \, & \,  6.067(2) \, & \, \ 3.5(1) \, & \, -42(5) \, & \, -28(63) \, & \, 290(276) \, \\ 
 \hline
 $\omega$ \, & \,  66.06(2) \, & \, \ 69(2)  \, & \, -709(56) \, & \, -3573(718)  \, & \, 16375(3168) \, \\ 
 \hline
 \hline
\end{tabular}
\end{center}
\caption{Best-fit coefficients for critical parameters as a function of powers of $\eps = \Ek^{\,1/3}$, computed using spectral-element calculations of \eqss{full linear u}{full linear mid-plane bc} and plotted in fig.~\ref{fig: DNS R vs E}. \label{fit parameters} }
\end{table}

\subsection{\label{S: Finite cylinder} Finite cylinder}

In the finite cylinder case, we use polar coordinates, $(r,\phi,z)$, where the radius $0 \le r \le \Gamma / 2$ and azimuth angle $0 \le \phi < 2 \pi$, where $\Gamma$ denotes the non-dimensional diameter or aspect ratio.

In this case, we solve the bulk equation first and substitute the result into the boundary condition. The radial dependence uses modified Bessel functions of the 1st kind, $I_{m}$, with integer wavenumber, $m \in \mathbb{Z}$, for $\beta = \sqrt{\pi^{2} + i\, \omega}$, 
\begin{eqnarray}
\theta(r,\phi,z)  = I_{m}(\beta\, r) \, e^{i (m \phi + \omega  t)} \sin(\pi z) + \text{c.c.}. 
\label{eq:dispersion relationship with aspect}
\end{eqnarray} 

The dispersion relation follows from the boundary condition applied at the outer radius,
\begin{eqnarray}
R  = \pi \beta \left(1 + \frac{\pi \Gamma}{2 i m} \right)  \frac{I_{m}'(\beta\, \Gamma/2)}{I_{m}(\beta\, \Gamma/2)} \qquad \text{where} \qquad \mathfrak{Im}(R) = 0.
\label{cylinder linear R}
\end{eqnarray}
Given $m$ and $\Gamma$ we solve $\mathfrak{Im}(R) = 0 $ using Newton's method for real-valued $\omega$. We define $R_{\text{c}}$ and $\omega_{\text{c}}$ as their respective values when minimising $R$ over integer $m$ values.

\begin{figure}
\begin{center}
\vspace{+0.10in}
\includegraphics[width=12cm]{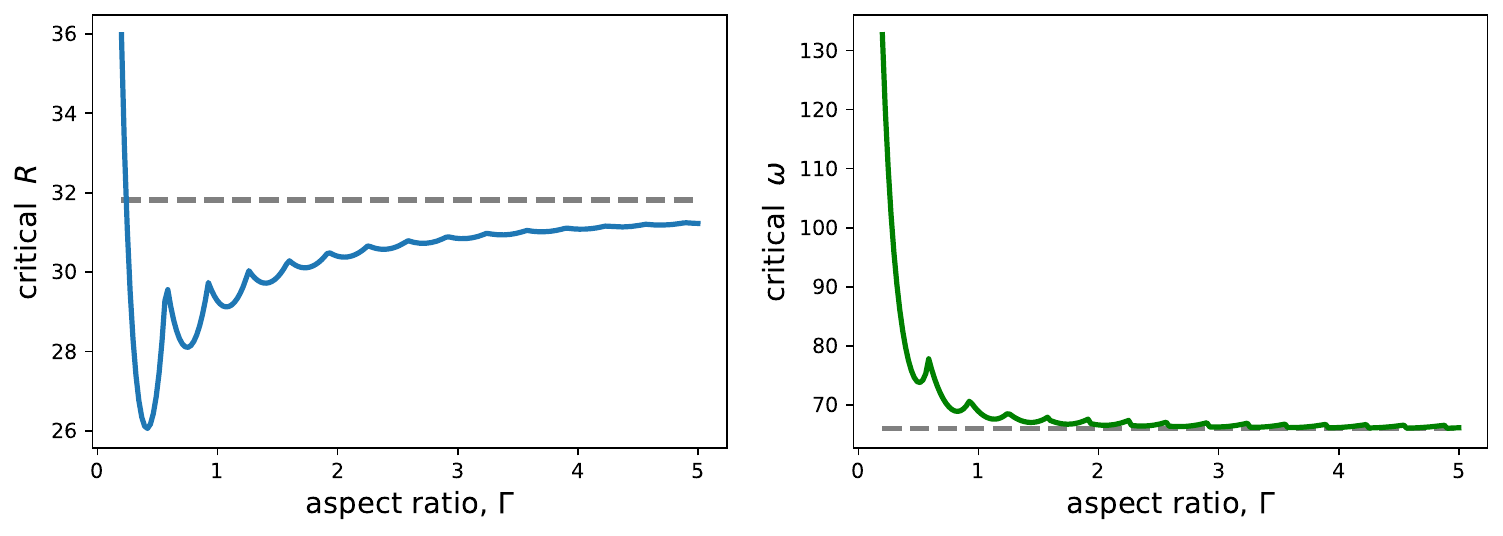}
\caption{Critical parameters for a finite cylinder with aspect ratio, $\Gamma$. In each case, the dashed lines show the value for the semi-infinite channel. The critical wavenumber is typically within one of $m \approx \lfloor \Gamma \,k_{\text{c}}/2  \rfloor$, where $k_{\text{c}}$ is the critical wavenumber from the semi-infinite case. Starting from $m=1$, for low aspect ratio, each cusp feature in $R_{\text{c}}$ indicates a unit increase in $m$. The overall minimum happens for $\Gamma \approx 0.366196$ and $m=1$ with $R \approx 26.3789$ and $\omega \approx 82.9854$.
\label{fig: critical cylinder parameters}}
\end{center}
\end{figure}

Fig.~\ref{fig: critical cylinder parameters} shows the dependence of $R_{\text{c}}$ and $\omega_{\text{c}}$ on aspect ratio, $\Gamma$. The cusp-like features in both plots correspond to discrete changes in the optimal $m$ value. At the lowest aspect ratios, the optimal $m=1$ and scales approximately as $m \sim \lfloor \Gamma \,k_{\text{c}}/2  \rfloor$. Compared to the various Ekman-number corrections the finite cylinder has a much stronger, $\Order{1}$, influence on the onset Rayleigh number.

\begin{figure}
\begin{center}
\vspace{+0.10in}
\includegraphics[width=12cm]{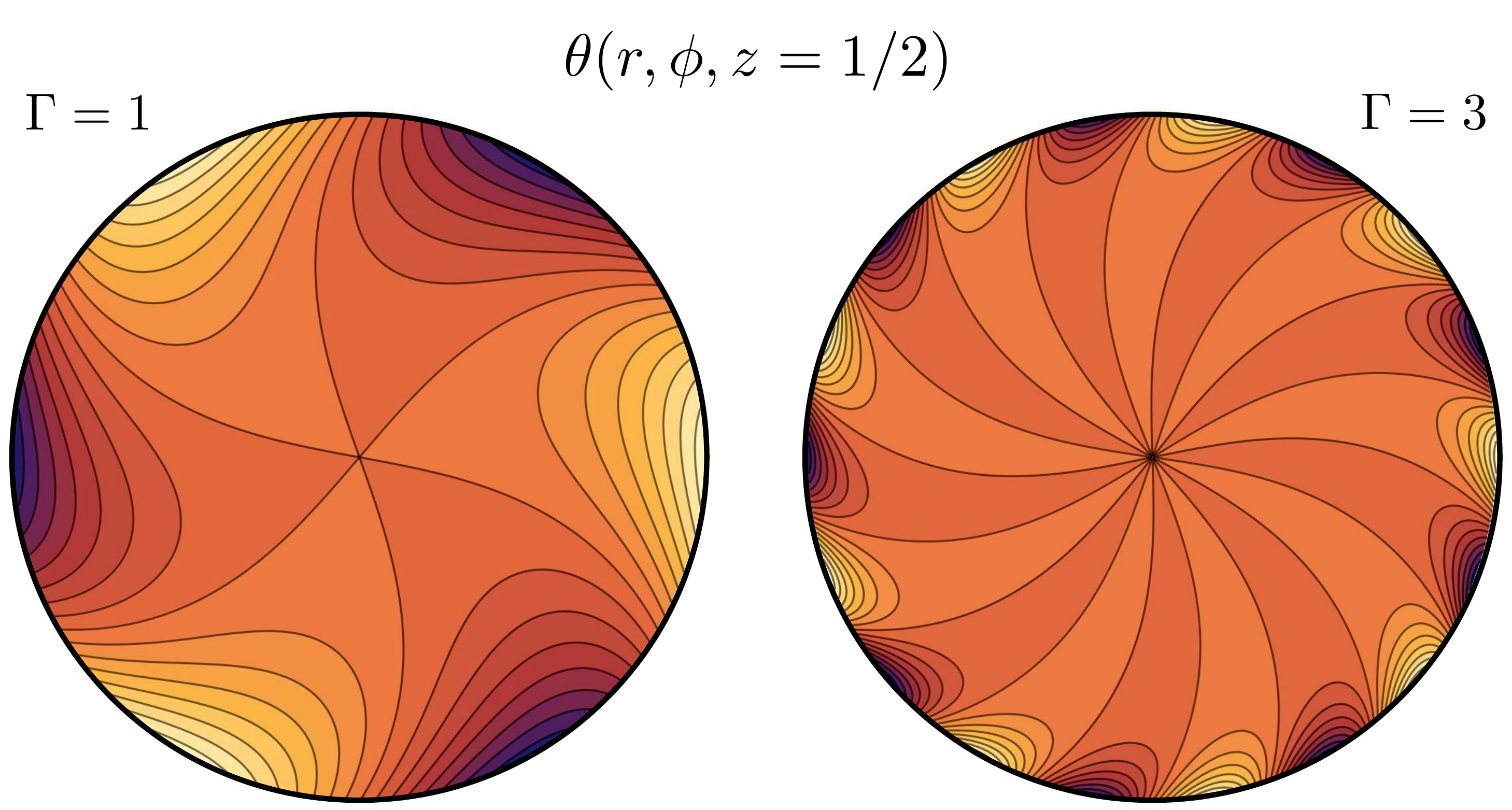}
\caption{The mid-plane linear temperature perturbations from two critical modes at aspect ratios $1$ and $3$, respectively. The $\Gamma=1$ case prefers and $m=3$ angular wavenumber, while $\Gamma = 3$ prefers $m=9$. 
\label{fig: temperature planforms}}
\end{center}
\end{figure}

Fig.~\ref{fig: temperature planforms} shows two critical mid-plane temperature perturbation profiles for respective aspect ratios, $\Gamma = 1, 3$.  We report the solutions from applying Newton's method to Equation~(\ref{eq:dispersion relationship with aspect}):

For $\Gamma = 1$, $R_{\text{c}} \approx 29.26899820255448$, $\omega_{\text{c}} \approx 68.93822823201539$ with $m_{\text{c}} = 3$. 

For $\Gamma = 3$, $R_{\text{c}} \approx 30.84620681276749$, $\omega_{\text{c}} \approx 66.29068835812032$ with $m_{\text{c}} = 9$. 

We also compare to finite $\Ek=10^{-6}$ calculations of \cite{zhang_etal_2024} with $\Gamma=1/2$, which find $R_{\rm c}\approx 28$ and $\omega_{\rm c} \approx  73.5$.
For asymptotically low $\Ek$, with $\Gamma = 1/2$, we find $R_{\text{c}} \approx 27.100728659192075$, $\omega_{\text{c}} \approx 73.83743415332742$ with $m_{\text{c}} = 1$.

\subsubsection{Tall aspect ratio}

In the past few years, multiple research groups \citep[e.g.,][]{julien_etal_2016, dewit_etal_2020, pandey_2024} have started using tall-aspect-ratio cylinders for rapidly rotating convection experiments; the reason being rotation's tendency to produce elongated vertical structures. The linear stability result in \eq{cylinder linear R} simplifies considerably as $\Gamma \to 0$. Assuming $\omega\,\Gamma \sim \mathcal{O}(1)$, 
\begin{align}
R  = \frac{2\pi  m}{\Gamma } - i \pi  \left(\pi -\frac{\omega\,\Gamma }{4(m+1)}\right) + \mathcal{O}(\Gamma).
\end{align}
Imposing $\mathfrak{Im}(R) = 0$ and minimising over $m$ implies 
\begin{align}
    R_{\rm c}\, \Gamma \, \sim \, 2\pi , \qquad \omega_{\rm c} \, \Gamma \, \sim \, 8\pi \qquad \text{as} \qquad \Gamma \, \to \, 0,
\end{align}
with $m_{\rm c} = 1$. Notably, the aspect ratio and scaled Rayleigh number exactly cancel the cylinder depth dependence in the onset condition,
\begin{align}
    \frac{g \alpha   \Delta T  D}{2\Omega\, \kappa  } \, = \, 2\pi,
\end{align}
where $D$ is the cylindrical diameter. Furthermore, as $
\Gamma \sim \Ek^{\,1/3}$ the critical Rayleigh number scales the same as for bulk onset, also with identical horizontal scales, $\mathcal{O}(\Ek^{\,1/3})$.

\subsection{\label{S: baroclinic instability} Local baroclinic instability}

When assembled in one place, it becomes apparent that the wall-mode system is closely akin to the well-known \textit{planetary geostrophy} (PG) equations, which typically model global-scale stratified ocean dynamics away from coastlines \cite[Ch. 5]{vallis_2017}. The PG system has rich dynamical structure. However, a common puzzle is the lack of self-consistent forcing and eventual rundown \citep{schonbek_vallis_1999}. Several works have considered coupling to smaller-scale \textit{quasi-geostrophic} (QG) dynamics as an intriguing alternative \citep[see][]{grooms_etal_2011}. 

This context clarifies that the wall-mode convection equations are a sideways forced PG-QG system that relies on non-hydrostatic QG \citep[similar to][]{sprague_etal_2006,julien_knobloch_2007} near lateral boundaries and interacts with a barotropic QG system in the stress-free case. However, we can also study the wall-mode equations without the lateral boundary interactions. 

In a highly reduced setting, the combined baroclinic temperature and barotropic vorticity equations have a rudimentary form of baroclinic instability. To demonstrate, we neglect all diffusion and boundary conditions. Hence,
\begin{eqnarray}
\pd{t} \pd{z} \psi + J( \psi, \pd{z}\psi) = 0, \qquad
\pd{t} \btropic{\nabla_{\!\bot}^{2} \psi} + \btropic{J(\psi,\nabla_{\!\bot}^{2} \psi )} = 0.
\end{eqnarray}
The baroclinic equation has a family of exact nonlinear solutions,
\begin{eqnarray}
\psi = P_{0}(z) + x\, V_{0}(z) + (c + V_{0}(z))\, A(y + c\, t) + \text{c.c.} ,
\end{eqnarray}
where $V_{0}(z)$ is an arbitrary $y$-direction velocity profile (without loss of generality, $\btropic{V_{0}} = 0$), $P_{0}(z)$ is and arbitrary hydrostatic pressure, $A(y+c\,t)$ is an arbitrary complex-valued amplitude function, and and $c$, is a complex-valued phase speed. 

The barotropic equation requires the solvability condition,
\begin{eqnarray}
\btropic{ (c + V_{0})^{2} } = 0 \qquad \implies \qquad c = \pm \,i\, \| V_{0} \|.
\end{eqnarray}
For example, $V_{0} = z-1/2$, $c = \pm i / \sqrt{12}$, or $V_{0} = \cos(\pi z)$,  $c = \pm i/\sqrt{2}$. When $A(y) = A_{0}\,e^{iky}$ for wavenumebr $k$, then the growth rate is $k \| V_{0} \|$.

\section{\label{S: Simulations} Nonlinear simulations}

Using Dedalus \citep{burns_etal_2020}, we simulate the full nonlinear system of reduced wall-mode equations summarised in \S\ref{S: Summary}. We use a full cylinder with radius $0 \le r \le \Gamma/2$, azimuth $0 \le \phi < 2 \pi$ and height $0 \le z \le 1$.  We use a spectral basis of generalised Zernike polynomials that naturally represent scalars, vectors and tensors in the full unit disk, including $r=0$ with no singularities; see \cite{vasil_etal_2016} for details.

\subsection{\label{sec: numerical setup} Setup for numerical implementation}

Our Dedalus implementation corresponds to the following description. 
We use the same non-dimensional form as the asymptotic analysis with control parameters $R$, $\sigma$ and $\Gamma$. We use $p(r,\phi,z)$ as the primary dynamical variable for the bulk. The respective thermal wind relations follow,
\begin{eqnarray}
\Theta = \pd{z} p, \qquad u_{\bot} = \mathcal{J} \cdot  \nabla_{\! \bot\, } p \qquad \text{where} \qquad 
\mathcal{J} = \hat{\phi} \otimes \hat{r} - \hat{r} \otimes \hat{\phi}.
\end{eqnarray}
We represent $p(r,\phi,z)$ in a vertical cosine series, $\cos(n \pi z)$. The temperature perturbations, $\Theta(r,\phi,z)$, are naturally a sine series $\sin(n \pi z)$. For $n \ge 1$, the baroclinic evolution equation 
\begin{eqnarray}
\pd{t}\Theta - ( \nabla_{\! \bot\, }^{2} + \pd{z}^{2}) \Theta = - u_{\bot} \cdot \nabla_{\! \bot\, } \Theta. \label{theta evolution}
\end{eqnarray}
We use a second-order accurate multi-stage, mixed implicit-explicit time-stepping scheme \citep[SBDF scheme from][]{wang_ruuth_2008}, with linear terms on the left-hand side treated implicitly and the nonlinear right-hand side treated explicitly. The time-step size follows a CFL criterion with a 0.2 safety factor. Because of the fine radial grid spacing near the walls, normal flow through the outer boundary principally determines the time-step size.

For the outer boundary conditions at $r=\Gamma/2$, we use $q_{z}(\phi,z)$ as an additional variable and define $q_{\phi} = -\Hilbert{q_{z}}$. The boundary conditions conditions are 
\begin{eqnarray}
\nabla_{\!\phi\,} q_{\phi} + \pd{z} q_{z} - \hat{r}\cdot u_{\bot} &=& 0, \\ 
R\, q_{z} - \hat{r}\cdot \nabla_{\! \bot\, } \Theta &=&  (q_{\phi} \nabla_{\!\phi\,} + q_{z} \pd{z}  ) \Theta, \label{theta BCs}
\end{eqnarray}
where $\nabla_{\!\phi\,} = \hat{\phi} \cdot \nabla_{\!\bot\,} = r^{-1} \pd{\phi}$. As before, all linear terms are time-implicit and nonlinear terms are explicit. 

The $n=0$ barotropic pressure mode has different dynamics depending on the velocity boundary conditions on the top and bottom. In both cases, we define the vorticity, $\zeta = \nabla_{\!\bot\,}^{2} p$.

For stress-free, we evolve the $n=0$ vorticity equation,
\begin{eqnarray}
\pd{t} \btropic{\zeta} - \sigma \nabla_{\!\bot\,}^{2} \btropic{\zeta} = - \nabla_{\!\bot\,}  \cdot \btropic{u_{\bot} \zeta},
\end{eqnarray}
with, again, the usual linear-nonlinear implicit-explicit time splitting. For boundary conditions at $r =\Gamma /2$,
\begin{eqnarray}
\btropic{p} = 0, \qquad - 2\sigma \, \hat{r} \cdot \nabla_{\!\bot\,} \btropic{p} = 3 \btropic{q_{\phi} \nabla_{\!\phi\,} q_{\phi} } - \btropic{q_{\phi} \nabla_{\!\phi\,} p }.
\end{eqnarray}

For no-slip, the $n=0$ mode involves a bit of subtlety. The Ekman boundary condition directly mixes barotropic and baroclinic variables, and parity mixes sine and cosine series. \cite{vasil_etal_2008a,vasil_etal_2008b}  discusses issues like this detail. The problem is that we need to represent expressions like \eq{eq: mean P - Phi} spectrally using sine/cosine series. We fix the barotropic pressure via
\begin{eqnarray}
\nabla_{\!\bot\,}^{2} \btropic{p} = \nabla_{\!\bot\,}^{2} \btropic{K_{\!N}  \Theta}, \label{NS-K-barotropic}
\end{eqnarray}
with the boundary condition $\btropic{p}=0$ at $r = \Gamma/2$.
We use the finite spectral representation of $1/2 -z$,
\begin{eqnarray}
K_{\!N}(z) = \sum_{n=1}^{N/2} \frac{\sin(2\pi n z)}{n\pi }  \quad \to \quad  \frac{1}{2} -z \qquad \text{as} \qquad N \to \infty.
\end{eqnarray}
A sufficiently large resolution parameter, $N$, captures the integral exactly. That is, 
\begin{eqnarray}
\int_{0}^{1} \! K_{N}(z) \, \sin(n \pi z) \dd{z} = \int_{0}^{1}\! \left(\frac{1}{2}-z \right)\,\sin(n \pi z) \dd{z} \quad \textit{(exactly)} 
\end{eqnarray}
for all $n \le N$. The advantage is that $K_{N}$ is a natural sine series having a natural inner product with $\Theta$.

\subsection{ \label{sec: Weakly nonlinear theory} Weakly nonlinear theory}

Weakly nonlinear theory \citep[see, e.g.,][]{hoyle_2006} is a useful diagnostic for validating calculations slightly above onset. We define the perturbation parameter, 
\begin{align}
    \delta = \frac{R}{R_{\rm c}} - 1. 
\end{align}
As $\delta \to 0$ we expect pressure perturbations of the form, 
\begin{align}
    p &= A(t)\, P_{1,1}(r) \,e^{i(m\phi + \omega_{\rm c} t)} \cos(\pi z) \, +  \nonumber \\ 
    & \left( |A(t)|^{2} \,P_{2,0}(r) + A(t)^{2}\,  P_{2,2}(r) \,e^{2i(m\phi + \omega_{\rm c} t)}  \right)(\cos(2\pi z) - 1) \, + \nonumber \\
    & \left( |A(t)|^{2} P_{2,0}(a) + A(t)^{2}\,P_{2,0}(a) (\tfrac{r}{a})^{2m} e^{2i(m\phi + \omega_{\rm c} t)}  \right) \, + \text{c.c.} + \mathcal{O}(\delta^{3/2}),
\end{align}
with a complex-valued amplitude varying on a slow timescale
\begin{align}
A(t) = \mathcal{O}(\delta^{1/2}), \qquad A'(t) = \mathcal{O}(\delta^{3/2}).
\end{align}
The pressure ansatz contains the leading-order convective modes derived from linear theory and the next-order convective feedback. At a given aspect ratio, we pick the most unstable wavenumber, $m$.  defining the non-dimensional radius, $a=\Gamma/2$. The radial functions are normalised modified Bessel functions, 
\begin{align}
    P_{n,j}(r) \, = \, \frac{I_{m}(r\beta_{n,j})}{ C_{n,j} }, \qquad \beta_{n,j} = \sqrt{(n\pi)^{2} + j\, \sqrt{-1} \, \omega}.
\end{align}
with the coefficients, 
\begin{align}
C_{1,1} &\, = \, I_{m}(a\beta_{1,1}), \\ 
C_{2,0} &\, = \, \frac{2 \left(\pi^{2} a^{2}+m^{2}\right)\beta_{2,0} I_{m}'(a\beta_{2,0})}{\pi m^{2}}, \\ 
C_{2,2} &\, = \, \frac{4 \pi  a (m - i \pi  a) \beta_{2,2} I_{m}'(a\beta_{2,2}) -2 a m R_{\rm c}I_{m}(a\beta_{2,2}) }{\pi  m (\pi  a+i m)},
\end{align} 
with $P_{n,-j}(r)= P^{*}_{n,j}(r)$.
The normalisations ensure the solution satisfies all horizontal boundary conditions, along with the no-slip barotropic condition \eq{NS-K-barotropic}; the stress-free barotropic condition would be considerably more complicated. 

The weakly nonlinear amplitude satisfies a complex Ginzburg-Landau equation, 
\begin{align}
    A'(t) =  (c_{0}\,\delta  - c_{1}\, |A(t)|^{2})\, A(t), \label{amplitude eq}
\end{align}
where 
\begin{align} 
    c_{0} \, = \, &  \tfrac{2 i \pi   m \left(\pi^{2}+i \omega \right) (\pi  a+i m) R_{\rm c}}{a^{2}
   m^{2} R_{\rm c}^{2}+\pi^{2} (\pi  a+i m)^{2} \left(m^{2}+a^{2} \left(\pi^{2}+i \omega 
   \right)\right)}, \\ 
c_{1} \, = \, & \tfrac{\frac{4 \pi m a}{m-i \pi  a}P_{2,0}(a)-2i m P_{2,2}(a) + m \int_{0}^{a} ( (4 \Phi_{2,2} - 10 P_{2,2}) \mathfrak{Im}(P_{1,1} P_{1,-1}') - 5i P_{1,1}^{2}P_{1,1}' )\dd{r}  }{2\int_{0}^{a} P_{1,1}^{2} r\dd{r} },
\end{align}
for $\Phi_{2,2} = P_{2,2}(a) (r/a)^{2m}$.
Linear theory is sufficient to determine $c_{0}$. The nonlinear coefficient, $c_{1}$ requires a messy perturbation series that closes at $\mathcal{O}(\delta^{3/2})$. There is no spatial modulation because of the discrete critical wavenumber, $m$.   

\begin{figure}
\begin{center}
\vspace{+0.10in}
\includegraphics[width=\textwidth]{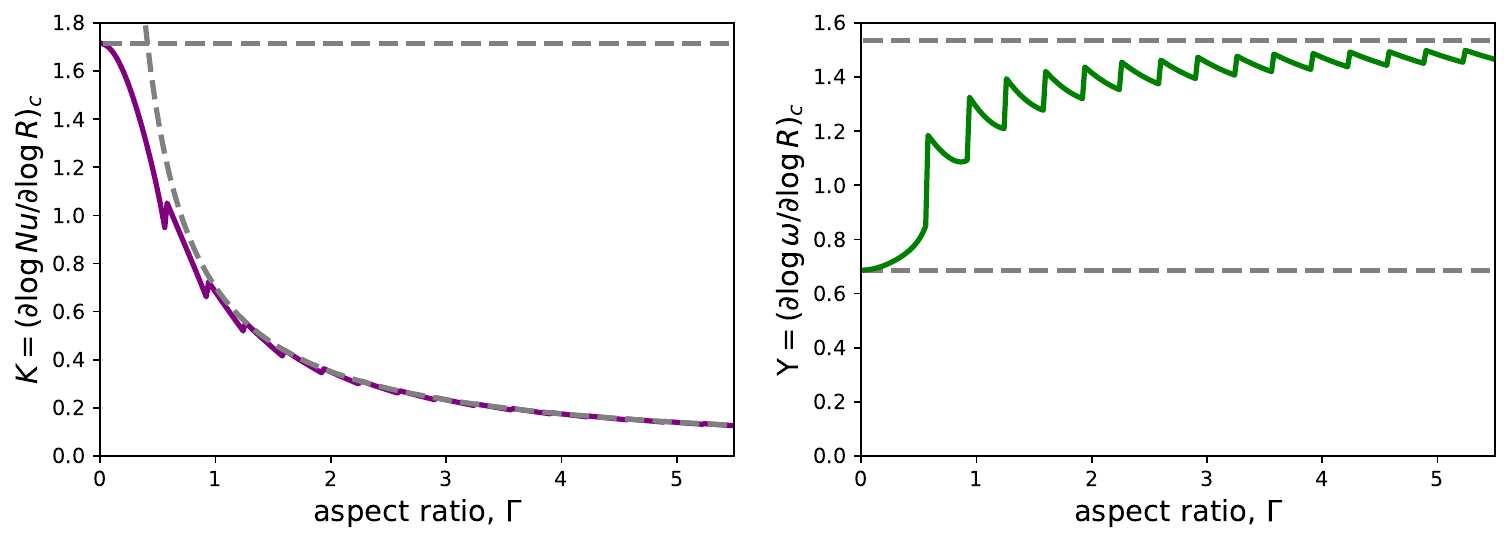}
\caption{The Nusselt number and relative frequency slopes as a function of aspect ratio in a finite cylinder. For the Nusselt number, the horizontal dashed grey line indicates the analytical value of $12/7$ as $\Gamma \to 0$ and the dashed grey curve shows the asymptotic scaling $\sim 0.701822/\Gamma$ as $\Gamma \to \infty$ derived from a flat Cartesian wall bounding an infinite bulk interior. For the relative frequency slope, the dashed grey curves indicate the analytical value $173/252$ as $\Gamma \to 0$ and the Cartesian result $\sim 1.533634$ as $\Gamma \to \infty$. The jagged behaviour in both plots results from jumps in the critical mode number, $m_{\rm c}$, as a function of aspect ratio.}
\label{fig: Nu-Om-crit}
\end{center}
\end{figure}

The amplitude in \eq{amplitude eq} converges to a travelling wave, 
\begin{align}
    A(t) \, \to \, \mathcal{A} \,e^{ i \Omega \, t},
\end{align}
with time-independent parameters,
\begin{align} 
    |\mathcal{A}|^{2} \, = \, \frac{\mathfrak{Re}(c_{0})  \,}{\mathfrak{Re} (c_{1}) } \, \delta,\qquad \Omega \, = \,   \frac{\mathfrak{Im}(c_{0})\, \mathfrak{Re} (c_{1}) - \mathfrak{Im}(c_{1})\, \mathfrak{Re}(c_{0}) }{\mathfrak{Re} (c_{1}) }  \,\delta.
\end{align}  
The weakly nonlinear solution also gives the Nusselt number near onset. In particular, 
\begin{align}
    \Nu - 1 \, = \, - \frac{\pd{z} \Theta|_{z=0}} {R} \, = \, K\, \delta, 
\end{align}
where we define the \textit{convective conductivity}, 
\begin{align}
K = \left. \frac{\partial \log \Nu}{\partial \log R}\right|_{R=R_{\rm c}} \, = \,  \frac{2 \pi  m^2}{a \left(\pi ^2 a^2+m^2\right)} \frac{\mathfrak{Re}(c_{0}) \,}{\mathfrak{Re} (c_{1}) } .
\end{align}
As the aspect ratio $\Gamma \to 0$ the calculation simplifies considerably and $K \to 12/7$. As the aspect ratio, $\Gamma \to \infty$, we can use a derivation along a flat Cartesian wall such that, 
\begin{align}
    K \, \sim \, \frac{0.7018222861264536}{\Gamma} \qquad \text{as} \qquad \Gamma \, \to \, \infty.
\end{align}
Like the conductivity, we find similar behaviour for the nonlinear frequency, $\omega = \omega_{\rm c} + \Omega$. We define the \textit{convective Doppler shift}, 
\begin{align}
     \Upsilon \, = \, \left. \frac{\partial \log \omega}{\partial \log R}\right|_{R=R_{\rm c}}.
\end{align}
As $\Gamma \to 0$, $\Upsilon \to 173/252$. Likewise, 
\begin{align}
\Upsilon \to 1.5336344674049933 \qquad \text{as} \qquad  \Gamma \to \infty.
\end{align}
Fig.~\ref{fig: Nu-Om-crit} shows plots of $K$ and $\Upsilon$ as a function of aspect ratio.

\subsection{\label{sec: Fully nonlinear} Fully nonlinear results}

We first use the methods described above to simulate the reduced wall-mode equations with no-slip boundary conditions, $R=2R_{\text{c}}$, and $\Gamma=4/5$.
For this choice of $\Gamma$, the most unstable mode at $R=R_{\text{c}}$ has an azimuthal wavenumber $m=2$ (fig.~\ref{fig: critical cylinder parameters}).
We initialise the simulation with random low-amplitude noise, which triggers the wall-mode instability.
At $R=2R_{\text{c}}$, the modes $m=2,3,4,5,6$ all have positive growth rates, and $m=3,4$ have the largest growth rates with $\gamma/\omega_{\text{c}} \approx 2.3$ and $\omega/\omega_{\text{c}} \approx 3.1, 2.7$ respectively.

\begin{figure}
\begin{center}
\vspace{+0.10in}
\includegraphics[width=\textwidth]{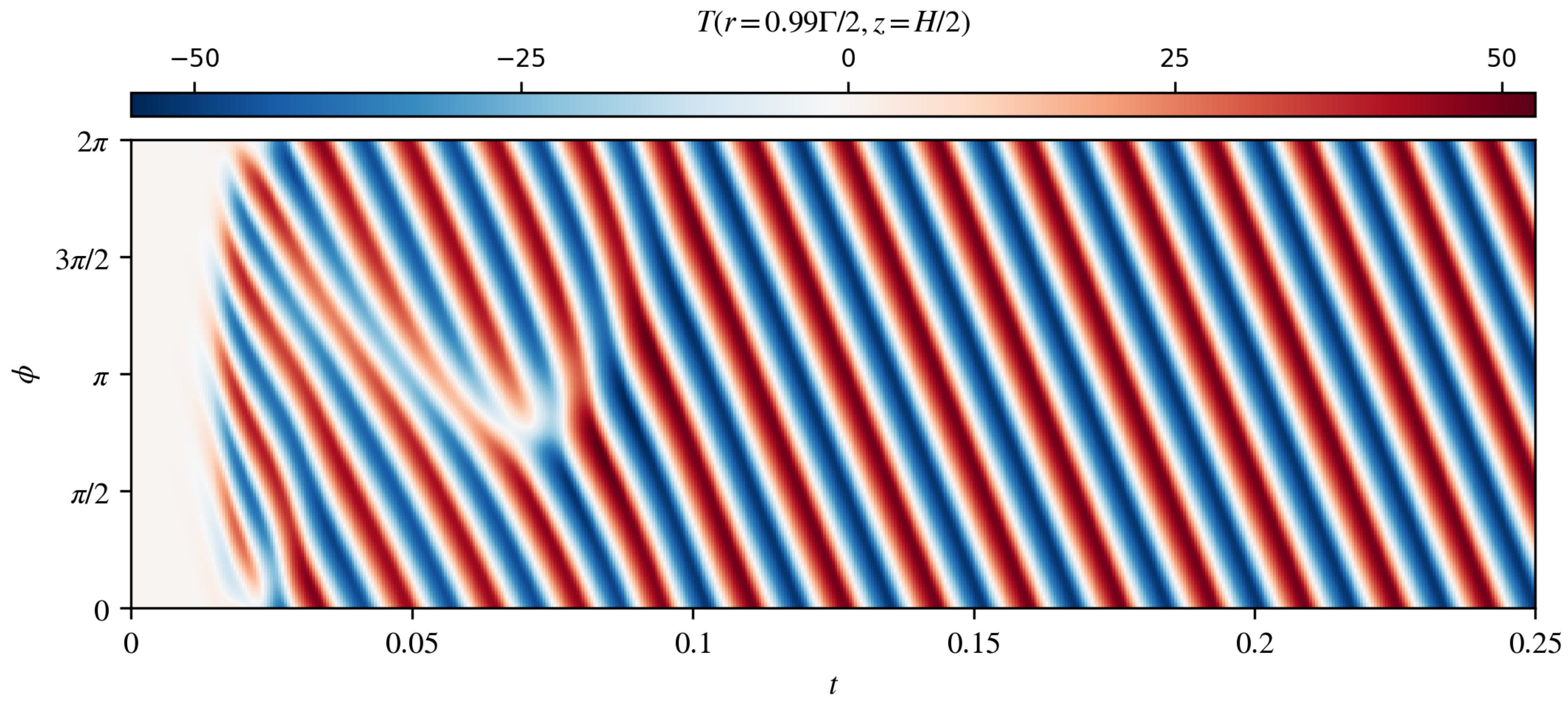}
\caption{Temperature space-time propagation diagram as a function of azimuth, $\phi$, and time, $t$. The pattern shows the $R/R_{\text{c}} = 2$, no-slip case. The simulation starts with low-amplitude noise. The modes $m=2,3,4,5,6$ are all unstable, with $m=3, 4$ having the largest growth rates.
The pattern starts with approximately three full wavelengths, undergoes a coarsening defect, and settles into a consistent two-wave pattern.}
\label{fig: hov}
\end{center}
\end{figure}

The instability saturates nonlinearly, producing characteristic patterns which propagate retrograde.
In fig.~\ref{fig: hov}, we plot the temperature near the wall and at mid-height as a function of $\phi$ and $t$.
The downward propagation of the constant temperature stripes shows the retrograde motion.
Initially, the dominant mode has $m=3$, but a coarsening event at $t\approx 0.075$ results in a steady $m=2$ pattern that persists for the rest of the simulation.
Others have also found coarsening dynamics in similar systems \citep{ecke_etal_1992,plaut_2003,scheel_etal_2003,choi_etal_2004,lopez_etal_2007,favier_knobloch_2020}.
In particular, \cite{ning_ecke_1993, liu_ecke_1997,liu_ecke_1999} study the coarsening dynamics for wall modes in great detail. They find that increasing the Rayleigh number slowly (as we do below) maintains $m$, but a rapid change in Rayleigh (e.g., starting a simulation with $R/R_{\text{c}}=2$ from noise) usually results in a decrease in $m$.
We chose $\Gamma=4/5$ because the simulations yield an $m=2$ pattern, implying more azimuthal resolution elements per wavelength than simulations with the same resolution but higher $m$.

Having found a steady propagating solution for no-slip boundary conditions and $R=2R_{\text{c}}$, we bootstrap simulations to progressively higher and lower $R$ and simulations with stress-free boundary conditions.
We run each simulation until it reaches a statistically-steady state.
Starting from the saturated state with a similar $R$ means that our subsequent simulations run much shorter than fig.~\ref{fig: hov}.
In each case, we find the $m=2$ structure persists.
Table~\ref{tab:sims} lists all simulations described in this paper and their spatial resolution.

\begin{table}
\begin{center}
\begin{tabular}{ |c|c|c|c|c|c|c|c|c| } 
   \ $R/R_{\text{c}}$ \ & \ \text{BC} \ &\ $N_{r}$ \ & \ $N_{\phi}$ \ & \ $N_{z}$ \ & \ $\Gamma$ \ & \ $\sigma$ \ & \ $\omega/\omega_{\text{c}}$ \ & \ $\Nu-1$ \ \\ 
 \hline
  \hline
1.01 & \textit{NS} & \ 32 \  & \ 96 \ & \ 48 \ & \ 4/5 \  &   & \ \ 1.01 \  &\  0.0081 \  \\ 
1.02 & \textit{NS} & 32      & 96 & 48 & 4/5  &  & \ 1.02 & 0.016 \\ 
1.05 & \textit{NS} & 32      & 96 & 48 & 4/5  &  & \ 1.05 & 0.041 \\ 
1.1  & \textit{NS} & 32      & 96 & 48 & 4/5  &    & \ 1.10 & 0.084 \\ 
1.2  & \textit{NS} & 32      & 96 & 48 & 4/5  &  & \ 1.19 & 0.18 \\ 
1.5  & \textit{NS} & 48      & 128 & 64 & 4/5  &  & \ 1.36 & 0.50 \\ 
2    & \textit{NS} & 64      & 128 & 64 & 4/5  &  & \ \ 1.39 \  & 1.1 \\  
3    & \textit{NS} & 128     & 256 & 64 & 4/5  &  & \ 0.88 & 2.1 \\ 
4    & \textit{NS} & 192     & 384 & 96 & 4/5  &  & \ 0.41 & 2.6 \\ 
5    & \textit{NS} & \ 256 \ & 512 & \ 128 \  & 4/5  &  & \ 0.04 & 2.8 \\ 
6    & \textit{NS} & 192 &  \ 384 \ & 96 & 4/5  &  & -0.30 & \ 2.9 \  \\
\hline 
2    & \textit{SF} & 64 & 128 & 64 & 4/5  & 1 & \ 1.58 & 0.79 \\ 
3    & \textit{SF} & 128 & 256 & 64 & 4/5 & 1 & \ 2.22 & 1.2 \\
4    & \textit{SF} & 192 & 512 & 96 & 4/5 & 1 & \ 3.12 & 1.5 \\ 
 \hline
 \hline
\end{tabular}
\end{center}
\caption{Summary of all simulation parameters. In every case, for $\Gamma = 4/5$, the critical scaled Rayleigh number, $R_{\text{c}} = 28.237851887421087$, and the critical frequency $\omega_{\text{c}} = 69.02735982770973$. For the weakly nonlinear parameters, 
$K = 0.8028404111646181$.
$\Upsilon = 1.0933189545586768$.
By convention, $\omega >0$ corresponds to \textit{retrograde} propagation, consistent with the linear instability calculation. For top/bottom boundary conditions, \textit{NS} is no-slip, and \textit{SF} is stress-free. While $R,\Gamma,\sigma$ are input control parameters, fig.~\ref{heat-transport plots} shows the output parameters, $\omega$ and $\Nu$. Note that the Prandtl number, $\sigma$ drops out of the system in the no-slip case.\label{tab:sims}}
\end{table}

\begin{figure} 
\begin{center} 
\includegraphics[width=\textwidth]{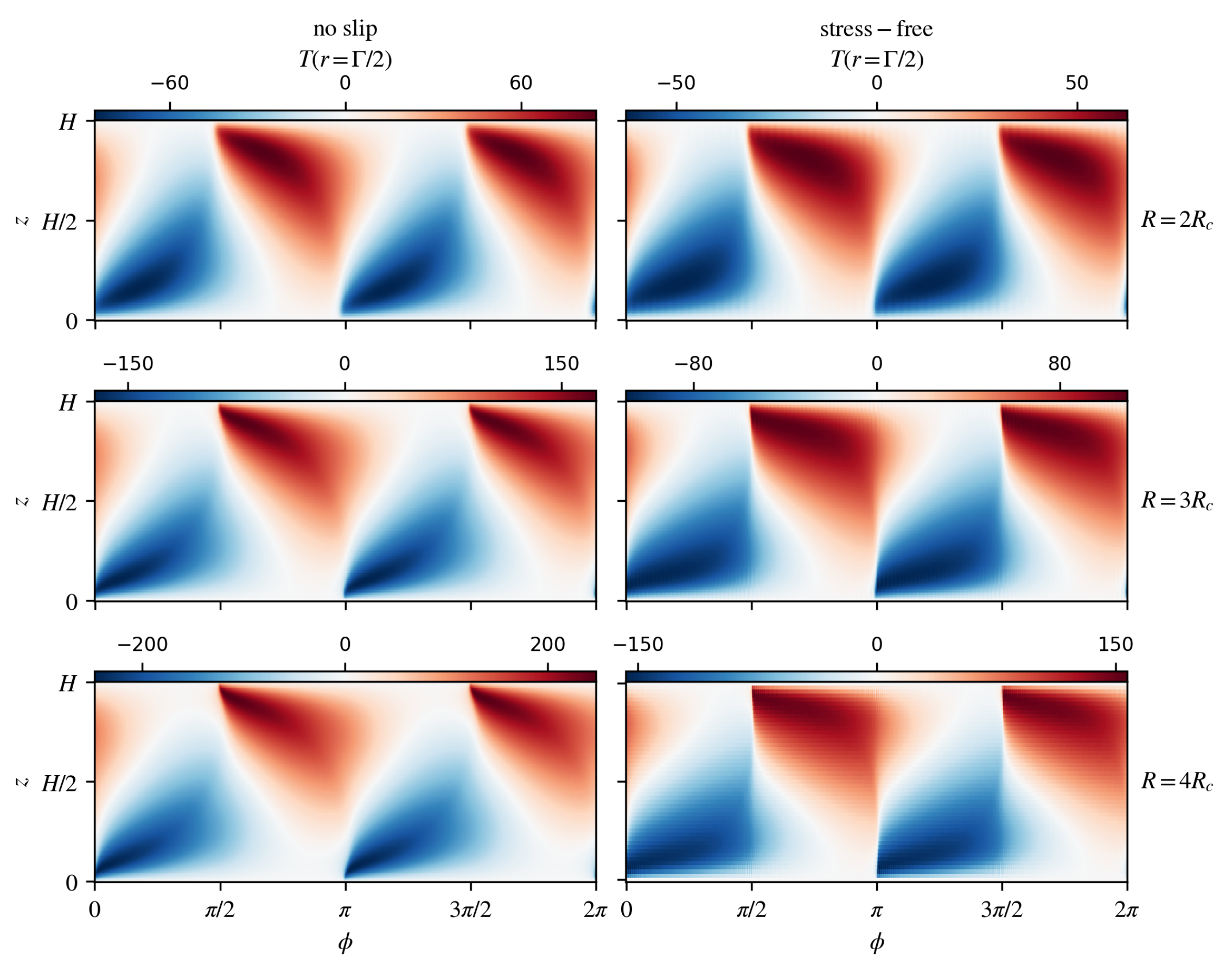}
\caption{Temperature patterns at the sidewall: The left column shows the no-slip case, and the right column shows the stress-free case. The rows show $R/R_{\text{c}} = 2,3,4$ respectively.}
\label{fig: temperature}
\end{center}
\end{figure}

Fig.~\ref{fig: temperature} illustrates the sidewall temperature pattern for different boundary conditions and $R$.
In all cases, alternating hot/cold spots exist near the top/bottom of the domain.
These patterns extend toward the mid-plane of the cylinder as $\phi$ increases past the location of the hot/cold spot.
The height of these patterns is smaller for no-slip boundary conditions than for stress-free boundary conditions.
As $R$ increases, the edges of the hot/cold spots sharpen,
producing front-like features that require higher and higher azimuthal resolution for increased $R$.
The stress-free simulations have sharper, more vertically aligned fronts.
For this reason, we could only run simulations up to $R=4R_{\text{c}}$ for stress-free boundary conditions, while we were able to reach higher values $R=6R_{\text{c}}$ in the no-slip simulations.

\begin{figure}
\begin{center}
\includegraphics[width=\textwidth]{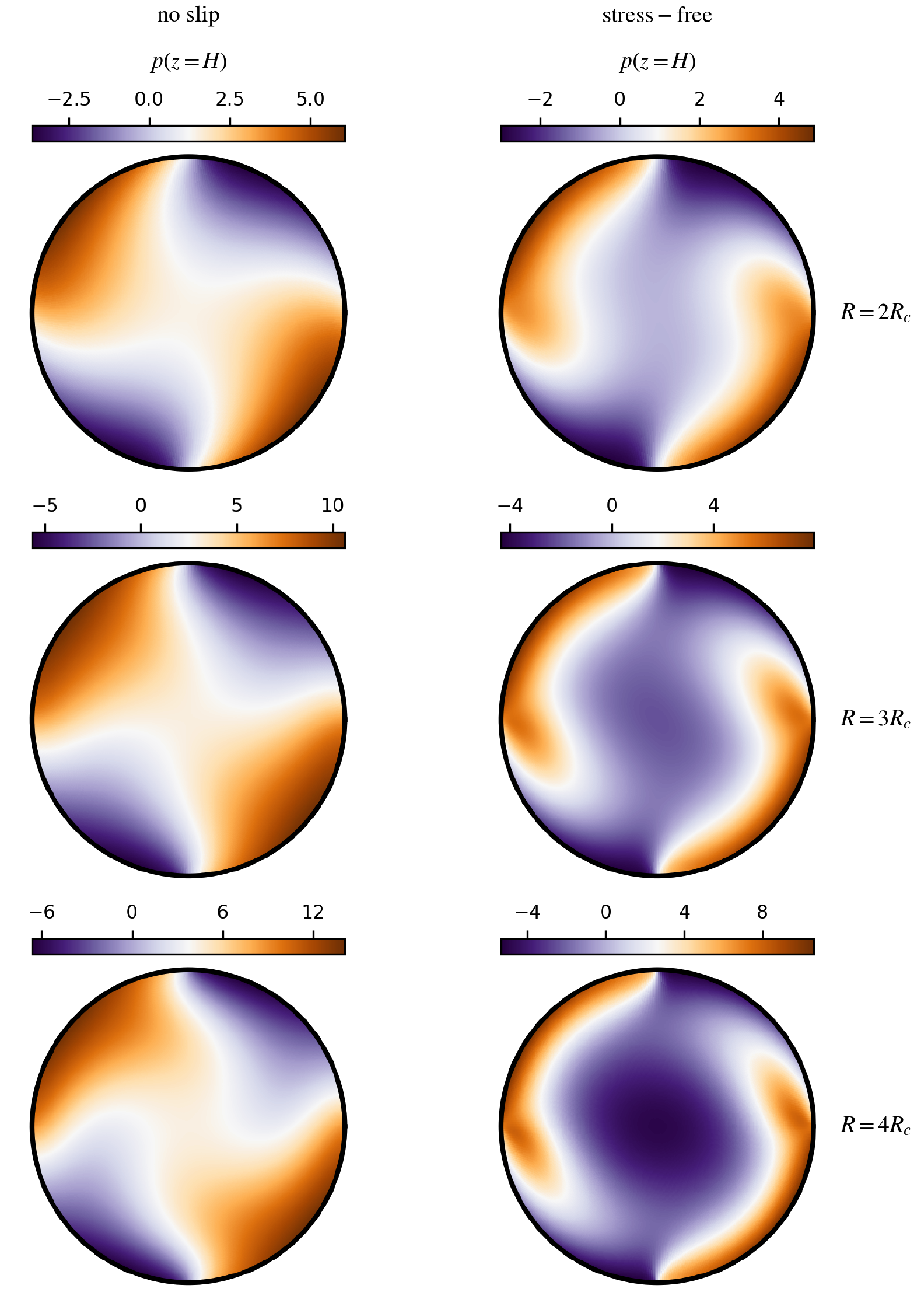}
\caption{Pressure patterns at the top: The left column shows the no-slip case, and the right column shows the stress-free case. The rows show $R/R_{\text{c}} = 2,3,4$ respectively.
}
\label{fig: pressure}
\end{center}
\end{figure}

To show the horizontal structure of the modes, we plot the pressure at the top of the cylinder in fig.~\ref{fig: pressure}.
The pressure perturbations extend significantly into the cylinder, although the features become increasingly localised to the walls at $R$ increases.
Recall that these visualisations do not directly show the Stewartson layers, which are infinitely thin in this analysis.
For stress-free boundary conditions, the pressure fields do not change substantially as $R$ increases, other than becoming more localised near the boundary.
However, there are more interesting differences between different $R$ for the no-slip boundary conditions.
For low $R$, the contours of constant pressure slope away from the walls in the counter-clockwise direction, as for the stress-free simulations.
However, for higher $R$, the no-slip simulations have contours of constant pressure that slope away from the walls in the clockwise direction.
This change in morphology may be related to the differences in propagation direction between the no-slip and stress-free simulations at higher $R$.

\begin{figure}
\begin{center}
\includegraphics[width=\textwidth]{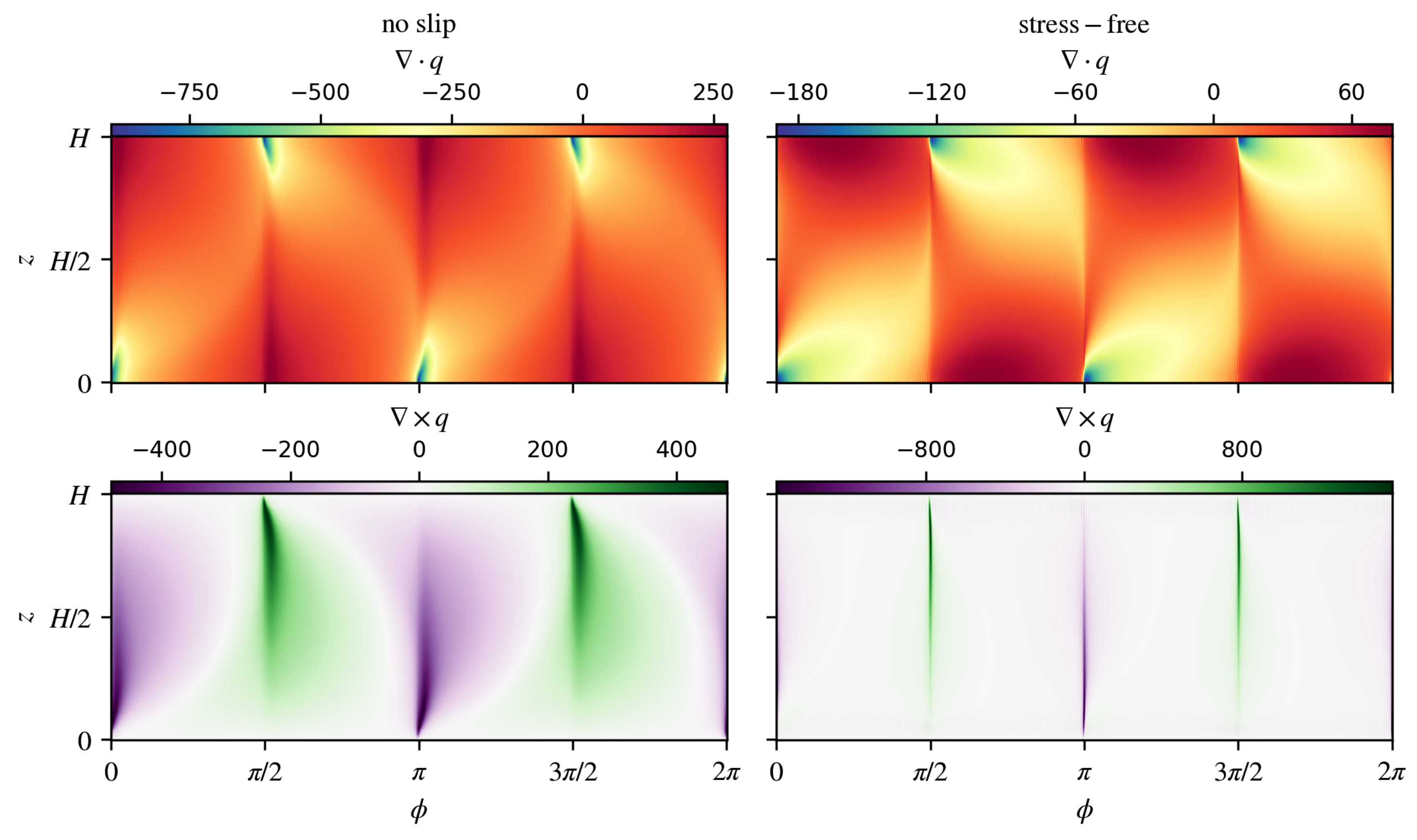}
\caption{Boundary-layer divergence and curl patterns at the sidewall: The left column shows the no-slip case, and the right column shows the stress-free case. Both cases correspond to $R/R_{\text{c}} =4$.
}
\label{fig: q}
\end{center}
\end{figure}

Fig.~\ref{fig: q} visualises the momentum fluxes within the Stewartson layers by plotting $\nabla\cdot q$ and $\nabla\times q$.
For the divergence, $\nabla\cdot q = u_{n}$, the normal velocity at the sidewall. For the curl, $\nabla\times q = \pd{\ell} q_{z} - \pd{z}q_{\ell}$.
The temperature at the sidewalls shows sharper azimuthal features in the stress-free simulations than in the no-slip simulations.
In particular, $\nabla \times q$ is very sharp and vertically aligned in the stress-free simulation.
In the no-slip simulation, $\nabla\cdot q$ and $\nabla\times q$ are much wider and show the same types of sloped features near the top and bottom of the cylinder as the temperature field.

Next, we focus on heat transport.
In the reduced wall-mode equations, the bulk has no vertical velocity.
Within the asymptotic approximation, the diffusive heat transport happens entirely in the bulk, while the convective heat transport happens entirely within the Stewartson layer.
In polar coordinates,
\begin{eqnarray}
F_{c} = \frac{1}{|\mathcal{A}|} \int_{0}^{2\pi}  \! q_{z} \Theta   \dd{\phi}, \qquad 
F_{d} = -\frac{1}{|\mathcal{A}|}\int_{0}^{2\pi}\!\!\int_{0}^{\frac{\Gamma}{2}} \! \pd{z} T \,r\dd{r}\dd{\phi},
\end{eqnarray}
where $|\mathcal{A}| = \pi \Gamma^{2}/4 = 4\pi/25 \approx 0.502$, and $\Theta = T - \overline{T}$.
These fluxes sum to the total $F_{t} = F_{c}+F_{d}$. The dynamical equations \eqss{theta evolution}{theta BCs} have the mean linear conduction temperature profile subtracted out, \textit{i.e.}, $- \pd{z} T_{0} = R$, therefore the Nussult number $\Nu = F_{t}/R$.

\begin{figure}
\begin{center}
\vspace{+0.10in}
\includegraphics[width=\textwidth]{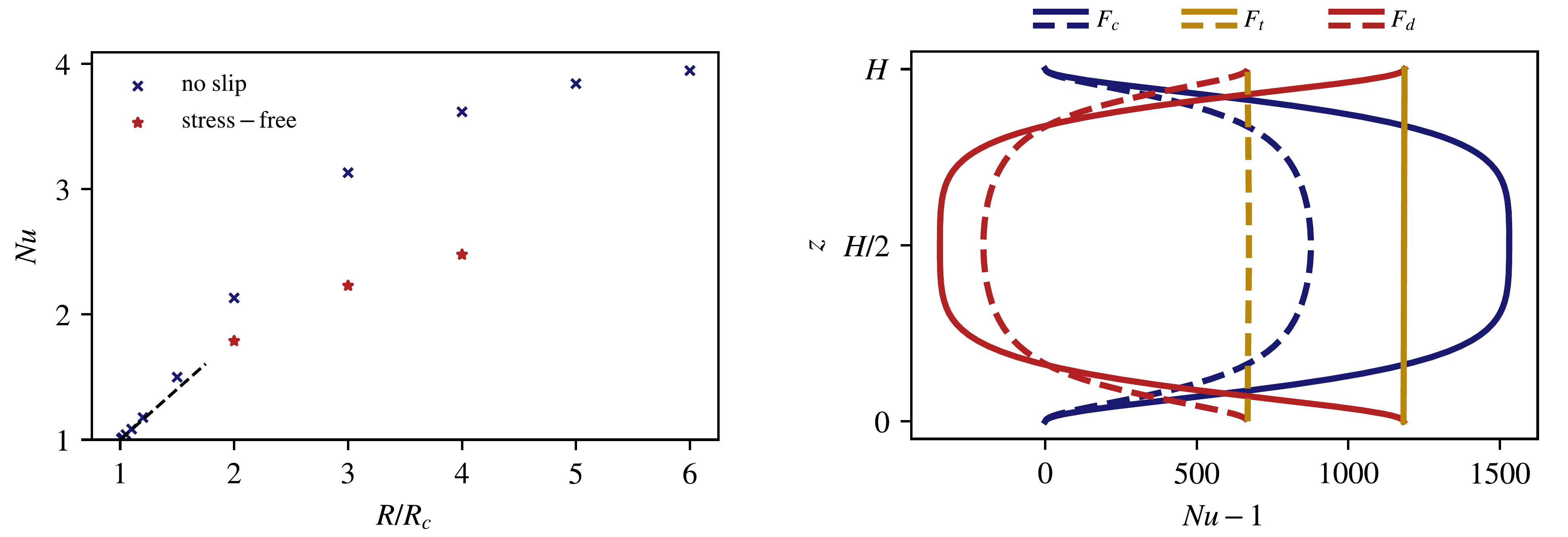}
\caption{Diagrams of heat transport: The left plot shows the Nusselt number, $\Nu$, versus the supercriticality, $R/R_{\text{c}}$; by definition $\Nu_{\,\text{c}} = 1$ at onset. The dashed line shows the approximate linear dependence $\Nu-1 \approx 0.807\,  (R/R_{\rm c}-1)$ near onset; weakly nonlinear predicts a slope of $\approx\,$0.803 (an $\approx\,$0.5\% difference). The right plot shows the depth-dependent diffusive ($F_{d}$, blue) and convective fluxes ($F_{c}$, red), along with their total ($F_{t} = F_{c} + F_{d}$, gold) for simulations with $R=4R_{\text{c}}$. The solid lines correspond to the no-slip case, and the dashed lines correspond to stress-free. The independence of $F_{t}$ on depth, $z$, implies the systems are in statistically steady state with a well-defined $\Nu$.
}
\label{heat-transport plots}
\end{center}
\end{figure}

The left panel of fig.~\ref{heat-transport plots} shows $\Nu$ as a function of $R/R_{\text{c}}$ for both no-slip and stress-free simulations.
As $R$ increases from $R_{\rm c}$, we find that $\Nu$ also increases linearly, as $\Nu - 1 \approx 0.807\,(R/R_{\rm c}-1)$, shown as a dashed line in the plot. Weakly nonlinear theory predicts a coefficient $\approx 0.8028404111646181$; see \S\ref{sec: Weakly nonlinear theory} and fig.~\ref{fig: Nu-Om-crit}.
As $R$ increases further, $\Nu$ appears to level off, although we have not been able to run at sufficiently high $R/R_{\ rmc}$ to probe the asymptotic behaviour as $R$ becomes large.

The right panel of fig.~\ref{heat-transport plots} shows the different depth-dependent heat fluxes for the no-slip and stress-free simulations with $R=4R_{\text{c}}$.
The convective heat flux is zero at the top and bottom boundaries ($q_z=0$) but becomes significant and roughly constant away from the horizontal boundaries.
The diffusive heat flux is large and positive in the top and bottom thermal boundary layers but intriguingly becomes negative near the mid-plane.
The (vertical) middle of the cylinder becomes stably stratified due to the wall modes.
All of our simulations are thermally equilibrated, where the total heat flux is constant with height.

\begin{figure}
\begin{center}
\vspace{+0.10in}
\includegraphics[width=9cm]{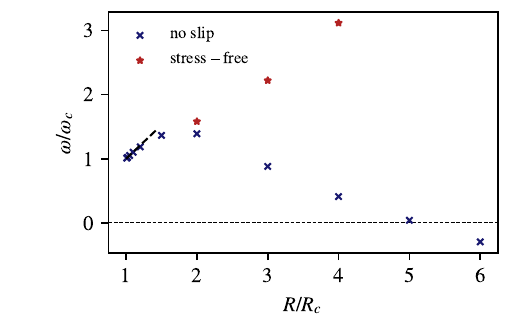}
\caption{The propagation rate, $\omega / \omega_{\text{c}}$ versus the supercriticality, $R/R_{\text{c}}$.  The dashed line shows the approximate linear dependence $\omega/\omega_c-1 \approx 1.05\,  (R/R_{\rm c}-1)$ near onset; compared to a predicted slope of $\approx\,$1.09. The rates continue to increase up to $R/R_{\text{c}} \approx 2$. At that point, the stress-free case continues the trend while the no-slip rate declines and eventually becomes prograde beyond $R/R_{\text{c}} \approx 5$.}
\label{fig: waves}
\end{center}
\end{figure}

Fig.~\ref{fig: hov} shows the instability of the wall modes saturates into nonlinear travelling waves.
At onset, we calculate the propagation rate via linear theory; for $\Gamma=4/5$, $\omega_{\text{c}} \approx 69$ in the retrograde direction.
Previous work has shown the propagation rate varies with $R$ as expected for a supercritical Hopf bifurcation \citep[e.g.][]{zhong_etal_1991,liu_ecke_1999,favier_knobloch_2020}.
Fig.~\ref{fig: waves} shows the propagation rate as a function of $R/R_{\text{c}}$ for no-slip and stress-free simulations.
As $R$ increases from $R_{\text{c}}$, the propagation rate $\omega$ also increases approximately linearly.
At $R=2R_{\text{c}}$, the propagation rates are similar for no-slip and stress-free simulations.
However, for larger $R$,  the propagation rate increases monotonically for stress-free simulations, whereas the propagation rate decreases for simulations with no-slip boundary conditions.
The experiments of \cite{zhong_etal_1993} at $\Ek \approx 10^{-4}$ similarly found a maximum propagation rate for $R/R_{\rm c} \approx \text{2-3}$.
At $R\approx5R_{\text{c}}$, there is a steady nonlinear solution with no propagation, and for higher $R$, the solution propagates in the prograde direction.
This switch in propagation direction may relate to the change in the azimuthal structure of the pressure perturbations shown in fig.~\ref{fig: pressure}.
\citet{horn_schmid_2017,ravichandran_wettlaufer_2024,zhang_etal_2024} all find mixtures of both prograde and retrograde propagation, depending on parameter values. 
Note that the structure of the temperature perturbations at the boundary (e.g., shown up to $R=4R_{\text{c}}$ in fig.~\ref{fig: temperature}) does not change significantly as $R$ increases, other than becoming increasingly sharp.

\begin{figure}
\begin{center}
\includegraphics[width=10cm]{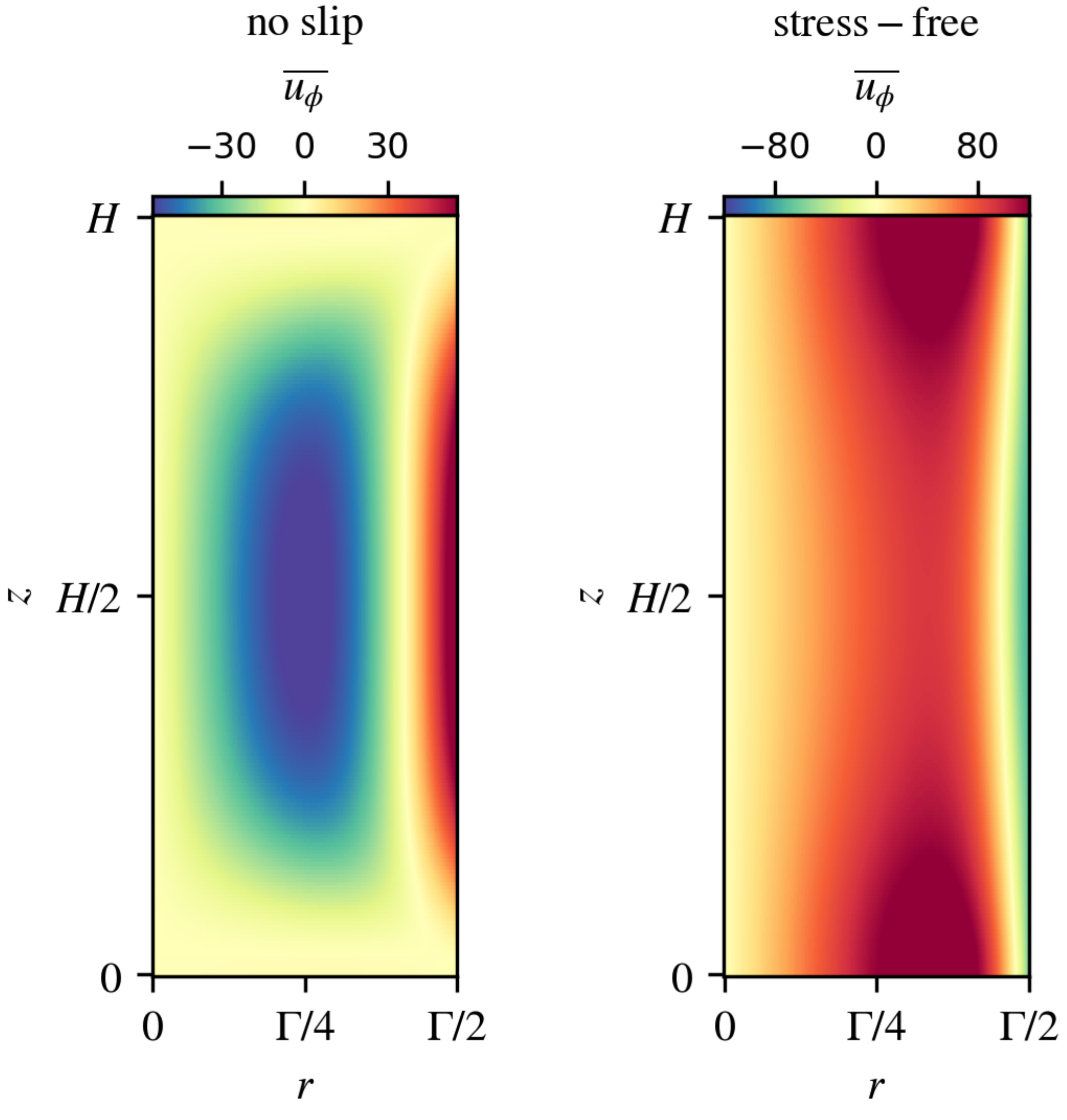}
\caption{Mean zonal flow in meridional planes: The left shows the no-slip case, and the right shows the stress-free case. Both cases correspond to $R/R_{\text{c}} =4$.
}
\label{fig: uphi}
\end{center}
\end{figure}

Finally, fig.~\ref{fig: uphi} shows the mean zonal flow for no-slip and stress-free simulations with $R=4R_{\text{c}}$.
At this $R$, the modes propagate retrograde in both simulations.
However, the mean zonal flow at the boundary is positive in the no-slip simulation, while it is (as expected) negative in the stress-free simulation.
The change in propagation direction for simulations with no-slip boundary conditions may be related to the development of a prograde mean zonal flow.

\section{\label{S: Conclusions} Discussion and Conclusions}

This work derives a balanced set of reduced equations governing the nonlinear development of convective wall-mode instabilities in rapidly rotating systems. Focusing on the essential dynamics within the bulk interior and the Stewartson boundary layers at the sidewalls captures the multiscale nature of wall-mode convection. The dynamics of the bulk interior diagnostically determine the small-scale behaviour within the boundary layers. In contrast, the sidewall layers feedback onto the interior through a nonlinear lateral heat-flux boundary condition, providing a closed and self-consistent system. The final reduced dynamics occupies the bulk interior and comprises geometrically prescribed scales. All boundary layers exist behind the scenes, their presence felt via a series of nontrivial boundary conditions that include nonlinear transport and non-local integral transforms. 

The bulk-boundary connection offers a clearer understanding of the interplay between multiscale dynamics, often obscured in full models due to their complexity. Systematically eliminating secondary effects isolates the dominant physical processes, enhancing explanatory power and providing avenues for investigating wall-mode convection in the strongly nonlinear regime. A key difference between our studies and previous high-resolution wall-mode simulations is that we find no evidence for secondary instability of the sharp propagating thermal structures. We do not know if this is because we have not reached high enough supercriticality (other work found secondary instabilities for $R/R_{\rm c}\gtrsim 10$; \citealt{favier_knobloch_2020,dewit_etal_2023,zhang_etal_2024}) or because the equations filter baroclinic inertial dynamics. We imagine future work methodically reintroducing dynamic ingredients that model various aspects of vortex separation, which can populate the interior with pseudo-convective structures. The first additional effect worth including is the $\Order{\Ek^{\,1/6}}$ Ekman flux correction in \eq{eq: ekman flux correction}.

We also reiterate that the final system of equations resembles boundary-forced planetary geostrophic baroclinic dynamics coupled with barotropic quasi-geostrophic vorticity. These connections point to ways systems can include more general effects, such as container topography. We also expect to include centrifugal buoyancy effects in future studies. We have also not considered the effect of a low Prandtl number, which would much more strongly emphasise inertia. Furthermore, these equations can incorporate additional physical effects, as described below.

\subsection{Double diffusion \label{S: Double diffusion}}

Given the simple physical principles involved, we can, in one way, generalise the thermal wall-mode instability almost by inspection. In the case of thermal-solutal buoyancy, we simply add an equation for the ``salt'' concentration, $s(t,x,y,z)$. The updated hydrostatic balance,
\begin{eqnarray}
    \theta - s = \pd{z}p.
\end{eqnarray}
The relevant linear bulk equation and boundary condition,
\begin{eqnarray}
\pd{t} s = \textit{Le}^{-1}\nabla^{2} s,\qquad \text{with}\qquad -q_{z}R_{S} = \textit{Le}^{-1} \pd{x}s\big|_{x=0}.
\end{eqnarray}
The total set of control parameters includes the solutal scaled Rayleigh number and the Lewis number,  
\begin{eqnarray}
R_{T} = \frac{g \,\alpha_{T}\, \Delta T \, H}{\kappa}, \qquad R_{S} = -\frac{g \, \alpha_{S} \, \Delta S\, H}{\kappa}, \qquad \textit{Le} = \frac{\kappa}{\lambda},
\end{eqnarray}
where $\lambda$ is the solute diffusivity, $\alpha_{S}$ is solutal the expansion coefficient, and $\Delta S$ is the total imposed concentration jump. As defined, $R_{T} > 0$ and $R_{S} > 0 $ both typically imply destabilising configurations. For a neutral-density background state, $R_{T} + R_{S} = 0$. In many common situations, $\textit{Le} \gg 1$; e.g. for salt and water $\textit{Le} \approx 10^{3}$.

Carrying out a nearly identical analysis to the purely thermal case, the condition for marginal stability is
\begin{eqnarray}
    \frac{R_{T}}{\sqrt{k^{2} + \pi^{2} + i \omega} } + \frac{R_{S} \textit{Le} }{ \sqrt{k^{2} + \pi^{2} + i \omega\textit{Le} } } = \frac{\pi  (ik + \pi)}{ik}, 
    \label{RT-RS-omega}
\end{eqnarray}
where $k$ is the $y$-direction Fourier wavenumber and $\omega$ is the complex-valued frequency. In this case, the dispersion relation shows similarities and differences with both the fingering and oscillatory regimes of double-diffusive convection \citep{garaud_2018}.
We leave the exploration of the critical $R_{T}, R_{S}$ curve for later work.

\subsection{Magnetism \label{S: Magnetism}}

Finally, we point out that magnetic wall modes have attracted significant recent attention  \citep{busse_2008,liu_etal_2018,grannan_etal_2022,xu_etal_2023,mccormack_etal_2023,teimurazov_etal_2024}. There are many ways these dynamics will interact with rapid rotational effects. 

In magnetohydrodynamics (MHD), the Chandrasekhar number \citep{chandrasekhar_1961} measures the influence of an imposed background magnetic field, 
\begin{eqnarray}
Q = \frac{B_{0}^{2} H^{2}}{\mu_{0} \rho_{0} \nu \eta},
\end{eqnarray}
where $B_{0}$ is the vertical field strength, $\mu_{0}$ is the permeability, $\rho_{0}$ is the average density, and $\eta$ is the magnetic diffusivity. 
In the large-$Q$ limit, several interesting overlaps occur with rapidly rotating parameters that point to extremely rich unexplored dynamics.

In particular, the equivalent MHD sidewall interactions will couple with the rotating version, provided the boundary layers have the same scaling in the same Rayleigh number regime. Indeed, \citet{busse_2008} shows wall-mode convection exists when
\begin{eqnarray}
\Ra \sim Q^{3/4} , & \qquad & \dd{x} \sim Q^{-1/4}, \\
\Ra  \sim \Ta^{-1/2} , & \qquad & \dd{x} \sim  \Ta^{-1/6}.
\end{eqnarray}
In both cases, the velocity amplitudes in the boundary layers scale $v,w \sim \dd{x}^{-1}$. 
Therefore, when $Q \sim \Ta^{2/3}$, we expect interesting nontrivial interactions with $\Order{1}$ bulk dynamics coupled to the sidewall via nonlinear and non-local boundary conditions. We should also expect magnetic double-diffusive effects \citep{silvers_etal_2009}.

However, even more fascinating complications can occur in the ``magnetostrophic" regime \citep{horn_aurnou_2022}. In this case, convection can arise purely in the bulk on $\mathcal{O}(1)$ length scales without reference to sidewalls. This regime requires a joint balance 
\begin{eqnarray}
\Ra \sim Q \sim \Ta^{1/2}.
\end{eqnarray}
Convection sets in when $Q^{2} \sim 3 \Ta$ and $\Ra \Ta^{-1/2} \sim 3 \sqrt{3}\, \pi^{2} \approx 51.284$, or $R/R_{\text{c}} \approx 1.6$ in terms of the wall modes.
The implication is that magnetostrophic convection will interact with the sidewall layers with a lower magnetic field strength than purely sidewall-catalysed magnetoconvection. If the wall modes can drive a kinematic dynamo, this is likely the regime the system would end up in. 

\bigskip

{\bf Acknowledgements.} The authors thank Jon Aurnou, Bob Ecke, Benjamin Favier, Ian Grooms, Susanne Horn, Eric King, Edgar Knobloch, Moritz Linkmann and Matthew McCormack for many helpful and insightful conversations on wall modes and related topics.  

We dedicate this work to our dear friend, Keith Julien --- you’ve done so much for so many in far too short a time. 

\bigskip

{\bf Code.} We use the Dedalus code and additional analysis tools written in Python and Mathematica. Beyond the main Dedalus installation, all scripts are available at GitHub (\url{https://github.com/geoffvasil/rapidly_rotating_wall_modes})

\bigskip

{\bf Funding.} K.J.B., D.L., J.S.O. and B.P.B. acknowledge support from the NASA HTMS grant 80NSSC20K1280 and the NASA OSTFL grant 80NSSC22K1738. D.L. is partially supported by Sloan Foundation grant FG-2024-21548, and Simons Foundation grant SFI-MPS-T-MPS-00007353.

\bigskip

{\bf Declaration of Interests.} The authors report no conflict of interest.

\pagebreak


\begin{thebibliography}{108}
\expandafter\ifx\csname natexlab\endcsname\relax\def\natexlab#1{#1}\fi

\bibitem[Ahlers {\em et~al.\/}(2009)Ahlers, Grossmann \&
  Lohse]{ahlers_etal_2009}
{\sc Ahlers, G., Grossmann, S. \& Lohse, D.} 2009 Heat transfer and large scale
  dynamics in turbulent {R}ayleigh-{B}{\'e}nard convection. {\em Reviews of
  Modern Physics\/} {\bf 81}~(2), 503–537.

\bibitem[Albaiz \& Hart(1990)]{albaiz_etal_1990}
{\sc Albaiz, A.O. \& Hart, J.E.} 1990 An experimental study of {S}tewartson
  layer separation. {\em Geophysical and Astrophysical Fluid Dynamics\/} {\bf
  54}~(3–4), 127–144.

\bibitem[Aurnou {\em et~al.\/}(2018)Aurnou, Bertin, Grannan, Horn \&
  Vogt]{aurnou_etal_2018}
{\sc Aurnou, J.M., Bertin, V., Grannan, A.M., Horn, S. \& Vogt, T.} 2018
  Rotating thermal convection in liquid gallium: multi-modal flow, absent
  steady columns. {\em Journal of Fluid Mechanics\/} {\bf 846}, 846–876.

\bibitem[Barcilon \& Pedlosky(1967{\natexlab{{\em
  a\/}}})]{barcilon_pedlosky_1967a}
{\sc Barcilon, V. \& Pedlosky, J.} 1967{\natexlab{{\em a\/}}} Linear theory of
  rotating stratified fluid motions. {\em Journal of Fluid Mechanics\/} {\bf
  29}~(1), 1–16.

\bibitem[Barcilon \& Pedlosky(1967{\natexlab{{\em
  b\/}}})]{barcilon_pedlosky_1967b}
{\sc Barcilon, V. \& Pedlosky, J.} 1967{\natexlab{{\em b\/}}} A unified linear
  theory of homogeneous and stratified rotating fluids. {\em Journal of Fluid
  Mechanics\/} {\bf 29}~(3), 609–621.

\bibitem[Benjamin(1967)]{benjamin_1967}
{\sc Benjamin, T.B.} 1967 Internal waves of permanent form in fluids of great
  depth. {\em Journal of Fluid Mechanics\/} {\bf 29}~(3), 559–592.

\bibitem[Boubnov \& Golitsyn(1986)]{boubnov_golitsyn_1986}
{\sc Boubnov, B.M. \& Golitsyn, G.S.} 1986 Experimental study of convective
  structures in rotating fluids. {\em Journal of Fluid Mechanics\/} {\bf
  167}~(1), 503.

\bibitem[Boubnov \& Golitsyn(1990)]{boubnov_golitsyn_1990}
{\sc Boubnov, B.M. \& Golitsyn, G.S.} 1990 Temperature and velocity field
  regimes of convective motions in a rotating plane fluid layer. {\em Journal
  of Fluid Mechanics\/} {\bf 219}~(1), 215.

\bibitem[Buell \& Catton(1983)]{buell_catton_1983}
{\sc Buell, J.C. \& Catton, I.} 1983 Effect of rotation on the stability of a
  bounded cylindrical layer of fluid heated from below. {\em The Physics of
  Fluids\/} {\bf 26}~(4), 892–896.

\bibitem[Burns {\em et~al.\/}(2022)Burns, Fortunato, Julien \&
  Vasil]{burns_etal_2022}
{\sc Burns, K.J., Fortunato, D., Julien, K. \& Vasil, G.M.} 2022 Corner cases
  of the tau method: symmetrically imposing boundary conditions on hypercubes.

\bibitem[Burns {\em et~al.\/}(2020)Burns, Vasil, Oishi, Lecoanet \&
  Brown]{burns_etal_2020}
{\sc Burns, K.J., Vasil, G.M., Oishi, J.S., Lecoanet, D. \& Brown, B.P.} 2020
  Dedalus: {A} flexible framework for numerical simulations with spectral
  methods. {\em Physical Review Research\/} {\bf 2}~(2).

\bibitem[Busse(2008)]{busse_2008}
{\sc Busse, F.H.} 2008 Asymptotic theory of wall-attached convection in a
  horizontal fluid layer with a vertical magnetic field. {\em Physics of
  Fluids\/} {\bf 20}~(2).

\bibitem[Chandrasekhar(1953)]{chandrasekhar_1953}
{\sc Chandrasekhar, S.} 1953 The instability of a layer of fluid heated below
  and subject to {C}oriolis forces. {\em Proceedings of the Royal Society of
  London. Series A. Mathematical and Physical Sciences\/} {\bf 217}~(1130),
  306–327.

\bibitem[Chandrasekhar(1961)]{chandrasekhar_1961}
{\sc Chandrasekhar, S.} 1961 {\em Hydrodynamic and Hydromagnetic Stability\/}.
  {\em International Series of Monographs on Physics\/} . Oxford.

\bibitem[Chassignet {\em et~al.\/}(2012)Chassignet, Cenedese \&
  J.]{chassignet_cenedese_verron_2012}
{\sc Chassignet, E.P., Cenedese, C. \& J., Verron} 2012 {\em Buoyancy-Driven
  Flows\/}. Cambridge University Press.

\bibitem[Choi {\em et~al.\/}(2004)Choi, Prasad, Camassa \&
  Ecke]{choi_etal_2004}
{\sc Choi, W., Prasad, D., Camassa, R. \& Ecke, R.E.} 2004 Traveling waves in
  rotating {R}ayleigh-{B}{\'e}nard convection. {\em Physical Review E\/} {\bf
  69}~(5).

\bibitem[Clune \& Knobloch(1993)]{clune_knobloch_1993}
{\sc Clune, T. \& Knobloch, E.} 1993 Pattern selection in rotating convection
  with experimental boundary conditions. {\em Physical Review E\/} {\bf
  47}~(4), 2536–2550.

\bibitem[Dawes(2001)]{dawes_2001}
{\sc Dawes, J.H.P.} 2001 Rapidly rotating thermal convection at low {P}randtl
  number. {\em Journal of Fluid Mechanics\/} {\bf 428}, 61–80.

\bibitem[Doering {\em et~al.\/}(2019)Doering, Toppaladoddi \&
  Wettlaufer]{doering_etal_2019}
{\sc Doering, C.R., Toppaladoddi, S. \& Wettlaufer, J.S.} 2019 Absence of
  evidence for the ultimate regime in two-dimensional {R}ayleigh-{B}{\'e}nard
  convection. {\em Physical Review Letters\/} {\bf 123}~(25).

\bibitem[Ecke \& Shishkina(2023)]{ecke_shishkina_2023}
{\sc Ecke, R.E. \& Shishkina, O.} 2023 Turbulent rotating
  {R}ayleigh–{B}{\'e}nard convection. {\em Annual Review of Fluid
  Mechanics\/} {\bf 55}~(1), 603–638.

\bibitem[Ecke {\em et~al.\/}(2022)Ecke, Zhang \& Shishkina]{ecke_etal_2022}
{\sc Ecke, R.E., Zhang, X. \& Shishkina, O.} 2022 Connecting wall modes and
  boundary zonal flows in rotating {R}ayleigh-{B}{\'e}nard convection. {\em
  Physical Review Fluids\/} {\bf 7}~(1).

\bibitem[Ecke {\em et~al.\/}(2024)Ecke, Zhang \& Shishkina]{ecke_etal_2024}
{\sc Ecke, R., Zhang, X. \& Shishkina, O.} 2024 {Wall modes for conducting
  sidewall boundary conditions in rotating {R}ayleigh-{B}{\'e}nard convection}.
  In {\em APS Division of Fluid Dynamics Meeting Abstracts\/}, {\em APS Meeting
  Abstracts\/} 005, p. C16.005.

\bibitem[Ecke {\em et~al.\/}(1992)Ecke, Zhong \& Knobloch]{ecke_etal_1992}
{\sc Ecke, R.E, Zhong, Fang \& Knobloch, E} 1992 {H}opf bifurcation with broken
  reflection symmetry in rotating {R}ayleigh-{B}{\'e}nard convection. {\em
  Europhysics Letters (EPL)\/} {\bf 19}~(3), 177–182.

\bibitem[Favier \& Knobloch(2020)]{favier_knobloch_2020}
{\sc Favier, B. \& Knobloch, E.} 2020 Robust wall states in rapidly rotating
  {R}ayleigh–{B}{\'e}nard convection. {\em Journal of Fluid Mechanics\/} {\bf
  895}.

\bibitem[Fein(1978)]{fein_1978}
{\sc Fein, J.S.} 1978 {\em Boundary Layers in Homogenous and
  Stratified-Rotating Fluids\/}. Gainesville, Florida: University Presses of
  Florida.

\bibitem[Garaud(2018)]{garaud_2018}
{\sc Garaud, P.} 2018 Double-diffusive convection at low {P}randtl number. {\em
  Annual Review of Fluid Mechanics\/} {\bf 50}~(1), 275–298.

\bibitem[Goldstein {\em et~al.\/}(1993)Goldstein, Knobloch, Mercader \&
  Net]{goldstein_etal_1993}
{\sc Goldstein, H.F., Knobloch, E., Mercader, I. \& Net, M.} 1993 Convection in
  a rotating cylinder. {P}art 1. {L}inear theory for moderate {P}randtl
  numbers. {\em Journal of Fluid Mechanics\/} {\bf 248}, 583–604.

\bibitem[Goldstein {\em et~al.\/}(1994)Goldstein, Knobloch, Mercader \&
  Net]{goldstein_etal_1994}
{\sc Goldstein, H.F., Knobloch, E., Mercader, I. \& Net, M.} 1994 Convection in
  a rotating cylinder. {P}art 2. {L}inear theory for low {P}randtl numbers.
  {\em Journal of Fluid Mechanics\/} {\bf 262}, 293–324.

\bibitem[Grannan {\em et~al.\/}(2022)Grannan, Cheng, Aggarwal, Hawkins, Xu,
  Horn, S{\'a}nchez-{\'A}lvarez \& Aurnou]{grannan_etal_2022}
{\sc Grannan, A.M., Cheng, J.S., Aggarwal, A., Hawkins, E.K., Xu, Y., Horn, S.,
  S{\'a}nchez-{\'A}lvarez, J. \& Aurnou, J.M.} 2022 Experimental pub crawl from
  {R}ayleigh–{B}{\'e}nard to magnetostrophic convection. {\em Journal of
  Fluid Mechanics\/} {\bf 939}.

\bibitem[Greenspan(1969)]{greenspan_1969}
{\sc Greenspan, H.} 1969 {\em The Theory of Rotating Fluid\/}. London:
  Cambridge.

\bibitem[Grooms {\em et~al.\/}(2011)Grooms, Julien \&
  Fox-Kemper]{grooms_etal_2011}
{\sc Grooms, I., Julien, K. \& Fox-Kemper, B.} 2011 On the interactions between
  planetary geostrophy and mesoscale eddies. {\em Dynamics of Atmospheres and
  Oceans\/} {\bf 51}~(3), 109–136.

\bibitem[Hart {\em et~al.\/}(2002)Hart, Kittelman \& Ohlsen]{hart_etal_2002}
{\sc Hart, J.E., Kittelman, S. \& Ohlsen, D.R.} 2002 Mean flow precession and
  temperature probability density functions in turbulent rotating convection.
  {\em Physics of Fluids\/} {\bf 14}~(3), 955–962.

\bibitem[Hashimoto(1976)]{hashimoto_1976}
{\sc Hashimoto, K.} 1976 On the stability of the {S}tewartson layer. {\em
  Journal of Fluid Mechanics\/} {\bf 76}~(2), 289–306.

\bibitem[van Heijst(1983)]{heijst_1983}
{\sc van Heijst, G.J.F.} 1983 The shear-layer structure in a rotating fluid
  near a differentially rotating sidewall. {\em Journal of Fluid Mechanics\/}
  {\bf 130}~(1), 1.

\bibitem[Herrmann \& Busse(1993)]{herrmann_busse_1993}
{\sc Herrmann, J. \& Busse, F.H.} 1993 Asymptotic theory of wall-attached
  convection in a rotating fluid layer. {\em Journal of Fluid Mechanics\/} {\bf
  255}, 183–194.

\bibitem[Hester \& Vasil(2023)]{hester_vasil_2023}
{\sc Hester, E.W. \& Vasil, G.M.} 2023 Orthogonal signed-distance coordinates
  and vector calculus near evolving curves and surfaces. {\em Proceedings of
  the Royal Society A: Mathematical, Physical and Engineering Sciences\/} {\bf
  479}~(2277).

\bibitem[Horn \& Aurnou(2022)]{horn_aurnou_2022}
{\sc Horn, S. \& Aurnou, J.M.} 2022 The {E}lbert range of magnetostrophic
  convection. {I}. {L}inear theory. {\em Proceedings of the Royal Society A:
  Mathematical, Physical and Engineering Sciences\/} {\bf 478}~(2264).

\bibitem[Horn \& Schmid(2017)]{horn_schmid_2017}
{\sc Horn, S. \& Schmid, P.J.} 2017 Prograde, retrograde, and oscillatory modes
  in rotating {R}ayleigh–{B}{\'e}nard convection. {\em Journal of Fluid
  Mechanics\/} {\bf 831}, 182–211.

\bibitem[Howard(1963)]{howard_1963}
{\sc Howard, L.N.} 1963 Heat transport by turbulent convection. {\em Journal of
  Fluid Mechanics\/} {\bf 17}~(03), 405.

\bibitem[Hoyle(2006)]{hoyle_2006}
{\sc Hoyle, R.} 2006 {\em {P}attern {F}ormation: {A}n {I}ntroduction to
  {M}ethods\/}. Cambridge University Press.

\bibitem[Julien {\em et~al.\/}(2016)Julien, Aurnou, Calkins, Knobloch, Marti,
  Stellmach \& Vasil]{julien_etal_2016}
{\sc Julien, K., Aurnou, J.M., Calkins, M.A., Knobloch, E., Marti, P.,
  Stellmach, S. \& Vasil, G.M.} 2016 A nonlinear model for rotationally
  constrained convection with {E}kman pumping. {\em Journal of Fluid
  Mechanics\/} {\bf 798}, 50–87.

\bibitem[Julien \& Knobloch(2007)]{julien_knobloch_2007}
{\sc Julien, K. \& Knobloch, E.} 2007 Reduced models for fluid flows with
  strong constraints. {\em Journal of Mathematical Physics\/} {\bf 48}~(6).

\bibitem[Julien {\em et~al.\/}(2012)Julien, Knobloch, Rubio \&
  Vasil]{julien_etal_2012}
{\sc Julien, K., Knobloch, E., Rubio, A.M. \& Vasil, G.M.} 2012 Heat transport
  in low-{R}ossby-number {R}ayleigh-{B}{\'e}nard convection. {\em Physical
  Review Letters\/} {\bf 109}~(25).

\bibitem[Julien {\em et~al.\/}(1996)Julien, Legg, Mcwilliams \&
  Werne]{julien_etal_1996}
{\sc Julien, K., Legg, S., Mcwilliams, J. \& Werne, J.} 1996 Rapidly rotating
  turbulent {R}ayleigh-{B}{\'e}nard convection. {\em Journal of Fluid
  Mechanics\/} {\bf 322}, 243–273.

\bibitem[Kerswell \& Barenghi(1995)]{kerswell_barenghi_1995}
{\sc Kerswell, R.R. \& Barenghi, C.F.} 1995 On the viscous decay rates of
  inertial waves in a rotating circular cylinder. {\em Journal of Fluid
  Mechanics\/} {\bf 285}, 203–214.

\bibitem[Kevorkian \& Cole(1981)]{kevorkian_cole_1981}
{\sc Kevorkian, J. \& Cole, J.D.} 1981 {\em Perturbation Methods in Applied
  Mathematics\/}. {\em Applied Mathematics Sciences Series\/} .
  Springer-Verlag.

\bibitem[King {\em et~al.\/}(2009)King, Stellmach, Noir, Hansen \&
  Aurnou]{king_etal_2009}
{\sc King, E.M., Stellmach, S., Noir, J., Hansen, U. \& Aurnou, J.M.} 2009
  Boundary layer control of rotating convection systems. {\em Nature\/} {\bf
  457}~(7227), 301–304.

\bibitem[Kunnen {\em et~al.\/}(2013)Kunnen, Clercx \& van
  Heijst]{kunnen_etal_2013}
{\sc Kunnen, R.P.J., Clercx, H.J.H. \& van Heijst, G.J.F.} 2013 The structure
  of sidewall boundary layers in confined rotating {R}ayleigh–{B}{\'e}nard
  convection. {\em Journal of Fluid Mechanics\/} {\bf 727}, 509–532.

\bibitem[Kunnen {\em et~al.\/}(2009)Kunnen, Geurts \& Clercx]{kunnen_etal_2009}
{\sc Kunnen, R.P.J., Geurts, B.J. \& Clercx, H.J.H.} 2009 Experimental and
  numerical investigation of turbulent convection in a rotating cylinder. {\em
  Journal of Fluid Mechanics\/} {\bf 642}, 445–476.

\bibitem[Kunnen {\em et~al.\/}(2011)Kunnen, Stevens, Overkamp, Sun, van Heijst
  \& Clercx]{kunnen_etal_2011}
{\sc Kunnen, R.P.J., Stevens, R.J.A.M., Overkamp, J., Sun, C., van Heijst, G.F.
  \& Clercx, H.J.H.} 2011 The role of {S}tewartson and {E}kman layers in
  turbulent rotating {R}ayleigh–{B}{\'e}nard convection. {\em Journal of
  Fluid Mechanics\/} {\bf 688}, 422–442.

\bibitem[Kuo \& Cross(1993)]{kuo_cross_1993}
{\sc Kuo, E.Y. \& Cross, M.C.} 1993 Traveling-wave wall states in rotating
  {R}ayleigh-{B}{\'e}nard convection. {\em Physical Review E\/} {\bf 47}~(4),
  R2245–R2248.

\bibitem[Li {\em et~al.\/}(2008)Li, Liao, Chan \& Zhang]{li_etal_2008}
{\sc Li, L., Liao, X., Chan, K.H. \& Zhang, K.} 2008 Linear and nonlinear
  instabilities in rotating cylindrical {R}ayleigh-{B}{\'e}nard convection.
  {\em Physical Review E\/} {\bf 78}~(5).

\bibitem[Liao {\em et~al.\/}(2006)Liao, Zhang \& Chang]{liao_etal_2006}
{\sc Liao, X., Zhang, K. \& Chang, Y.} 2006 On boundary-layer convection in a
  rotating fluid layer. {\em Journal of Fluid Mechanics\/} {\bf 549}~(1), 375.

\bibitem[Liu {\em et~al.\/}(2018)Liu, Krasnov \& Schumacher]{liu_etal_2018}
{\sc Liu, W., Krasnov, D. \& Schumacher, J.} 2018 Wall modes in
  magnetoconvection at high {H}artmann numbers. {\em Journal of Fluid
  Mechanics\/} {\bf 849}.

\bibitem[Liu \& Ecke(1997)]{liu_ecke_1997}
{\sc Liu, Y. \& Ecke, R.E.} 1997 {E}ckhaus-{B}enjamin-{F}eir instability in
  rotating convection. {\em Physical Review Letters\/} {\bf 78}~(23),
  4391–4394.

\bibitem[Liu \& Ecke(1999)]{liu_ecke_1999}
{\sc Liu, Y. \& Ecke, R.E.} 1999 Nonlinear traveling waves in rotating
  {R}ayleigh-{B}\'enard convection: {S}tability boundaries and phase diffusion.
  {\em Physical Review E\/} {\bf 59}, 4091--4105.

\bibitem[Liu \& Ecke(2011)]{liu_ecke_2011}
{\sc Liu, Y. \& Ecke, R.E.} 2011 Local temperature measurements in turbulent
  rotating {R}ayleigh-{B}{\'e}nard convection. {\em Physical Review E\/} {\bf
  84}~(1).

\bibitem[Lohse \& Shishkina(2024)]{lohse_shishkina_2024}
{\sc Lohse, D. \& Shishkina, O.} 2024 Ultimate {R}ayleigh-{B}énard turbulence.
  {\em Reviews of Modern Physics\/} {\bf 96}~(3).

\bibitem[Lopez {\em et~al.\/}(2007)Lopez, Marques, Mercader \&
  Batiste]{lopez_etal_2007}
{\sc Lopez, J.M., Marques, F., Mercader, I. \& Batiste, O.} 2007 Onset of
  convection in a moderate aspect-ratio rotating cylinder:
  {E}ckhaus–{B}enjamin–{F}eir instability. {\em Journal of Fluid
  Mechanics\/} {\bf 590}, 187–208.

\bibitem[Marques \& Lopez(2008)]{marques_lopez_2008}
{\sc Marques, F. \& Lopez, J.M.} 2008 Influence of wall modes on the onset of
  bulk convection in a rotating cylinder. {\em Physics of Fluids\/} {\bf
  20}~(2).

\bibitem[McCormack {\em et~al.\/}(2023)McCormack, Teimurazov, Shishkina \&
  Linkmann]{mccormack_etal_2023}
{\sc McCormack, M., Teimurazov, A., Shishkina, O. \& Linkmann, M.} 2023 Wall
  mode dynamics and transition to chaos in magnetoconvection with a vertical
  magnetic field. {\em Journal of Fluid Mechanics\/} {\bf 975}.

\bibitem[Nakagawa \& Frenzen(1955)]{nakagawa_frenzen_1955}
{\sc Nakagawa, Y. \& Frenzen, P.} 1955 A theoretical and experimental study of
  cellular convection in rotating fluids. {\em Tellus\/} {\bf 7}~(1), 1–21.

\bibitem[Niemela {\em et~al.\/}(2010)Niemela, Babuin \&
  Sreenivasan]{niemela_etal_2010}
{\sc Niemela, J.J., Babuin, S. \& Sreenivasan, K.R.} 2010 Turbulent rotating
  convection at high {R}ayleigh and {T}aylor numbers. {\em Journal of Fluid
  Mechanics\/} {\bf 649}, 509–522.

\bibitem[Niiler \& Bisshopp(1965)]{niiler_bisshopp_1965}
{\sc Niiler, P.P. \& Bisshopp, F.E.} 1965 On the influence of {C}oriolis force
  on onset of thermal convection. {\em Journal of Fluid Mechanics\/} {\bf
  22}~(04), 753.

\bibitem[Ning \& Ecke(1993)]{ning_ecke_1993}
{\sc Ning, L. \& Ecke, R.E.} 1993 Rotating {R}ayleigh-{B}{\'e}nard convection:
  {A}spect-ratio dependence of the initial bifurcations. {\em Physical Review
  E\/} {\bf 47}~(5), 3326–3333.

\bibitem[Ono(1975)]{ono_1975}
{\sc Ono, H.} 1975 Algebraic solitary waves in stratified fluids. {\em Journal
  of the Physical Society of Japan\/} {\bf 39}~(4), 1082–1091.

\bibitem[Pandey \& Sreenivasan(2024)]{pandey_2024}
{\sc Pandey, A. \& Sreenivasan, K.R.} 2024 Turbulent convection in rotating
  slender cells. {\em Journal of Fluid Mechanics\/} {\bf 999}.

\bibitem[Pedlosky(1987)]{pedlosky_1987}
{\sc Pedlosky, J.} 1987 {\em Geophysical Fluid Dynamics\/}, 2nd edn. New York:
  Springer-Verlag.

\bibitem[Pedlosky(2009)]{pedlosky_2009}
{\sc Pedlosky, J.} 2009 The response of a weakly stratified layer to buoyancy
  forcing. {\em Journal of Physical Oceanography\/} {\bf 39}~(4), 1060–1068.

\bibitem[Pfotenhauer {\em et~al.\/}(1987)Pfotenhauer, Niemela \&
  Donnelly]{pfotenhauer_1987}
{\sc Pfotenhauer, J.M., Niemela, J.J. \& Donnelly, R.J.} 1987 Stability and
  heat transfer of rotating cryogens. {P}art 3. {E}ffects of finite cylindrical
  geometry and rotation on the onset of convection. {\em Journal of Fluid
  Mechanics\/} {\bf 175}~(1), 85.

\bibitem[Plaut(2003)]{plaut_2003}
{\sc Plaut, E.} 2003 Nonlinear dynamics of traveling waves in rotating
  {R}ayleigh-{B}{\'e}nard convection: {E}ffects of the boundary conditions and
  of the topology. {\em Physical Review E\/} {\bf 67}~(4).

\bibitem[Ravichandran \& Wettlaufer(2023)]{ravichandran_wettlaufer_2024}
{\sc Ravichandran, S. \& Wettlaufer, J.S.} 2023 Prograde and meandering wall
  modes in rotating {R}ayleigh-{B}{\'e}nard convection with conducting walls.

\bibitem[Rossby(1969)]{rossby_1969}
{\sc Rossby, H.T.} 1969 A study of {B}{\'e}nard convection with and without
  rotation. {\em Journal of Fluid Mechanics\/} {\bf 36}~(2), 309–335.

\bibitem[Rubio {\em et~al.\/}(2009)Rubio, Lopez \& Marques]{rubio_etal_2009}
{\sc Rubio, A., Lopez, J.M. \& Marques, F.} 2009 Interacting oscillatory
  boundary layers and wall modes in modulated rotating convection. {\em Journal
  of Fluid Mechanics\/} {\bf 625}, 75–96.

\bibitem[Sakai(1997)]{sakai_1997}
{\sc Sakai, S.} 1997 The horizontal scale of rotating convection in the
  geostrophic regime. {\em Journal of Fluid Mechanics\/} {\bf 333}, 85–95.

\bibitem[Scheel {\em et~al.\/}(2003)Scheel, Paul, Cross \&
  Fischer]{scheel_etal_2003}
{\sc Scheel, J.D., Paul, M.R., Cross, M.C. \& Fischer, P.F.} 2003 Traveling
  waves in rotating {R}ayleigh-{B}{\'e}nard convection: {A}nalysis of modes and
  mean flow. {\em Physical Review E\/} {\bf 68}~(6).

\bibitem[Schonbek \& Vallis(1999)]{schonbek_vallis_1999}
{\sc Schonbek, M. \& Vallis, G.K.} 1999 Energy decay of solutions to the
  {B}oussinesq, primitive, and planetary geostrophic equations. {\em Journal of
  Mathematical Analysis and Applications\/} {\bf 234}~(2), 457–481.

\bibitem[Schumacher \& Sreenivasan(2020)]{schumacher_sreenivasan_2020}
{\sc Schumacher, J. \& Sreenivasan, K.R.} 2020 Colloquium: {U}nusual dynamics
  of convection in the {S}un. {\em Reviews of Modern Physics\/} {\bf 92}~(4).

\bibitem[Silvers {\em et~al.\/}(2009)Silvers, Vasil, Brummell \&
  Proctor]{silvers_etal_2009}
{\sc Silvers, L.J., Vasil, G.M., Brummell, N.H. \& Proctor, M.R.E.} 2009
  Double-diffusive instabilities of a shear-generated magnetic layer. {\em The
  Astrophysical Journal Letters\/} {\bf 702}~(1), L14–L18.

\bibitem[Spiegel \& Veronis(1960)]{spiegel_veronis_1960}
{\sc Spiegel, E.A. \& Veronis, G.} 1960 On the {B}oussinesq approximation for a
  compressible fluid. {\em The Astrophysical Journal\/} {\bf 131}, 442.

\bibitem[Sprague {\em et~al.\/}(2006)Sprague, Julien, Knobloch \&
  Werne]{sprague_etal_2006}
{\sc Sprague, M., Julien, K., Knobloch, E. \& Werne, J.} 2006 Numerical
  simulation of an asymptotically reduced system for rotationally constrained
  convection. {\em Journal of Fluid Mechanics\/} {\bf 551}~(1), 141.

\bibitem[Stellmach {\em et~al.\/}(2014)Stellmach, Lischper, Julien, Vasil,
  Cheng, Ribeiro, King \& Aurnou]{stellmach_etal_2014}
{\sc Stellmach, S., Lischper, M., Julien, K., Vasil, G.M., Cheng, J.S.,
  Ribeiro, A., King, E.M. \& Aurnou, J.M.} 2014 Approaching the asymptotic
  regime of rapidly rotating convection: {B}oundary layers versus interior
  dynamics. {\em Physical Review Letters\/} {\bf 113}~(25).

\bibitem[Stevens {\em et~al.\/}(2011)Stevens, Overkamp, Lohse \&
  Clercx]{stevens_2011}
{\sc Stevens, R.J.A.M., Overkamp, J., Lohse, D. \& Clercx, H.J.H.} 2011 Effect
  of aspect ratio on vortex distribution and heat transfer in rotating
  {R}ayleigh-{B}{\'e}nard convection. {\em Physical Review E\/} {\bf 84}~(5).

\bibitem[Stewartson(1957)]{stewartson_1957}
{\sc Stewartson, K.} 1957 On almost rigid rotations. {\em Journal of Fluid
  Mechanics\/} {\bf 3}~(1), 17–26.

\bibitem[Teimurazov {\em et~al.\/}(2024)Teimurazov, McCormack, Linkmann \&
  Shishkina]{teimurazov_etal_2024}
{\sc Teimurazov, A., McCormack, M., Linkmann, M. \& Shishkina, O.} 2024
  Unifying heat transport model for the transition between buoyancy-dominated
  and {L}orentz-force-dominated regimes in quasistatic magnetoconvection. {\em
  Journal of Fluid Mechanics\/} {\bf 980}.

\bibitem[Terrien {\em et~al.\/}(2023)Terrien, Favier \&
  Knobloch]{terrien_etal_2023}
{\sc Terrien, L., Favier, B. \& Knobloch, E.} 2023 Suppression of wall modes in
  rapidly rotating {R}ayleigh-{B}{\'e}nard convection by narrow horizontal
  fins. {\em Physical Review Letters\/} {\bf 130}~(17).

\bibitem[Vallis(2017)]{vallis_2017}
{\sc Vallis, Geoffrey~K.} 2017 {\em {A}tmospheric and {O}ceanic {F}luid
  {D}ynamics: {F}undamentals and {L}arge-{S}cale {C}irculation\/}. Cambridge
  University Press.

\bibitem[Vasil {\em et~al.\/}(2008{\natexlab{{\em a\/}}})Vasil, Brummell \&
  Julien]{vasil_etal_2008a}
{\sc Vasil, G.M., Brummell, N.H. \& Julien, K.} 2008{\natexlab{{\em a\/}}} A
  new method for fast transforms in parity-mixed {PDE}s: {P}art {I}.
  {N}umerical techniques and analysis. {\em Journal of Computational Physics\/}
  {\bf 227}~(17), 7999–8016.

\bibitem[Vasil {\em et~al.\/}(2008{\natexlab{{\em b\/}}})Vasil, Brummell \&
  Julien]{vasil_etal_2008b}
{\sc Vasil, G.M., Brummell, N.H. \& Julien, K.} 2008{\natexlab{{\em b\/}}} A
  new method for fast transforms in parity-mixed {PDE}s: {P}art {II}.
  {A}pplication to confined rotating convection. {\em Journal of Computational
  Physics\/} {\bf 227}~(17), 8017–8034.

\bibitem[Vasil {\em et~al.\/}(2016)Vasil, Burns, Lecoanet, Olver, Brown \&
  Oishi]{vasil_etal_2016}
{\sc Vasil, G.M., Burns, K.J., Lecoanet, D., Olver, S., Brown, B.P. \& Oishi,
  J.S.} 2016 Tensor calculus in polar coordinates using {J}acobi polynomials.
  {\em Journal of Computational Physics\/} {\bf 325}, 53–73.

\bibitem[Vasil {\em et~al.\/}(2021)Vasil, Julien \&
  Featherstone]{vasil_etal_2021}
{\sc Vasil, G.M., Julien, K. \& Featherstone, N.A.} 2021 Rotation suppresses
  giant-scale solar convection. {\em Proceedings of the National Academy of
  Sciences\/} {\bf 118}~(31).

\bibitem[Veronis(1968)]{veronis_1968}
{\sc Veronis, G.} 1968 Large-amplitude {B}{\'e}nard convection in a rotating
  fluid. {\em Journal of Fluid Mechanics\/} {\bf 31}~(1), 113–139.

\bibitem[Wang \& Ruuth(2008)]{wang_ruuth_2008}
{\sc Wang, D \& Ruuth, S~J} 2008 {Variable step-size implicit-explicit linear
  multistep methods for time-dependent partial differential equations}. {\em
  Journal of Computational Mathematics\/} .

\bibitem[Wedi {\em et~al.\/}(2022)Wedi, Moturi, Funfschilling \&
  Weiss]{wedi_etal_2022}
{\sc Wedi, M, Moturi, V.M., Funfschilling, D. \& Weiss, S.} 2022 Experimental
  evidence for the boundary zonal flow in rotating {R}ayleigh–{B}{\'e}nard
  convection. {\em Journal of Fluid Mechanics\/} {\bf 939}.

\bibitem[Weiss \& Ahlers(2011)]{weiss_ahlers_2011}
{\sc Weiss, S. \& Ahlers, G.} 2011 Heat transport by turbulent rotating
  {R}ayleigh–{B}{\'e}nard convection and its dependence on the aspect ratio.
  {\em Journal of Fluid Mechanics\/} {\bf 684}, 407–426.

\bibitem[de~Wit {\em et~al.\/}(2020)de~Wit, Aguirre~Guzmán, Madonia, Cheng,
  Clercx \& Kunnen]{dewit_etal_2020}
{\sc de~Wit, X.M., Aguirre~Guzmán, A.J., Madonia, M., Cheng, J.S., Clercx,
  H.J.H. \& Kunnen, R.P.J.} 2020 Turbulent rotating convection confined in a
  slender cylinder: {T}he sidewall circulation. {\em Physical Review Fluids\/}
  {\bf 5}~(2).

\bibitem[de~Wit {\em et~al.\/}(2023)de~Wit, Boot, Madonia, Aguirre~Guzmán \&
  Kunnen]{dewit_etal_2023}
{\sc de~Wit, X.M., Boot, W.J.M., Madonia, M., Aguirre~Guzmán, A.J. \& Kunnen,
  R.P.J.} 2023 Robust wall modes and their interplay with bulk turbulence in
  confined rotating {R}ayleigh-{B}{\'e}nard convection. {\em Physical Review
  Fluids\/} {\bf 8}~(7).

\bibitem[Xu {\em et~al.\/}(2023)Xu, Horn \& Aurnou]{xu_etal_2023}
{\sc Xu, Y., Horn, S. \& Aurnou, J.M.} 2023 Transition from wall modes to
  multimodality in liquid gallium magnetoconvection. {\em Physical Review
  Fluids\/} {\bf 8}~(10).

\bibitem[Zhan {\em et~al.\/}(2009)Zhan, Liao, Zhu \& Zhang]{zhan_etal_2009}
{\sc Zhan, X., Liao, X., Zhu, R. \& Zhang, K.} 2009 Convection in rotating
  annular channels heated from below. {P}art 3: {E}xperimental boundary
  conditions. {\em Geophysical and Astrophysical Fluid Dynamics\/} {\bf 103},
  443--466.

\bibitem[Zhang \& Liao(2009)]{zhang_liao_2009}
{\sc Zhang, K. \& Liao, X.} 2009 The onset of convection in rotating circular
  cylinders with experimental boundary conditions. {\em Journal of Fluid
  Mechanics\/} {\bf 622}, 63--73.

\bibitem[Zhang {\em et~al.\/}(2007)Zhang, Liao \& Busse]{zhang_etal_2007}
{\sc Zhang, K., Liao, X. \& Busse, F.H.} 2007 Asymptotic theory of inertial
  convection in a rotating cylinder. {\em Journal of Fluid Mechanics\/} {\bf
  575}, 449–471.

\bibitem[Zhang \& Roberts(1998)]{zhang_roberts_1998}
{\sc Zhang, K. \& Roberts, P.H.} 1998 A note on stabilising and destabilising
  effects of {E}kman boundary layers. {\em Geophysical and Astrophysical Fluid
  Dynamics\/} {\bf 88}, 215--223.

\bibitem[Zhang {\em et~al.\/}(2021)Zhang, Ecke \& Shishkina]{zhang_etal_2021}
{\sc Zhang, X., Ecke, R.E. \& Shishkina, O.} 2021 Boundary zonal flows in
  rapidly rotating turbulent thermal convection. {\em Journal of Fluid
  Mechanics\/} {\bf 915}.

\bibitem[Zhang {\em et~al.\/}(2020)Zhang, van Gils, Horn, Wedi, Zwirner,
  Ahlers, Ecke, Weiss, Bodenschatz \& Shishkina]{zhang_etal_2020}
{\sc Zhang, X., van Gils, D.P.M., Horn, S., Wedi, M., Zwirner, L., Ahlers, G.,
  Ecke, R.E., Weiss, S., Bodenschatz, E. \& Shishkina, O.} 2020 Boundary zonal
  flow in rotating turbulent {R}ayleigh-{B}\'enard convection. {\em Physical
  Review Letters\/} {\bf 124}, 084505.

\bibitem[Zhang {\em et~al.\/}(2024)Zhang, Reiter, Shishkina \&
  Ecke]{zhang_etal_2024}
{\sc Zhang, X., Reiter, P., Shishkina, O. \& Ecke, R.E.} 2024 Wall modes and
  the transition to bulk convection in rotating {R}ayleigh-{B}{\'e}nard
  convection. {\em Physical Review Fluids\/} {\bf 9}~(5).

\bibitem[Zhong {\em et~al.\/}(1991)Zhong, Ecke \& Steinberg]{zhong_etal_1991}
{\sc Zhong, F., Ecke, R.E. \& Steinberg, V.} 1991 Asymmetric modes and the
  transition to vortex structures in rotating {R}ayleigh-{B}{\'e}nard
  convection. {\em Physical Review Letters\/} {\bf 67}, 2473--2476.

\bibitem[Zhong {\em et~al.\/}(1993)Zhong, Ecke \& Steinberg]{zhong_etal_1993}
{\sc Zhong, F., Ecke, R.E. \& Steinberg, V.} 1993 Rotating
  {R}ayleigh-{B}{\'e}nard convection: {A}symmetri modes and vortex states. {\em
  Journal of Fluid Mechanics\/} {\bf 249}, 135--159.

\bibitem[Zhong \& Ahlers(2010)]{zhong_ahlers_2010}
{\sc Zhong, J.-Q. \& Ahlers, G.} 2010 Heat transport and the large-scale
  circulation in rotating turbulent {R}ayleigh-{B}{\'e}nard convection. {\em
  Journal of Fluid Mechanics\/} {\bf 665}, 300--333.

\end{thebibliography}
\end{document}